\documentclass[12pt]{article}
 \usepackage{xcolor} 

\usepackage{lipsum}
\usepackage[margin=1in]{geometry}

\usepackage{amsmath, amssymb, amsthm}
\usepackage{hyperref}
\usepackage{dsfont}
\usepackage[utf8]{inputenc}
\usepackage{natbib}
\usepackage{graphicx}
\usepackage{caption}
\usepackage{comment}
\usepackage{subcaption}
\usepackage{soul}
\usepackage[vlined,ruled,resetcount]{algorithm2e}
\allowdisplaybreaks

\usepackage{color}
\usepackage{float}
\usepackage{placeins}
\newcommand{\V}{V}
\newcommand{\E}{E}

\renewcommand{\k}{k}

\newcommand{\xeq}{x^{*}}
\newcommand{\qeq}{q^{*}}

\newcommand{\partition}{z}

\newcommand{\T}{\mathcal{T}}
\renewcommand{\S}{\mathcal{S}}

\newcommand{\price}{p}
\newcommand{\K}{\mathcal{K}}
\newcommand{\signal}{t}
\newcommand{\Signal}{\mathcal{T}}
\newcommand{\rate}{r}
\newcommand{\Vn}{\V_n}
\newcommand{\state}{s}
\newcommand{\Kplus}{K}
\newcommand{\dl}{\underline{d}}
\renewcommand{\dh}{\bar{d}}
\newcommand{\nh}{\bar{n}}
\newcommand{\zl}{\underline{z}}
\newcommand{\zh}{\bar{z}}
\newcommand{\zmean}{z^*}
\renewcommand{\nl}{\underline{n}}
\newcommand{\Kminus}{\tilde{K}}
\newcommand{\Treeminus}{\tilde{\Gamma}}
\newcommand{\qeqp}{q^{*'}}

\newcommand{\zkl}{\underline{z}_{\ell}}
\newcommand{\zkph}{\bar{z}_{\ell'}}
\newcommand{\kltangent}{k_{\ell}^{*}}
\newcommand{\klptangent}{k_{\ell'}^{*}}
\newcommand{\tangentk}{\barstate_0[\kltangent]}
\newcommand{\intervalj}{I_j}
\newcommand{\px}{p_x}
\newcommand{\py}{p_y}
\newcommand{\zjp}{z_j^{'}}
\newcommand{\zjpp}{z_j^{"}}
\newcommand{\zjhatp}{z_{\jhat}^{'}}
\newcommand{\zjhatpp}{z_{\jhat}^{"}}
\newcommand{\jhat}{j}
\newcommand{\intervaljhat}{I_{\jhat}}
\newcommand{\zja}{z_{j,a}}
\newcommand{\zjb}{z_{j,b}}
\newcommand{\zjhata}{z_{\jhat,a}}
\newcommand{\zjhatb}{z_{\jhat,b}}
\newcommand{\J}{J}
\newcommand{\zdag}{z^{\dagger}}
\newcommand{\zddag}{z^{\ddagger}}
\newcommand{\epa}{\epsilon_a}
\newcommand{\epb}{\epsilon_b}
\renewcommand{\j}{j}
\newcommand{\tangentkp}{\barstate_0[\klptangent]}

\newcommand{\Treedag}{\Gamma^{\dagger}}
\newcommand{\Tree}{\Gamma}
\newcommand{\Treeplus}{\Tree}

\newcommand{\revenue}{R}

\newcommand{\qwe}{q^{*}}

\newcommand{\barrevenue}{R}

\newcommand{\distance}{d}
\newcommand{\peq}{\price^{*}}
\newcommand{\barstate}{\state}
\usepackage{setspace}

\newtheorem{theorem}{Theorem}
\newtheorem{lemma}{Lemma}
\newtheorem{proposition}{Proposition}
\newtheorem{corollary}{Corollary}

\newtheorem{assumption}{Assumption}
\theoremstyle{definition}

\newtheorem{definition}{Definition}

\DeclareMathOperator*{\argmin}{arg\,min}

\DeclareMathOperator*{\argmax}{arg\,max}

\usepackage{xcolor}

\SetKw{KwBy}{by}

\newcommand{\changed}[1]{{\color{blue} {#1}}}

\newlength{\bibitemsep}
\newlength{\bibparskip}
\let\oldthebibliography\thebibliography
\renewcommand\thebibliography[1]{%
  \oldthebibliography{#1}%
  \setlength{\parskip}{\bibitemsep}%
  \setlength{\itemsep}{\bibparskip}%
}

\providecommand{\keywords}[1]
{\smallskip 
  \small	
  \noindent\textbf{\textit{Keywords: }} #1
}

\begin{document}
\title{Information Design for Spatial Resource Allocation}
%
%
\author{Ozan Candogan \\
  \footnotesize{University of Chicago, Booth School of Business, ozan.candogan@chicagobooth.edu} \and
Manxi Wu \\
  \footnotesize{Cornell University, Operations Research and Information Engineering, manxiwu@cornell.edu} \\}
\date{}
\maketitle     
 \begin{abstract}
In this paper, we study platforms where resources and jobs are spatially distributed, and resources have the flexibility to strategically move to different locations for better payoffs. The price of the service at each location depends on the number of resources present and the market size, which is modeled as a random state. Our focus is on how the platform can utilize information about the underlying state to influence resource repositioning decisions and ultimately increase commission revenues. We establish that in many practically relevant settings a simple monotone partitional information disclosure policy is optimal. This policy reveals state realizations below a threshold and above a second (higher) threshold, and pools all states in between and maps them to a unique signal realization. We also provide algorithmic approaches for obtaining (near-)optimal information structures that are monotone partitional in  general settings.

{\keywords{Information design, spatial resource allocation, monotone partitional information mechanisms.}}
\end{abstract}
\section{Introduction}


In many operational settings, resources that serve jobs are spatially distributed, and mismatch between the locations of jobs and resources causes inefficiencies. To complicate things further, in recent years, platforms where resources are self-interested independent contractors have become prevalent. In these platforms the resources decide whether or not to provide service and how to relocate from one region to another. This may further exacerbate the spatial frictions increasing the aforementioned inefficiencies.
Such dynamics are common and increasingly relevant in many platforms,  including
ride-sharing platforms (such as Uber and Lyft), 
online freight marketplaces (such as Convoy and Uber Freight),
food delivery services (such as DoorDash and GrubHub),
and other last mile delivery services (such as Amazon Flex, Gopuff, and Instacart).

\begin{figure}[h]
\centering
\begin{subfigure}[t]{0.3\textwidth}
\includegraphics[width=0.9\linewidth, height=8cm]{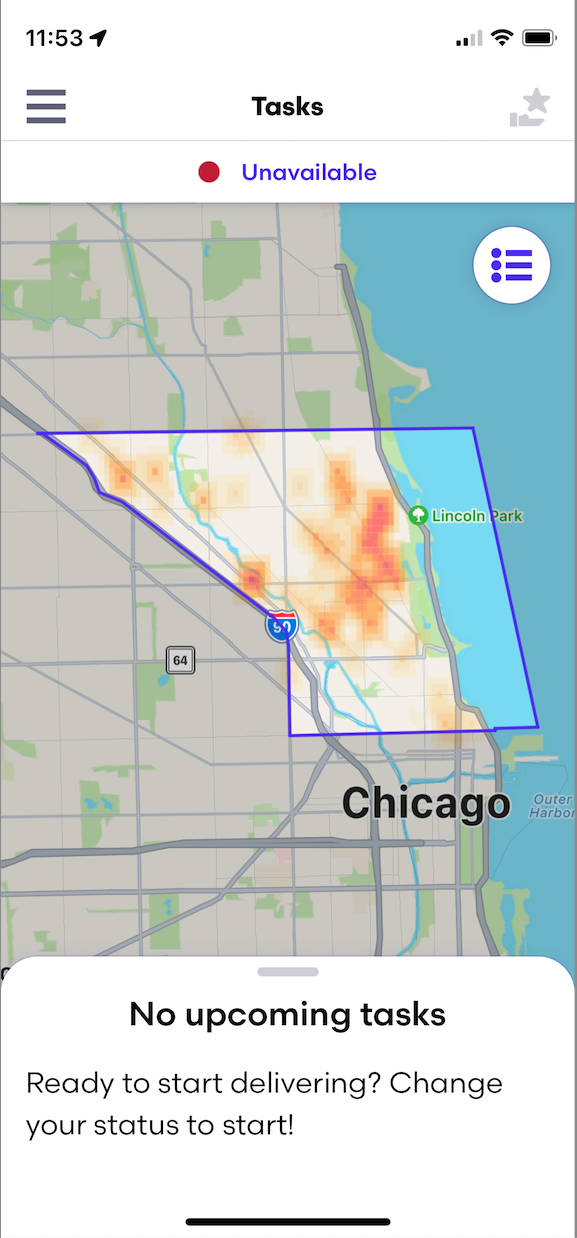} 
\caption{Grubhub}
\label{fig:subim1}
\end{subfigure}
\begin{subfigure}[t]{0.3\textwidth}
\includegraphics[width=0.9\linewidth, height=8cm]{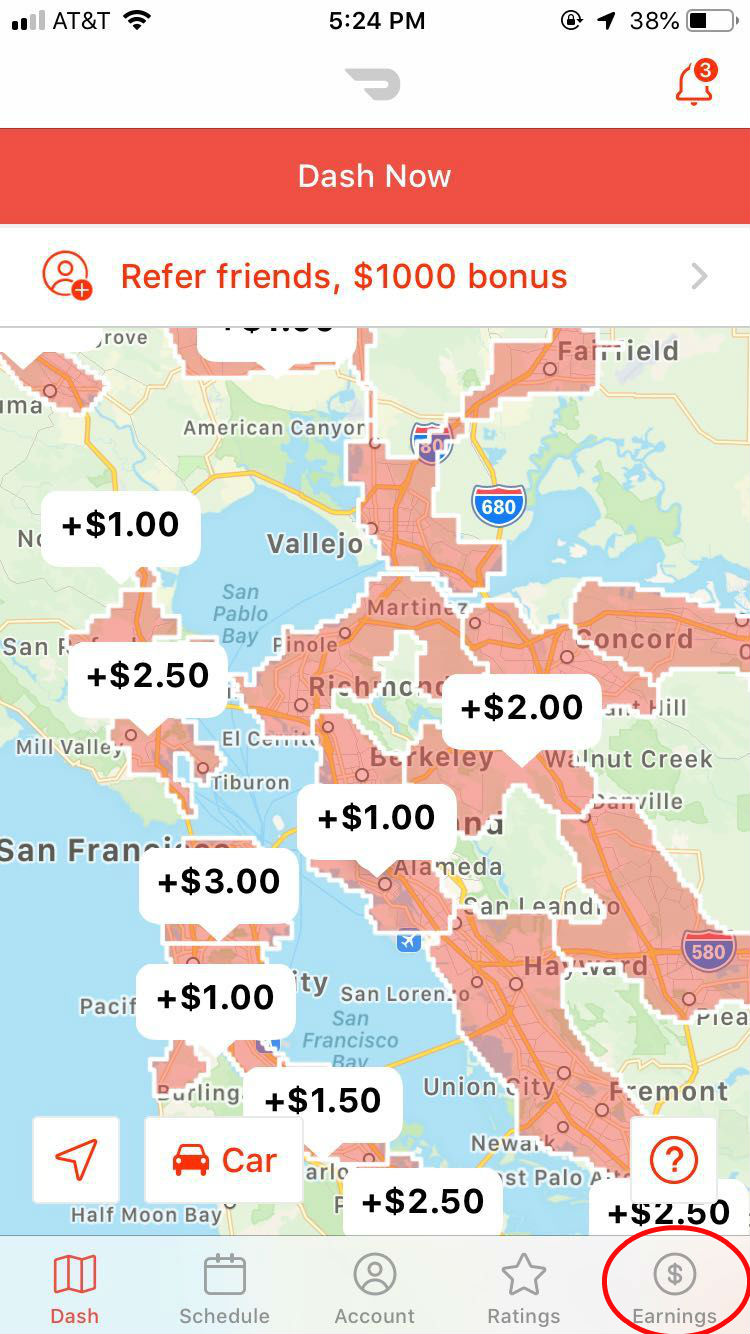}
\caption{Doordash}
\label{fig:subim3}
\end{subfigure}
\begin{subfigure}[t]{0.3\textwidth}
\includegraphics[width=0.9\linewidth, height=7cm]{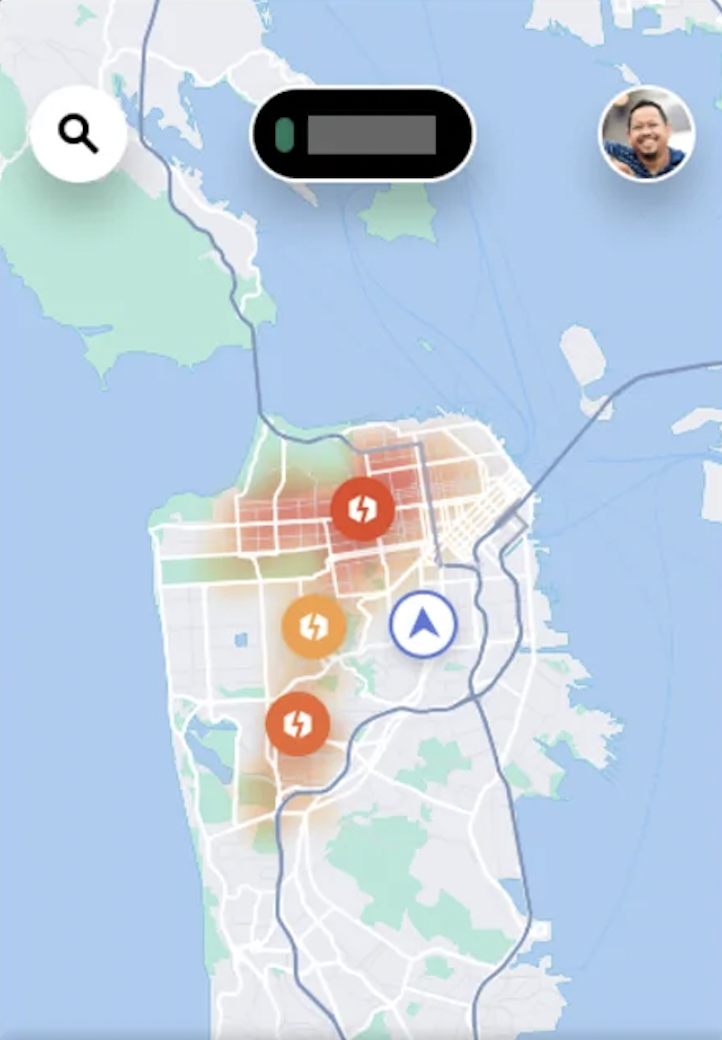}
\caption{Uber}
\label{fig:subim2}
\end{subfigure}
\caption{Examples of heat maps used by different platforms.}
\label{fig:heatmaps}
\end{figure}

To combat the mismatch of supply and demand in settings with self-interested resources, platforms have taken two main approaches. The first, ``market driven'', approach relies on offering different prices for services that take place in different locations. Surge pricing that is common in ride-sharing is an example,  and in fact such spatial price differences predate platforms and have been employed in freight brokerage industry while determining freight rates.
The second approach involves sharing information with the resources about the spatial distribution of jobs, so that the resources can reposition themselves to locations that are in need of additional resources. A common example of such a lever is an heat map that designates the regions that have the highest amount of unfilled jobs (see Figure \ref{fig:subim1}).  Another related version combines this approach with the first one, and highlights regions where the demand is high relative to supply, and 
service prices are surging, as well as how much they are surging (see \ref{fig:subim3}).
A third version does not explicitly state how much the prices are surging, but only designates the regions where the prices are surging, and in effect the high/low demand regions, leaving the inference on price changes to the resources (in this case the drivers; see Figure \ref{fig:subim2}).\footnote{It is worth noting that firms have experimented with different versions of these ideas, and not only that different versions of these heat maps are in use by different firms, but also over the years some firms have  switched back and forth between  different alternatives.
}

The recent literature has explored in depth how platforms should design spatial price discrimination policies to improve their profits, or mitigate inefficiencies (see, e.g., 
\cite{bimpikis2019spatial,banerjee2022pricing}).
However, the question of leveraging information to influence the decisions of self-interested resources has not received as much attention. The objective of this paper is to close this gap in the literature, 
offer a framework for using information as a lever for spatial resource allocation, 
and shed light on when different types of practically-relevant information structures are relevant.

To that end, we focus on an undirected network model where each node corresponds to a different location. 
Nodes are initially endowed with a set of resources, which can reposition from their original node to another node.
The edges connect pairs of nodes between which the resources can reposition, and are associated with cost terms that capture the cost a resource incurs when it repositions from one end point to the other end point of the edge.
The number of available resources in each region
after possible repositioning decisions,
 impacts the service price. 
In practice, the service price in a location is also influenced by the demand shocks
 in this location. Our model accommodates this, by allowing for random shocks 
 (which we refer to as the state)
that shift the price curve in a location. The platform can commit to a mechanism that shares information about the aforementioned shocks  once they are realized.
For instance, the realization of the shock can be fully revealed, or the platform can reveal whether the realization is below or above a threshold, or it can employ other alternatives.

The platform collects a commission 
for facilitating matches between the resources and the jobs,
which is modeled as a constant fraction of the generated revenues. 
The resources 
are self-interested and they
try to maximize their payoffs (payment received for providing service minus commission and relocation costs).  
Thus, revealing information appropriately, influences the resources' repositioning decisions, and, in turn, the induced service prices and the platform's revenue.
We investigate how the platform can maximize its revenue by using appropriate information structures.
In the context of the applications mentioned earlier, this question can be equivalently cast as the question of designing heat maps 
(which assign different demand levels to different signal realizations) that maximize  the platform's revenue.\footnote{In terms of the examples in Figure \ref{fig:heatmaps}, our setting is closer to (a) and (c) -- 
the platform sends signals about the demand shocks, without explicitly revealing the impact on prices. That said, in our equilibrium model, the drivers can infer the implied expected service price. Thus, revealing price information in a consistent way as in (b) could also be accommodated in our model.}


Our first contribution is a characterization of the equilibrium repositioning decisions of resources. We show that these can be obtained by solving a simple convex (in fact quadratic) optimization problem. This characterization is made possible through a connection to potential games,  and applies for any shock realization profile, including shocks that impact multiple nodes.


We then use this result to explore when \emph{monotone partitional mechanisms} are optimal. These mechanisms partition the state space into subintervals, and 
for each subinterval either  (i) pool all of the states in this subinterval and map them to a unique signal, or (ii)  reveal the state.
We focus on these mechanisms for two reasons. The first reason is their practical relevance. They allow for grouping states to low/medium/high regions and simply reveal the region to which the state realization belongs -- which is consistent with the way the heat map
examples in Figure
\ref{fig:heatmaps}
work.
Second, perhaps surprisingly, we show that this class of mechanisms is in fact optimal under fairly general and practically relevant assumptions.


To obtain the latter insight, we focus on a setting where there is a single node that is exposed to shocks, and impose two regularity conditions: The first requires an initial ``demand balance'' condition which ensures that pre-shock the supply is distributed in a way that induces identical prices at all locations. The second condition requires no market depletion: i.e.,   the shock is not so large (resp. small) that all the resources in a non-shock (resp. shock) location will leave their location leaving zero resource available in that location.

Our main theorem establishes the optimality of monotone partitional mechanisms  under practically relevant conditions on the market sizes of different locations (characterized in terms of the intercepts of the  price curves). 
Specifically, we study the change in the market sizes relative to the change in the distance from the shock center. If the ratio of these quantities, hereafter the market size change rate, is small, we say that the market sizes are similar relative to distances.
More formally, 
we require the absolute value of the aforementioned rate of change to be bounded by a constant given by the reciprocal of the net income rate of the resources (i.e., one minus the commission paid to the platform).
Conversely, if the rate of change is positive (resp. negative) and strictly larger (resp. smaller) in absolute value than the aforementioned constant, then we say that the market sizes   increase (resp. decrease) relative to distances.
Our result shows that when the market sizes  closest to or furthest away from the shock center have similar sizes, and they are increasing or decreasing in between, monotone partitional mechanisms 
with a special ``reveal-pool-reveal'' structure are optimal. More precisely, these mechanisms
reveal low and high states, while 
pooling the states in between and mapping them to 
a unique signal realization.
It is worth emphasizing that the initial or latter (or both) similarity regions may be empty, in which case the state pooling region will not be sandwiched in between two regions where the state is fully revealed. The region where the market size changes monotonically can be empty (i.e., all market sizes are similar), in which case always revealing the state becomes optimal. 

Our result yields an important insight: substantial changes in market sizes of nearby regions may require witholding some information from the resources to improve revenues (of both the resources and the platform). Intuitively, fully revealing information in such cases may lead to accumulation of too many 
resources in some locations, thereby leading to ``congestion'' and low prices.
Appropriately pooling some states, and in effect providing information with lower fidelity, can eliminate this problem.
Similarly, our result establishes that pooling information is an especially important lever in the regime where the platform's commission rate is small, whereas in the other extreme with very high commission rates (where the platform captures most of the revenues)  the optimal thing to do is full revelation.


Empirically, it has been observed that
the requests of ride-hailing services are often higher in regions close to the central business districts  \cite{dong2018empirical, dean2021spatial}. Thus, abstracting away the details, a city can be modeled
as a collection of (i) central locations with large market sizes, (ii) suburban locations with small market sizes, and (iii) a transition region around the central core 
where market sizes get progressively smaller as one  gets further away from the center.
Assuming that the sizes of the markets in (i) (resp. (ii)) are similar to each other, our results imply that both for shocks in central locations, and those in suburban locations it is optimal to use monotone information structures.


We also relax the  initial
demand balance and no market depletion assumptions. 
Without these assumptions the model is very rich, which in turn  makes it harder to provide intuitive certificates for the optimality of monotone partitions. Still, we are able to characterize the conditions under which monotone partitions are optimal. On the methodological side, these results contribute to the general theory of  monotone partitions, by providing novel tools for their study 
that apply when the designer's payoff is piecewise linear in the posterior mean her signal induces.


{Our final contributions address algorithmic aspects of our information design problem. When the aforementioned conditions do not hold, the optimal information structure may not be monotone partitional. We offer a convex optimization formulation to characterize the optimal information structure in such cases, noting that a double-interval structure may be required for maximizing the platform's payoff, as detailed in \cite{candogan2019persuasion}. In addition, we provide  a dynamic programming approach for obtaining near-optimal monotone information structures and demonstrate its applicability to practical scenario-based models providing a valuable toolkit for spatial resource allocation problems with multiple shocks.
Due to page limitations, the last set of contributions are relegated to Appendix  \ref{apx:discussion}.
}

\paragraph{Related literature:}
The question of information design for influencing agents' spatial 
distribution was previously studied in 
\cite{yang2019information}. 
In this paper, the authors focus on a setting with two locations and binary state, where
initially all agents are in one location.
The other location may have a resource, whose availability is stochastic, and is represented by a binary state.
If agents move to the latter location, they derive a utility that depends on the state as well as the number of agents who relocate there.\footnote{Note that in our paper agents are resources who serve customer demand, whereas in \cite{yang2019information} agents ``consume'' resources that possibly become available in one of the locations. Leaving aside the difference in nomenclature, the models capture similar frictions. For instance, in both models the utility the agents derive from being in one location decreases in the number of agents who are present there.}
The paper studies  the design of optimal public and private information structures (where all receivers see identical or possibly different signals, respectively). 
Two main differences from our work are worthwhile highlighting.
First, the paper leaves open the design of optimal information structures when there are more than two locations -- which is the setting we focus on.
Second, due to binary state, unlike the continuous state space in our paper, 
the question of monotone partitional mechanisms is outside the scope of that paper. By contrast, our focus
is on understanding when such mechanisms are optimal, and our main result shows this to be the case in fairly broad and practically-relevant settings.

\cite{dworczak2019simple}
introduces monotone partitional information mechanisms, and also provide 
necessary and sufficient condition on the sender’s payoff function under which a monotone partitional signal is optimal \emph{for any prior distribution of the state}. 
In our problem, the sender's payoff function has additional structure: It is piecewise linear.
Exploiting this observation we provide a refinement of 
the results of the aforementioned paper, and provide
novel conditions -- this time jointly on the payoff function and the distribution -- for the optimality of monotone partitions. In addition,
we  identify
regimes for the parameters of our problem where these conditions hold, and argue that a particular 
(reveal-pool-reveal type)
monotone partitional information 
structure  turns out to be optimal in many practically relevant cases.

Information design tools have been employed in different operational settings. \cite{vasserman2015implementing, das2017reducing, tavafoghi2017informational, wu2019information, meigs2020optimal} investigate the role of information design in managing equilibrium flows in incident-prone traffic networks. \cite{alizamir2020warning, de2021informing,shah2022optimal} study the optimal information disclosure for the risk of pandemic. 
 \cite{Ling} studies how a designer can reveal informative signals on the queue length to influence customers' decisions on whether to join a queue.  \cite{drakopoulos2018persuading} and \cite{lingenbrink2018signaling} explore how a seller can signal product availability to influence
 the purchase decisions of buyers.
  \cite{kuccukgul2022engineering}
 focuses on a dynamic setting where
 the seller can disclose information 
 to persuade new coming buyers to purchase the product. \cite{candogan2017optimal,candogan2022persuasion} study 
 how information design can be a useful tool to induce desired outcomes in social networks where agents' actions influence their neighbors' payoffs.
 {\cite{Yan18}, \cite{gur2019disclosure}, and \cite{johari2019quality} explore how platforms leverage information to achieve desired outcomes, focusing on incentivizing exploration, enabling dynamic pricing for third-party sellers, and signaling agent quality in two-sided platforms to boost transaction value.} 
\cite{candogan2023value} studies 
how downstream retailers can reveal demand information
to upstream suppliers in supply chains, so as to reduce supply chain costs. The present paper contributes to this growing literature by shedding light on practically relevant mechanisms for spatial resource allocation problems which are prevalent in many modern platforms.


\section{Model and preliminaries}\label{sec:model}
\subsection{Networks and agents}
We consider an \emph{undirected} network $(\V, \E)$, where $\V = \{0, 1, \dots, n\}$ is the set if nodes, and $\E$ is the set of edges. A set of {resources, modeled as nonatomic agent populations,} are distributed at nodes $i \in \V$ in the network with mass vector $m=(m_i)_{i \in \V}$, where $ m_i\geq 0$ is the mass of agent population originating from node $i$. The agent population at each node $i$ decides if they stay at their origin node or reposition to another node in the network. The cost of agents repositioning from node $i$ to $j$ is $c_{ij} \geq 0$, and $c_{ii}=0$ for all $i \in \V$. We denote agents' strategy distribution as $x = (x_{ij})_{i,j \in \V}$, where $x_{ij}$ is the mass of agents originating at node $i$ who choose the strategy of repositioning to node $j$ for $j \in V\setminus \{i\}$ or staying at their origin node $i$ for $j =i$. A repositioning strategy distribution $x \in X$ is feasible if 
\begin{align*}
    \sum_{j \in \V} x_{ij} = m_i, \quad \forall i \in \V, \quad x_{ij} \geq 0, \quad \forall i, j \in  \V. 
\end{align*}
The distribution of agents induced by $x$ is $q = (q_{i})_{i \in \V}$, where 
\begin{align}\label{eq:q_x}
q_i = \sum_{j \in \V}x_{ji}, \quad \forall i \in \V.
\end{align}
The service price at node $i$, denoted as $ \price_i(q_i)$, is a linear function of $q_i$:\footnote{All results generalize to piecewise linear price functions. In particular, a node with a piecewise price function composed of $n$ linear pieces can be equivalently represented as $n$ nodes at the same location of the network, each with a linear price functions.}
\begin{align*}
\price_i(q_i) = \state_i - \beta_i q_i, \quad \forall i \in \V. 
\end{align*}
where $\state_i\geq 0$ is the \emph{market size} at node $i$, and $\beta_i \geq 0$ is the \emph{price elasticity} at node $i$. For every transaction, the platform collects commission with a \emph{fixed} rate $\rate \in [0,1]$. Thus, the payoff received by an individual agent is $(1-\rate) \price_i(q_i)$, and the total commission (i.e. revenue) collected by the platform at node $i$ is $\rate \price_i(q_i) q_i$.

One or multiple nodes in the network may experience demand shocks that affect the market sizes. We denote the  state of the network as the vector of the realized market sizes $s= (s_i)_{i \in \V} \in \S$, where $\S=\prod_{i \in \V} \S_i$ such that each $\S_i$ is a continuous and closed interval of $\mathbb{R}$. The cumulative distribution of the state, referred as \emph{the prior}, is $F: \mathcal{\S} \to [0,1]$. The prior $F$ is common knowledge.
For most of our analysis, we assume that state realizations belong to a $1$-dimensional subspace of $\cal S$, and assume that $F$ restricted to this subspace is absolutely continuous.
The platform observes the realization $s$, but the agents do not.

\subsection{Platform's design problem}
The platform designs a \emph{public} information provision mechanism $(\Signal, \pi)$, where $\Signal$ is the set of possible signal realizations, denoted generically as $\signal$, and $\pi(\cdot|\state)$ is the probability density function of signal realization given state $\state$. We consider the setting where the state and the signal sets are continuous.
We assume that $\pi$ satisfies mild measurability conditions that ensure that the conditional expectations below are well defined.

The platform commits to their information mechanism $(\Signal, \pi)$ before observing the state realization. After observing the state $\state$, the platform generates a signal $\signal$ according to $\pi(\cdot|\state)$ and sends the signal to \emph{all} agents. Two special cases of information mechanisms are: (i) the platform provides \emph{full information} if $\Signal = \S$ and $\pi(\signal|\state)=1$ for all $\signal = \state$; and (ii) the platform provides \emph{no information} if $\pi(\cdot|\state)$ does not depend on $\state$. When $(\Signal, \pi)$ does not belong to (i) or (ii), the information mechanism provides partial information of the state.

After receiving the realized signal $\signal$, agents compute the expected value of the state $\mathbb{E}[\state|\signal]$, and make repositioning decisions based on the received signal, i.e. the strategy distribution $x(\signal): \Signal \to X$. The utility of agents who reposition from $i$ to $j$ equals to the expected payoff received at node $j$ minus the repositioning cost $c_{ij}$:
\begin{align*}
    u_{ij}(x|\signal) = (1-\rate)(\mathbb{E}[\state_j|\signal] - \beta_j q_j(\signal)) - c_{ij}, \quad \forall i, j \in \V, 
\end{align*}
where $q(\signal)$ is the distribution of agents induced by $x(\signal)$ as in \eqref{eq:q_x}. Agents are self-interested in that they make repositioning decisions to maximize their expected utility. Given signal $\signal$, we define the equilibrium strategy distribution $\xeq(\signal)$ as follows: 
\begin{definition}
For any $\signal \in \Signal$, a strategy profile $\xeq(\signal)$ is a Wardrop equilibrium if 
\begin{align*}
    \xeq_{ij}(\signal)>0, \quad \Rightarrow \quad u_{ij}(\xeq|\signal) \geq u_{ij'}(\xeq|\signal), \quad \forall j, j' \in \V, \quad \forall i \in \V. 
\end{align*}
\end{definition}That is, in equilibrium, the mass of agents repositioning from node $i$ to $j$ is nonzero if the expected utility $u_{ij}(\xeq|\signal)$ based on the observed signal is the maximum compared to staying at node $i$ or repositioning to any other node in the network.


The objective of the platform is to design the optimal information mechanism to maximize the total expected revenue -- the total commission collected at all nodes. For any signal $\signal$, the platform's expected revenue $\revenue(\signal)$ in equilibrium is given by: 
\begin{align}\label{eq:revenue_t}
\revenue(\signal)&= \rate \sum_{i \in \V}  (\mathbb{E}[\state_i|\signal] - \beta_i \qwe_i(\signal))\qwe_i(\signal).
\end{align}Thus, the total expected revenue of the platform given mechanism $(\Signal, \pi)$ is
\begin{align}\label{eq:expected_revenue}
\barrevenue &= \int_{\state \in \S}\int_{\signal \in \Signal}\revenue(\signal) \pi(\signal|\state) dF(\state) d\signal.
\end{align}

\subsection{Potential function of the repositioning game}
We first show that given any signal $t$, the induced repositioning game is a population potential game, and $\xeq(t)$ can be computed as the maximizer of a potential function. 
\begin{proposition}\label{prop:potential}
For any $\signal \in \Signal$, $\xeq(\signal)$ can be computed by maximizing the following potential function $\Phi(x|\signal)$: 
\begin{equation}\label{eq:potential}
\begin{split}
    \max_{x \in X} \quad &\Phi(x|\signal) :=  (1-\rate)\sum_{i \in \V} \int_{0}^{q_i(t)}(\mathbb{E}[\state_i|\signal] - \beta_j z) dz -  \sum_{i, j \in \V}c_{ij} x_{ij}(t), \\
    s.t. \quad &x(\signal) \in X, \quad \text{and $q(\signal)$ satisfies \eqref{eq:q_x}.}
    \end{split}
\end{equation}
Moreover, the agents' distribution in equilibrium $\qeq(\signal)$ is unique for all $\signal \in \Signal$. 
\end{proposition}
In \eqref{eq:potential}, the potential function $\Phi(x|\signal)$ is quadratic, and the feasibility constraints are linear. Thus, $\xeq(t)$ can be computed as the maximizer of the quadratic program in polynomial time. Proposition \ref{prop:potential} also demonstrates that equilibrium is essentially unique in that the agents' distribution $\qeq(\signal)$ is unique. As a result, the expected service price at each node and the revenue $R(t)$ is unique in equilibrium for any $\signal \in \Signal$.\changed{\footnote{The uniqueness of $\qeq(t)$ holds for any strictly increasing price function. In addition, Proposition \ref{prop:potential} does not require single dimensional shocks or absolute continuity of the state.}}

Furthermore, we know from Proposition \ref{prop:potential} and \eqref{eq:revenue_t} that both $\qeq(\signal)$ and $\revenue(\signal)$ only depend on the realized signal $\signal$ through the induced posterior mean estimate of the state $\mathbb{E}[\state|\signal]$, i.e. $\qeq(\signal) = \qeq(\signal')$ for any two signals $\signal, \signal' \in \Signal$ such that $\mathbb{E}[\state|\signal] = \mathbb{E}[\state|\signal']$. Therefore, we know that the platform's total expected revenue $\barrevenue$ depends on the distribution of the posterior mean of the state $\mathbb{E}[s|\signal]$ induced by the information mechanism $(\Signal, \pi)$. The set of all possible posterior means is the state set $\S$. We denote the cumulative distribution of the posterior mean as $G: \S \to [0,1]$.

With slight abuse of notation, we denote the strategy profile associated with any posterior mean that takes the value $\barstate \in S$ as $x(\barstate)$, agents' distribution as $q(\barstate)$, and the revenue function as $\revenue(\barstate)$. We re-write the total expected revenue $\barrevenue$ in \eqref{eq:expected_revenue} as follows: 
\begin{align}\label{eq:expected_revenue_rewrite}
\barrevenue &= \int_{\barstate \in \S}\revenue(\barstate) dG(\barstate)= \rate \int_{\barstate \in \S} \sum_{i \in \V}  (\barstate- \beta_i \qwe_i(\barstate))\qwe_i(\barstate) dG(\barstate).
\end{align}

In Sec. \ref{sec:single_shock} -- \ref{sec:single_general}, we focus on the single shock case, where only node $0$ is prone to demand shock. We denote the set of states as $\S_0$ with generic member $s_0$. Here, $\state_0$ is a real number and $\S_0$ is a closed interval of $\mathbb{R}$. In this case, a posterior mean distribution $G$ is feasible (i.e. induced by an information mechanism given the prior $F$) if and only $F$ is a mean-preserving spread of $G$, denoted as $G \preceq F$ (see \cite{blackwell1953equivalent, gentzkow2016rothschild, kolotilin2018optimal}; also a formal definition is given in Appendix \ref{apx:proof_4} for completeness). Therefore, the optimal posterior mean distribution $G^*$ can be solved as follows in a single shock case: 
\begin{align}\label{eq:opt_single}
\max_G \int_{z \in \S_0} \revenue(z)dG(z), \quad s.t. \quad G \preceq F. 
\end{align}
We define the value of information design $V^*_F$ as the difference between the revenue with optimal information mechanism $R_{G^*}$ minus the revenue with no information provision $R_{F}$, i.e. $V^*_F = R_{G^*} - R_{F}$. We show that the value of information design is higher for $F'$ that is a mean-preserving spread of $F$. This result builds on the fact that the feasible set of $G$ in \eqref{eq:opt_single} is larger with $F'$ than with $F$.\footnote{Proposition \ref{prop:ordering} holds for any information design problem with one-dimensional state space. } 
\begin{proposition}\label{prop:ordering}
    For any $F \preceq F'$, $V^*_F \leq V^*_{F'}$.  
\end{proposition}

\section{Single shock with homogeneously balanced markets}\label{sec:single_shock}

In this section, we focus on identifying the conditions under which optimal revenue can be achieved through information mechanisms with simple partitional structure. Such mechanism involves partitioning the state space $\S_0$ into intervals, and within each interval, the information mechanism either fully reveals the state realization or only provides information indicating that the realized state falls within that specific interval. Formally, this type of information mechanism is referred to as a \emph{monotone partitional information mechanism} (\cite{dworczak2019simple}): 

\begin{definition}[Monotone partitional information mechanism]
    An information mechanism is monotone partitional if there exists a finite partition of the state set $\S_0$ into intervals $\S_0= \cup_{\k \in \K} [\partition^{\k}, \partition^{\k+1}]$ such that for each $\k$, the information mechanism either (i) fully reveals the state for all $\barstate_0 \in [\partition^{\k}, \partition^{\k+1}]$ (full revelation); or (ii) only reveals that the realized state is in $[\partition^{\k}, \partition^{\k+1}]$ (pooling).
\end{definition}

Given a monotone partitional information mechanism, for any state realization $\state_0 \in [\partition^{\k}, \partition^{\k+1}]$, the corresponding signal is  $\signal = \state_0$ if interval $\k$ is a full revelation interval.
If  $k$ is a pooling interval, without loss of generality, we set the  signal realization to the associated posterior mean, i.e., $\signal = \mathbb{E}_F[S_0|S_0 \in [\partition^{\k}, \partition^{\k+1}]]$.
In either case, each state realization corresponds to a unique signal realization and higher realized state corresponds to a higher signal realization.
Therefore, monotone partitional information mechanisms have the advantage of sending deterministic and monotone signals. 
In theory, monotone partitional information mechanisms may or may not be optimal. In this section, we characterize verifiable conditions that guarantee the optimality of monotone partitional information mechanism. To better demonstrate the intuition and practical implications, we first present our results under two assumptions -- homogeneously balanced market condition and no market depletion condition. We will remove these two assumptions when generalizing our results in the next section.

For all $i \in \V$, we let $\distance_i = c_{i0}$, where $c_{i0}$ is the distance of the shortest path between node $0$ and node $i$ (recall that the underlying network is undirected). Assumption \ref{as:complete_market_balance} ensures that with no information provision, markets at all nodes have homogeneous prices so that agents have no incentive to move across nodes.\footnote{In practice, this assumption typically holds in the long run when agents' repeated repositioning decisions even out payoff differences across nodes. In the short term, two nodes may still have different prices especially when the distance between them is large.}
\begin{assumption}[Homogeneously balanced markets]\label{as:complete_market_balance} Given the initial distribution of agents $m$, the service price is identical across all nodes. Without loss of generality, we normalize the price to zero, i.e.  
\[\mathbb{E}_F[s_0] - \beta_0 m_0 = \state_i - \beta_i m_i=0, \quad \forall i \in \V.\]
\end{assumption}
Assumption \ref{as:no_depletion_max} imposes bounds on the maximum realization of demand shock, and rules out the possibility that agents at one node are all drawn away in response to a high demand realization at node $0$, or all agents leave node 0 when the demand realization is low (Lemma \ref{lemma:no_depletion}). 
\begin{assumption}[No market depletion]\label{as:no_depletion_max}
\[\sup \S_0 \leq \max_{i \in V\setminus \{0\}} \state_i + \frac{d_i}{1-\rate}, \quad \inf \S_0 \geq - \frac{m_0 + \sum_{i=1}^{\hat{k}} \frac{d_i}{(1-\rate)\beta_i}}{\sum_{i=1}^{\hat{k}} \frac{1}{\beta_i}},\]
where $\hat{k}$ satisfies: 
\[-d_{\hat{k}} \geq - \frac{m_0 + \sum_{i=1}^{\hat{k}} \frac{d_i}{(1-\rate)\beta_i}}{\sum_{i=1}^{\hat{k}} \frac{1}{\beta_i}} \geq - d_{\hat{k}+1}.\]
\end{assumption}
 \begin{lemma}\label{lemma:no_depletion}
     Under Assumption \ref{as:no_depletion_max}, the agents' equilibrium distribution $\qeq(\barstate_0)$ satisfies $\qeq_i(\barstate_0)>0$ for all $i \in \V$ and all $\barstate_0 \in \S_0$. 
 \end{lemma}

\subsection{Equilibrium characterization}
In this section, we provide a closed form characterization of the equilibrium strategy distributions and the platform's revenue. We show that the revenue is a piecewise linear function of the posterior mean of demand at node 0. We partition nodes in $\V$ into $\{0\} \cup \{\cup_{n=1}^{N}\Vn\}$, where all nodes in $\V_n$ have the same distance $d_n$ to node $0$. We label $n=1, \dots, N$ in increasing order of $d_n$, i.e. $0< \distance_1 < \distance_2 < \cdots < \distance_N$. Clearly, $N \leq |\V|$.

\begin{proposition}\label{prop:eq_strategy_simple}
Under Assumptions \ref{as:complete_market_balance} and \ref{as:no_depletion_max}, the equilibrium agent distribution $\qeq$ and platform's revenue $R$ are piecewise linear functions of the posterior state mean $\barstate_0$, and exhibit $\Kplus+ \Kminus+1$ regimes, where 
\begin{align*}
\Kplus &= \max \left\{k=1, \dots, N \left\vert ~ \barstate_0[k] \leq \sup \S_0 \right.\right\}, \quad 
\Kminus  =  \max \left\{k=1, \dots, N \left\vert ~ \barstate_0[-k] \geq \inf \S_0\right.\right\},
\end{align*}
and the regime thresholds are given by: 
\begin{subequations}
\begin{align}
    \barstate_0[k] &= \mathbb{E}_{F}[S_0]+ \frac{d_k }{1-\rate} + \sum_{i \in \cup_{n=1}^kV_n} \frac{\beta_0(d_k - \distance_i)}{\beta_i(1-\rate)}, \quad \forall k=1, \dots, \Kplus,\\ 
    \barstate_0[-k] &=\mathbb{E}_{F}[S_0] - \frac{d_k }{1-\rate} -\sum_{i \in \cup_{n=1}^kV_n}    \frac{\beta_0  (d_k-\distance_i)}{\beta_i (1-\rate)}, \quad \forall k=1, \dots, \Kminus.
\end{align}
\end{subequations}
\medskip 
\noindent \underline{Regime 0}: $\barstate_0[-1]\leq \barstate_0 < \barstate_0[1]$. \begin{align}\label{eq:q_zero_simple}
        \qwe_0 &= m_0, \quad \qwe_i =m_i, \quad \forall i \in V \setminus \{0\}. 
    \end{align}
\medskip
\noindent \underline{Regime $\{k\}_{k=1}^{\Kplus}$}: $\barstate_0[k] \leq \barstate_0 < \barstate_0[k+1]$.
\begin{subequations}\label{eq:q_positive_simple}
     \begin{align}
       \qeq_0 &= \frac{1}{\beta_0 \left(\sum_{i \in \cup_{n=1}^{k}V_n}\frac{1}{\beta_i} + \frac{1}{\beta_0}\right)}\left( m_0+  \sum_{i \in \cup_{n=1}^{k}V_n} \frac{1}{\beta_i} \left(\barstate_0 - \frac{\distance_i}{1-\rate}\right)\right), \\
       \qeq_i &= \frac{1}{\beta_i} \left( s_i - \barstate_0 + \frac{d_i}{1-\rate}\right)+ \frac{\beta_0}{\beta_i} \qeq_0, \quad \forall i\in \cup_{n=1}^k V_n, \quad \qeq_i = m_i, \quad \forall i\in \V \setminus ({\cup_{n=1}^k V_n \cup \{0\}}). 
    \end{align}
\end{subequations}

    \medskip 
    \noindent \medskip
\noindent \underline{Regime $\{-k\}_{k=1}^{\Kminus}$}: $\barstate_0[-k-1] \leq \barstate_0 < \barstate_0[-k]$.
\begin{subequations}\label{eq:q_negative_simple}
    \begin{align}
    \qeq_0 &= \frac{1}{\beta_0 \left(\sum_{i \in \cup_{n=1}^{k}V_n}\frac{1}{\beta_i} + \frac{1}{\beta_0}\right)}\left( m_0+ \sum_{i \in \cup_{n=1}^{k}V_n} \frac{1}{\beta_i} \left(\barstate_0 + \frac{\distance_i}{1-\rate}\right)\right), \\
    \qeq_i &= \frac{1}{\beta_i} \left( s_i - \barstate_0 - \frac{d_i}{1-\rate}\right)+ \frac{\beta_0}{\beta_i} \qeq_0, \quad \forall i\in \cup_{n=1}^k V_n, \quad \qeq_i = m_i, \quad \forall i\in\V \setminus (\cup_{n=1}^k V_n \cup \{0\}).
    \end{align}
\end{subequations}
 Moreover, the platform's revenue $\revenue(\barstate_0) = r \sum_{i \in \V} \qeq_i \peq_i$ is a continuous piecewise linear function of $\barstate_0$. In each regime $k$, $\revenue(\barstate_0)$ is a linear function of $\barstate_0$, and the derivative of the linear function, denoted as $\frac{d \revenue(\barstate_0)}{d \barstate_0}[k]$, is given by: 
\begin{align*}
    \frac{d \revenue(\barstate_0)}{d \barstate_0}[k] &= \left\{ \begin{array}{ll}
    r m_0, & \quad k=0,\\
    &\\
    \rate \frac{m_0+ \sum_{i \in \cup_{n=1}^kV_n}  \left(m_i + d_i/\beta_i (1-\rate) \right)}{\beta_0\left(\sum_{i \in \cup_{n=1}^kV_n}\frac{1}{\beta_i}+\frac{1}{\beta_0} \right)}, &\quad  \forall \k = 1, \dots, \Kplus,\\
    &\\
    \rate \frac{m_0+ \sum_{i \in \cup_{n=1}^kV_n}  \left(m_i - d_i/\beta_i (1-\rate) \right)}{\beta_0\left(\sum_{i \in \cup_{n=1}^kV_n}\frac{1}{\beta_i}+\frac{1}{\beta_0} \right)} , & \quad  \forall \k = -1, \dots, -\Kminus.\end{array}\right.
\end{align*}
\end{proposition}

Proposition \ref{prop:eq_strategy_simple} shows that when the demand realization exceeds the prior mean, agents from nodes in proximity to node 0 move to node 0, starting from nodes with the smallest distances $d_1$ and gradually progressing to more distant nodes as the demand realization further increases. The equilibrium price at node 0 is equal to the price at each node $i$ where agents move to node 0, plus the distance $d_i$. Conversely, when the demand realization decreases below the prior mean, agents from node 0 begin to move to other nodes. Initially, they move to nodes with small distances $d_1$, as the demand further decreases, they move to more distant nodes. In this scenario, the price at node 0 is equivalent to the price at each node $i$ where agents move to, minus the distance $d_i$. 

When $x_{i0}>0$ 
(resp. $x_{0i}>0$)
for some $i\in V$, we say that agents move from $i $ to $0$ 
(resp. from $0$ to $i$)
in equilibrium, due to a demand shock that yields a higher 
(resp. lower)
realization than the prior mean at node $0$.  
Proposition \ref{prop:eq_strategy_simple} demonstrates that equilibrium regimes are distinguished by the changes of the set of affected nodes -- the set of affected nodes is $\cup_{n=1}^k V_n$ in regimes $k$ and $-k$. Moreover, due to agents' movements in equilibrium, the price sensitivity at node 0 (with respect to the change of demand) depends on the price sensitivity of all of the nodes that are affected. As a result, the derivative of the platform's revenue with respect to the demand realization changes as the regime changes.

\subsection{Optimal partitional information mechanism}
In this section, we demonstrate the efficacy of monotone partitional information mechanisms by showing that under a set of practically relevant conditions on market size distributions, optimal revenue can be achieved by simple monotone partitional mechanisms with at most one pooling interval. We also provide an algorithm for computing such simple partitional mechanism. 

Before presenting the theorem, we first define similar and monotone market sizes. We say that the market sizes are \emph{similar relative to distances} for nodes $i$ and $j$ with $d_i \neq d_j$ if $s_i$ and $s_j$ satisfy:
\begin{align}\label{eq:similar}
\left\vert\frac{s_i - s_j}{d_{i}- d_{j}}\right\vert \leq \frac{1}{1-\rate},
\end{align}
This indicates that the proportion of the changes of market sizes between any two nodes $i$ and $j$ with respect to the differences of their distances to node $0$ is upper bounded by $1/(1-\rate)$, which increases as the commission rate $\rate$ increases (Recall that this rate is less than 1). Additionally, we say that market sizes \emph{increase relative to distances} (resp. \emph{decrease relative to distances}) for nodes $i$ and $j$ with $d_i > d_j$ if $s_i$ and $s_j$ satisfy \eqref{subeq:increase_size} (resp. \eqref{subeq:decrease_size}):
\begin{subequations}\label{eq:monotone}
    \begin{align}
    &\frac{s_i - s_j}{d_{i}- d_{j}} >  \frac{1}{1-\rate}, \label{subeq:increase_size} \\
&\frac{s_i - s_j}{d_{i}- d_{j}} < - \frac{1}{1-\rate}, \label{subeq:decrease_size}
    \end{align}
\end{subequations}
Equation \eqref{subeq:increase_size} (resp. \eqref{subeq:decrease_size}) indicates that the nodes that are further away (resp. closer) to node $0$ have larger (resp. smaller) market sizes, and the rate of increase (resp. decrease) with respect to the change of distances is larger than $1/(1-\rate)$.

We define $D$ as the maximum distance of nodes that can be affected by demand shock in equilibrium, i.e. $D= \max\{d_{-\Kminus}, d_{\Kplus}\}$. Agents from any nodes with distances higher than $D$ will not move in equilibrium given any demand realization, and thus the market sizes of these nodes are irrelevant for information design. The following theorem shows that a monotone partitional information mechanism is optimal if {there exists a monotone \emph{transition region}, where nodes with distances between $\underline{d}$ and $\bar{d}$ have monotonically increasing or decreasing market sizes relative to their distances to node $0$. } 
\begin{theorem}\label{theorem:simple_partitional}
Under Assumptions \ref{as:complete_market_balance} and \ref{as:no_depletion_max}, suppose that there exists $0 \leq \dl \leq \dh \leq D$ such that nodes with distances less than $\dl$ 
have similar market sizes,
nodes with distances in $[\dl, \dh]$ have decreasing or increasing market sizes,
and
nodes with distances higher than $\dh$ also have similar market sizes relative to distances. Given any prior distribution $F$, there exists an optimal monotone partitional information mechanism with thresholds $\inf \S_0 \leq \zl \leq \zh \leq \sup \S_0$ that fully reveals states $\state_0 \leq \zl$ and $\state_0 \geq \zh$, and pools states $\state_0 \in [\zl, \zh]$ with posterior mean $\zmean = \mathbb{E}_F[S_0|\zl \leq S_0\leq \zh]$.  
Specifically,  
\begin{enumerate}
\item[(i)] $\zl=\zh$ if $\dl=\dh$. That is, full information revelation is optimal if all nodes have similar market sizes relative to distances. 
\item[(ii)] $\zmean> \mathbb{E}_F[S_0]$ if nodes with distances in $[\dl, \dh]$ have decreasing market sizes relative to distances. 
\item[(iii)] $\zmean< \mathbb{E}_F[S_0]$ if nodes with distances in $[\dl, \dh]$ have increasing market sizes relative to distances.
\end{enumerate} 
\end{theorem}

\begin{figure}[ht]
    \centering
    \begin{subfigure}[b]{0.4\textwidth}
         \centering
         \includegraphics[width=\textwidth]{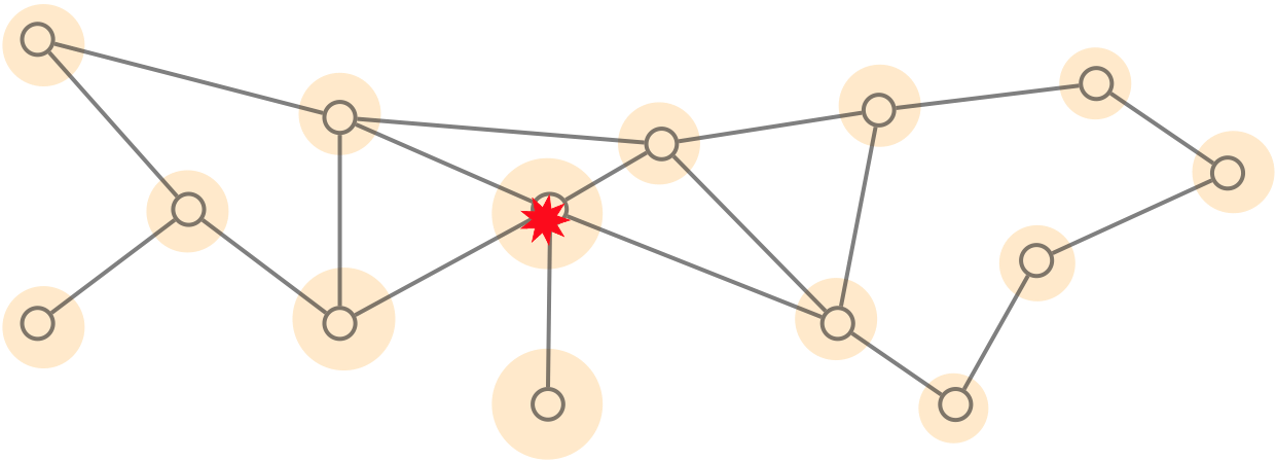}
         \caption{}
         \label{subfig:similar}
     \end{subfigure}
     ~
     \begin{subfigure}[b]{0.4\textwidth}
         \centering
         \includegraphics[width=\textwidth]{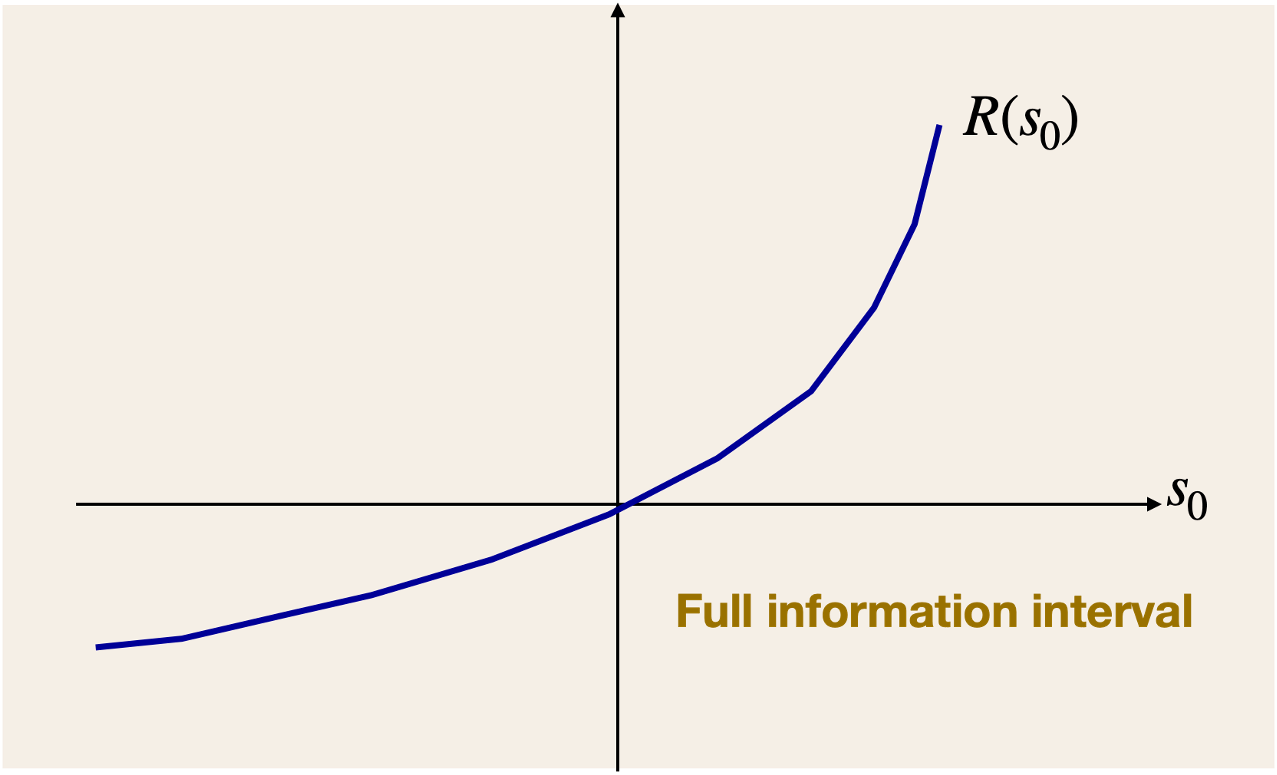}
         \caption{}
         \label{subfig:similar_function}
     \end{subfigure}\\
      \begin{subfigure}[b]{0.4\textwidth}
         \centering
         \includegraphics[width=\textwidth]{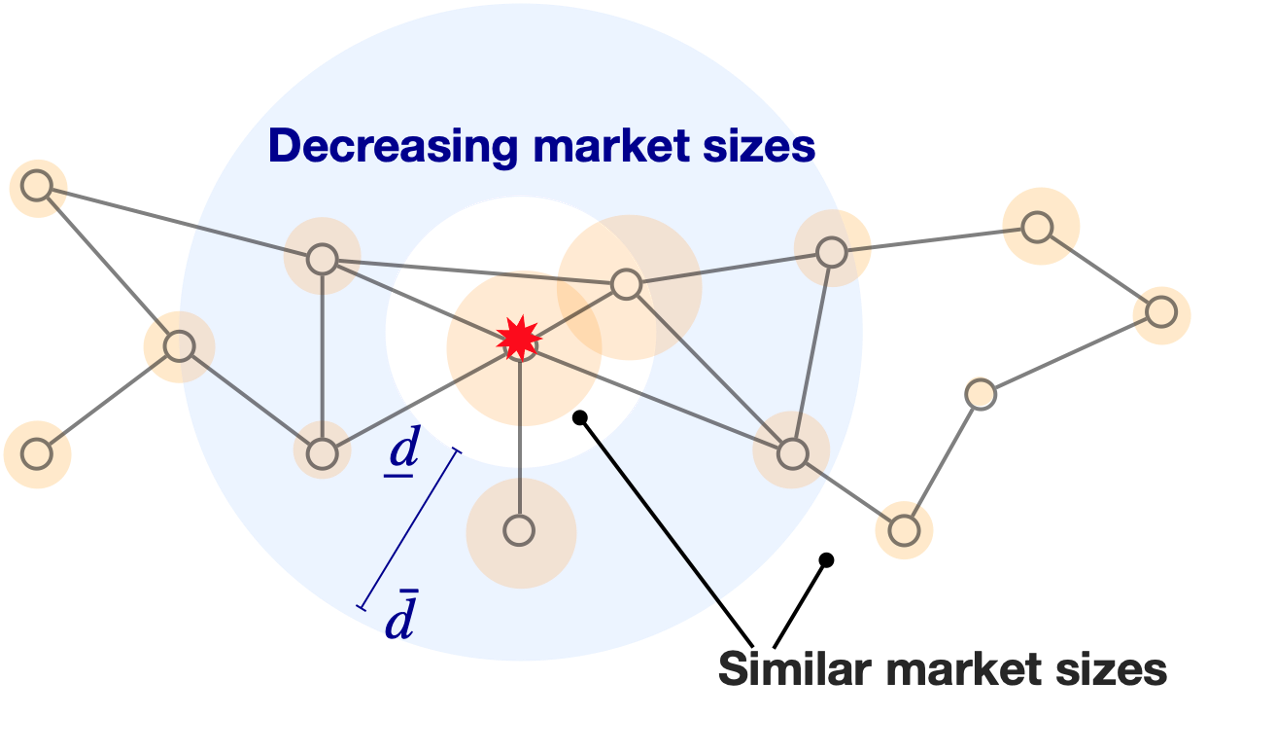}
         \caption{}
         \label{subfig:similar_decrease_similar}
     \end{subfigure}
     ~
     \begin{subfigure}[b]{0.4\textwidth}
         \centering
         \includegraphics[width=\textwidth]{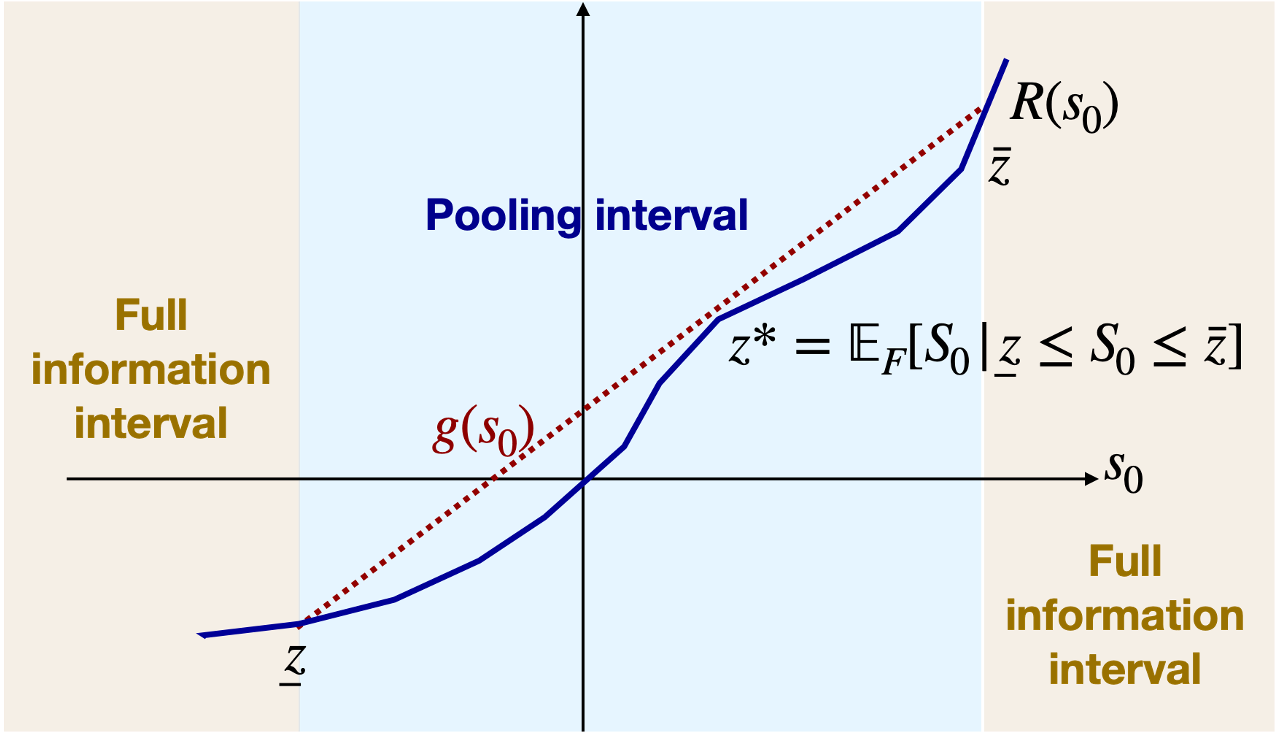}
         \caption{}
    \label{subfig:similar_decrease_similar_function}
     \end{subfigure}\\
     \begin{subfigure}[b]{0.4\textwidth}
         \centering
         \includegraphics[width=\textwidth]{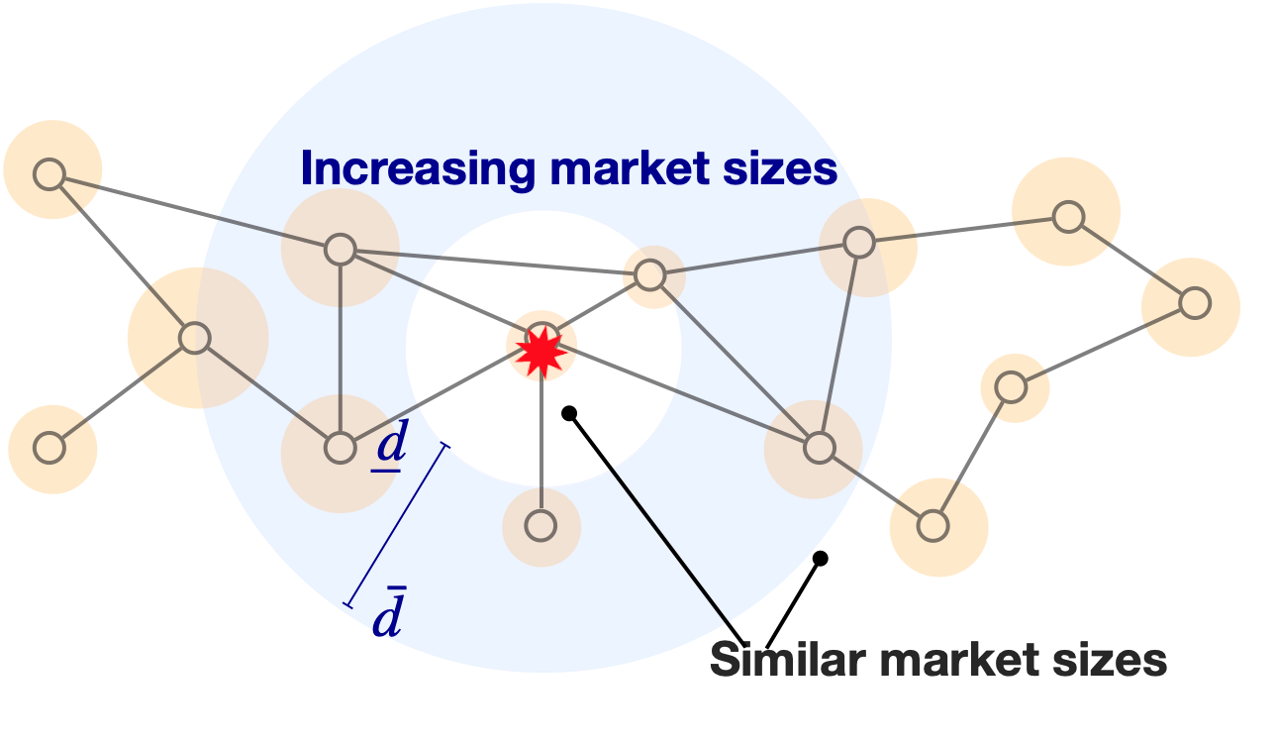}
         \caption{}
         \label{subfig:similar_increase_similar}
     \end{subfigure}
     ~
     \begin{subfigure}[b]{0.4\textwidth}
         \centering
         \includegraphics[width=\textwidth]{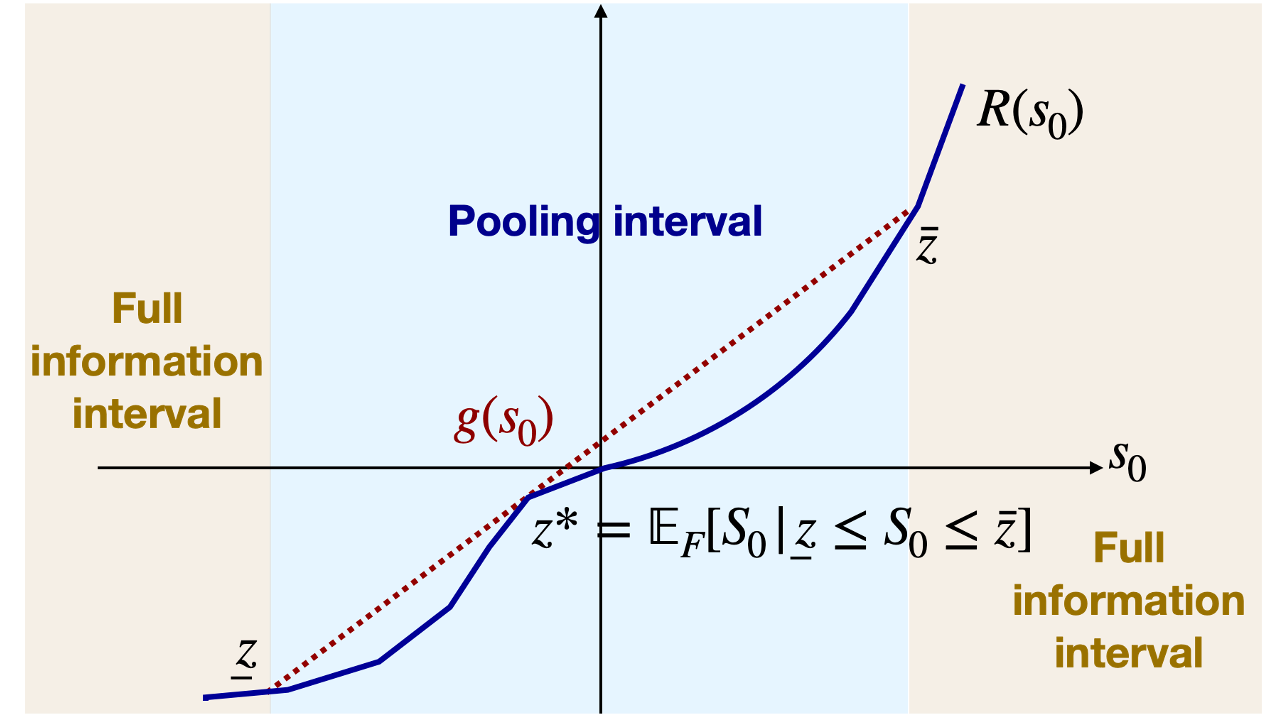}
         \caption{}
    \label{subfig:similar_increase_similar_function}
     \end{subfigure}
     \caption{(a)-(b) Nodes with similar market sizes and convex revenue function associated with full information revelation mechanism; (c)-(d) Nodes with decreasing transition region and revenue function associated with one pooling interval; (e)-(f) Nodes with increasing transition region and revenue function associated with one pooling interval. }
\end{figure}

\medskip 
Theorem \ref{theorem:simple_partitional} shows that the optimal information mechanism reveals at least partial demand information to agents and has a simple partitional structure with up to three intervals.\footnote{{Since the revenue function $\revenue(s_0)$ is piecewise linear, fully revealing state information is equivalent to revealing the regime interval in which the state realization resides.}} These intervals consist of one pooling interval between $\zl$ and $\zh$, and two full information revelation intervals for demands below $\zl$ and above $\zh$. When all market sizes are similar relative to distances to node 0 (case (i) as shown in Fig. \ref{subfig:similar}), maximum revenue is achieved by revealing all information, resulting in a degenerate pooling interval.  In both cases (ii) and (iii) as shown in Figures \ref{subfig:similar_decrease_similar_function}and \ref{subfig:similar_increase_similar_function}, the optimal information mechanism includes a pooling interval. In case (ii) (resp. case (iii)), the pooling interval generates a posterior mean that induces agents to move to node 0 (resp. move away from node 0) when the transition region has decreasing (resp. increasing) market sizes.



{
To understand the intuition behind cases (ii) and (iii) and the value of pooling more clearly, consider a simple example with three nodes on a line, where node $0$ is prone to demand shocks. Suppose nodes $1$ and $2$ with distances $d_1<d_2$ do not have similar market sizes, and one of them is small whereas the other one is large. Consider positive shocks at node $0$, which always improve the platform's revenues.  If the shock is large enough to influence the resources only in the small market, then this will have only marginal impact on the revenue change rate at node $0$, but if it influences the large market, then the revenue change rate becomes substantially larger (see Proposition \ref{prop:eq_strategy_simple}). This implies that if nodes $1$ and $2$ exhibit increasing market sizes, then the revenue function in the positive shock regime is convex, but in the decreasing market sizes case it is concave.   This (local) concavity of the revenue function implies that as opposed to revealing the state, with appropriate pooling the platform can ensure higher profits. Specifically, in the positive shock regime, with decreasing market sizes, when the state realization is relatively small, revealing the states induces a few resources to reposition to $0$ preserving a relatively high price there. However, when the realization is large, revealing the state leads to too many resources to reposition to $0$, and in some realizations it may even incentivize the resources in the further out small markets to move, thereby leading to excessively low prices. While in the former case, pooling may lower revenues, in the latter case, by eliminating low prices, it ensures higher revenues. Concavity of the revenue function ensures that the latter effect dominates, and the expected revenues are higher with pooling. With increasing market sizes, convexity in the positive shock regime may at first suggest that pooling should not help. However, it turns out that in this case, in the negative shock regime we have local concavity of the payoff function. Hence, appropriate pooling, once again, improves revenues.

}

The scenario of similar or monotone market sizes described in Theorem \ref{theorem:simple_partitional} are of practical interests. Empirical studies \cite{dong2018empirical, dean2021spatial} have demonstrated that the requests of ride-hailing services are often higher in regions close to the central business districts, which have more jobs, services and other economic activities, and decrease in regions far away from the central business districts. In Theorem \ref{theorem:simple_partitional}, case (i) corresponds to the scenario where all affected nodes are within or close to the central business district, and thus all nodes have similar market sizes. Additionally, case (ii) corresponds to the scenario where the shock affects a node within the central business district and affect nodes that are outside of the district, which have decreasing market sizes relative to distances. On the other hand, case (iii) corresponds to the scenario where the shock happens at a node that is far away from the central business district, and therefore nodes that are further away from node 0 have higher market sizes. Theorem \ref{theorem:simple_partitional} demonstrates that in both cases, a simple partitional information mechanism with one pooling interval is optimal.

\medskip 
\noindent\emph{Proof sketch.} The proof of Theorem \ref{theorem:simple_partitional} builds on the duality theory of optimal information design introduced in \cite{dworczak2019simple}. This duality theory demonstrates that the optimal posterior mean distribution $G^*$ can be constructed by finding an upper closure $\nu(\barstate_0)$ of the objective function $\revenue(\barstate_0)$ that satisfies (i) the function $\nu(\barstate_0)$ is convex and $\nu(\barstate_0) \geq \revenue(\barstate_0)$; (ii) the expected value of $\nu(\barstate_0)$ with respect to $G^*$ is the same as  that with $\revenue(\barstate_0)$; (iii) The optimal posterior $G^*$ is a mean preserving spread of the prior $F$, and
the support set of $G^*$ is a subset of $\{S_0|\nu(s_0) = \revenue(s_0)\}$ (Lemma \ref{lemma:duality} in Appendix \ref{apx:proof_three}). 

Building on the general theory, we further show that the construction of such convex upper closure function is associated with the second-order properties (convexity or concavity) of the revenue function $\revenue(\barstate_0)$. In our problem, the second order property of $\revenue(\barstate_0)$ is governed by how the derivative of the linear revenue function changes from one equilibrium regime to another. From Proposition \ref{prop:eq_strategy_simple}, we find that the changes of the derivatives depend on the market sizes of nodes that are affected. Since nodes are added to the affected set according to their distances to node $0$, the change of derivatives depend on the changes of market sizes relative to the distances. 

We show that under Assumptions \ref{as:complete_market_balance} -- \ref{as:no_depletion_max} and the condition that market sizes are similar or monotone with respect to their distances to node 0, there exists at most one sub-interval of states $[\underline{s}_0, \bar{s}_0]$ such that $\revenue(\barstate_0)$ is concave in $[\underline{s}_0, \bar{s}_0]$ 
and convex to the left or right of this interval (Lemma \ref{lemma:convex_concave} in Appendix \ref{apx:proof_three}). In particular, the sub-interval is empty and the function $\revenue(\barstate_0)$ is convex if all nodes have similar market sizes with respect to distances. Thus, $\nu(\barstate_0) = \revenue(\barstate_0)$, $G^*= F$ satisfy conditions (a) -- (c) indicating that the optimal information mechanism is to fully reveal the demand realization (Fig. \ref{subfig:similar_function}).

Additionally, in cases (ii) (resp.(iii)), the concave subinterval 
results in a higher (resp. lower) posterior mean than the prior mean.
We prove that in both cases, the convex closure function can be constructed as $\nu(\barstate_0) = \max\{g(\barstate_0), \revenue(\barstate_0)\}$, where $g(\barstate_0)$ is an affine function tangent to the concave piece of the revenue function $\revenue(\barstate_0)$. We show that the optimal information mechanism exhibits a simple partitional structure with one pooling interval that associated with the interval of states such that $\nu(\barstate_0) = g(\barstate_0)$, and the remaining states out of this interval are fully revealed. The posterior mean of the pooled state equals to tangent point $\zmean$, which is higher than (resp. lower than) the prior mean in case (ii) (resp. case (iii)). In Fig. \ref{subfig:similar_increase_similar_function} and \ref{subfig:similar_decrease_similar_function}, we illustrate the pooling and revealing intervals, and the affine function $f$ associated with the optimal information mechanism for cases (ii) -- (iii), respectively. 



The next corollary shows that when all nodes have monotone market sizes relative to their distances, one full information revelation interval becomes degenerate. 
\begin{corollary}\label{cor:two_regions}
    If all nodes have {monotonically} decreasing (resp. increasing) market sizes relative to distances, then given any prior distribution $F$, the optimal information mechanism pools states $\state_0\geq \zl$ (resp. $\state_0 \leq \zh$) such that $\zl < \mathbb{E}_F[S_0]$ (resp. $\zh > \mathbb{E}_F[S_0]$) and fully reveal states $\state_0 < \zh$ (resp. $\state_0 > \zh$).   
\end{corollary}
Corollary \ref{cor:two_regions} builds on Theorem \ref{theorem:simple_partitional}: When all nodes have decreasing (resp. increasing) market sizes, the revenue function is concave (resp. convex) for $\barstate_0 > \mathbb{E}_F[S_0]$ and convex (resp. concave) for $\barstate_0 < \mathbb{E}_F[S_0]$. Consequently, we show that one of the two full revelation region becomes empty, and the optimal information mechanism pool states above threshold $\zl$ (resp. below $\zh$). 

Furthermore, we show that when the commission rate $\rate$ exceeds a certain threshold, all market sizes are categorized as being similar as in \eqref{eq:similar}. Therefore, full information revelation is optimal with high commission rate following case (i) in Theorem \ref{theorem:simple_partitional}. 

\begin{corollary}\label{cor:high_rate}
Under Assumptions \ref{as:complete_market_balance} and \ref{as:no_depletion_max}, for any prior state distribution $F$ and any sequence of market sizes $\{s_i\}_{i \in \V}$, full information provision mechanism is optimal for any $\rate> \bar{\rate}$, where 
\[\bar{\rate} = 1- \min_{i, j \in \{\V| d_i \neq d_j\}}\left\vert \frac{d_{i}- d_j}{s_{i} - s_j}\right\vert.\]
\end{corollary}

 Finally, we present an algorithm to compute the thresholds $\zh$ and $\zl$ of the monotone partitional information mechanism. 
\begin{proposition}\label{prop:computation}
Under Assumptions \ref{as:complete_market_balance} and \ref{as:no_depletion_max}, suppose that there exists $0 \leq \dl \leq \dh \leq D$ such that nodes with distances less than $\dl$ or higher than $\dh$ have similar market sizes relative to distances, and nodes with distances in $[\dl, \dh]$ have increasing or decreasing market sizes relative to distances.  The thresholds of the pooling interval $\zl, \zh$ and the posterior mean $\zmean$ are computed by Algorithm \ref{alg:threshold_computation}.
\end{proposition}

{Proposition \ref{prop:computation} and Algorithm \ref{alg:threshold_computation} build on the proof of Theorem \ref{theorem:simple_partitional}. In particular, if the derivative of the linear revenue function in each regime is non-decreasing, then Algorithm \ref{alg:threshold_computation} returns identical $\zl$ and $\zh$ indicating that the pooling interval is degenerate. Otherwise, we identify a concave interval from regime $\underline{k}$ to $\bar{k}$. Building on Theorem \ref{theorem:simple_partitional}, Algorithm \ref{alg:threshold_computation} computes the pooling interval by searching for the affine function $g(s_0)$ that is tangent to $\revenue(s_0)$. In particular, the algorithm starts with the affine function being the revenue function associated with regime $\underline{k}$, and iteratively proceeds to regimes with increasing $k$ until $\bar{k}$. In each iteration, {the algorithm computes the interval thresholds as the intersection of the affine function and the revenue function, and computes the mean of the state within the corresponding  interval}. The algorithm terminates by finding the interval such that the state mean equals to the point, where the affine function is tangent to the revenue function. The proof of Theorem \ref{theorem:simple_partitional} ensures that such interval exists, and is the optimal pooling region. } 


\begin{algorithm}[ht]  
    \DontPrintSemicolon
    \KwIn{$\frac{d \revenue(\barstate_0)}{d \barstate_0}[k]$ for $k=-\Kminus, -\Kminus+1, \dots, -1, 0, 1, \dots, \Kplus-1, \Kplus$.}
    \eIf{$\frac{d \revenue(\barstate_0)}{d \barstate_0}[k]$ is non-decreasing for all $k$ }{$\zl \leftarrow \mathbb{E}_F[S_0]$, $\zh \leftarrow \mathbb{E}_F[S_0]$
    }{$\underline{k} \leftarrow \min_{k} \left\{k \left\vert \frac{d \revenue(\barstate_0)}{d \barstate_0}[k] > \frac{d \revenue(\barstate_0)}{d \barstate_0}[k+1]\right.\right\}$, $\bar{k} \leftarrow \max_{k} \left\{k \left\vert \frac{d \revenue(\barstate_0)}{d \barstate_0}[k-1] > \frac{d \revenue(\barstate_0)}{d \barstate_0}[k]\right.\right\}$\;
    $j \leftarrow \underline{k}$\;
    \While{$j \leq \bar{k}$}{
    \[g_j(z) = \frac{d \revenue(\barstate_0)}{d \barstate_0}[j] (z- \barstate_0[j]) + \revenue(\barstate_0[j]).\]
    $\zl[j] \leftarrow \max\left\{z \in S_0 \left\vert z\leq \barstate_0[j], \quad g_j(z) \geq \revenue(z) \right.\right\}$, $\zh[j] \leftarrow \min\left\{z \in S_0 \left\vert z\geq \barstate_0[j+1], \quad g_j(z) \geq \revenue(z) \right.\right\}$.\;
    \eIf{$\mathbb{E}_F[S_0 | \zl[j] \leq S_0 \leq \zh[j]] \leq \barstate_{0}[j+1]$}{break}{$j \leftarrow j+1$.}
    }}
    \eIf{$\mathbb{E}_F[S_0 | \zl[j] \leq S_0 \leq \zh[j]] \geq \barstate_{0}[j]$}{$\zl \leftarrow \zl[j]$, $\zh \leftarrow \zh[j]$, $\zmean \leftarrow \mathbb{E}_F[
S_0 | \zl[j] \leq S_0 \leq \zh[j]]$}{$\zmean \leftarrow \barstate_0[j]$, and $\zh, \zl$ are solutions of the following equations:
    \begin{align*}
        \zmean &= \mathbb{E}_F[S_0 | \zl \leq S_0 \leq \zh],~ 
        g_{\gamma}(z) = \gamma (z- \barstate_0[j])+ \revenue(\barstate_0[j]), ~
        \frac{d \revenue(\barstate_0)}{d \barstate_0}[j+1] \leq \gamma \leq \frac{d \revenue(\barstate_0)}{d \barstate_0}[j],\\
        \zl &= \max\left\{z \in S_0 \left\vert z<\barstate_0[j], ~ g_{\gamma}(z) \geq \revenue(z) \right.\right\}, ~
        \zh = \min\left\{z \in S_0 \left\vert z> \barstate_0[j], ~ g_{\gamma}(z) \geq \revenue(z) \right.\right\}.
    \end{align*}}
    \KwOut{$\zh, \zl, \zmean$.}
\caption{Computation of thresholds $\zl, \zh$ in the optimal monotone partitional information mechanism.}\label{alg:threshold_computation}
\end{algorithm}

\section{Generalized results for single shock}\label{sec:single_general}

\subsection{Generalization of optimal partitional information mechanism}
In this section, we drop Assumptions \ref{as:complete_market_balance} and \ref{as:no_depletion_max}, and instead make the following assumption on the initial market balance -- no agents have incentives to move across nodes. This assumption is without loss of generality since 
under any reasonable equilibrium model
agents will always make repositioning decisions so that $m$ satisfies the balancedness condition before the platform reveals any information of demand realization. 
\begin{assumption}[Initial market balance] \label{as:market_balance}The market is initially balanced in that no agents have incentive to move given the initial agent distribution $m$, i.e. 
\begin{align*}
  (1-\rate)(s_j - \beta_j m_j) - c_{ij}  \leq   (1-\rate)(s_i - \beta_i m_i)  \leq (1-\rate)(s_j - \beta_j m_j) + c_{ij}, \quad \forall i, j \in \V \setminus \{0\}, \\
  (1-\rate)(s_i - \beta_i m_i) - c_{i0}  \leq   (1-\rate)(\mathbb{E}_F(S_0) - \beta_0 m_0)  \leq (1-\rate)(s_i - \beta_i m_i) + c_{i0}, \quad \forall i \in \V \setminus \{0\}, 
\end{align*}
where $\mathbb{E}_F[S_0] = \int_{\state_0} \state_0 dF(\state_0)$ is the prior mean of the state $s_0$. 
\end{assumption}

We only describe the ideas of equilibrium characterization and regime computation in the main text. 
The formal algorithm of computing equilibrium regimes and complete characterization of equilibrium agent distribution and revenue function are included in Appendix \ref{apx:characterization}. 
Our characterization approach closely follows the one in the previous section, but we need to have slight modifications since we now impose a more permissive initial market balance condition.
In particular,
in the general case, the equilibrium regimes are jointly determined by the distance $d_i$ and the price with the initial agent mass $s_i - \beta_i m_i$ for all nodes. For each $\barstate_0 > \mathbb{E}_{F}[\state_0]$, there exists a set of nodes $\Treeplus$, where agents move to node 0, and the agents from the remaining nodes $V \setminus \Treeplus$ do not move. Moreover, agents from a subset of nodes $\Treedag \subseteq \Treeplus$ may entirely move to node 0 since the no depletion assumption is no longer imposed. 

In Appendix \ref{apx:characterization}, we develop Algorithm \ref{alg:plus} to compute equilibrium regimes for all $\barstate_0 \geq \mathbb{E}_F[S_0]$. The algorithm starts with regime 0, where both sets $\Treeplus[0]$ and $\Treedag[0]$ are empty, and all agents stay at their origin nodes. As $\barstate_0$ increases, agents from node $i \in V\setminus \{0\}$ with the maximum initial payoff plus distance (i.e. $(1-\rate) (s_i- \beta_i m_i) + d_i$) start to move to node 0 in equilibrium of regime $1$, and thus node $i$ is added to the set $\Treeplus[1]$. The regime threshold $\barstate_0[1]$ is computed as the state realization such that agents at node $i$ are indifferent between staying at $i$ and moving to node $0$. Similarly, Algorithm \ref{alg:plus} computes all possible equilibrium regimes $\k =1, \dots, \Kplus$, where the regime change from $k$ to $\k+1$ occurs when either a node $i\in \argmax_{\V\setminus \Treeplus[k]} (1-\rate) (s_i- \beta_i m_i) + d_i$ is added to the set $\Treeplus[k+1]$ or the price at node $0$ surpasses the maximum payoff plus the distance $(1-\rate) s_i +d_i$ of a node in $\Treeplus[k]$ so that all agents leave node $i$ (i.e. node $i$ is added into $\Treedag[k+1]$). The algorithm terminates with regime $\Kplus$ such that all agents move to node $0$. Since any node can only be added to $\Treeplus$ and $\Treedag$ once, the maximum number of regimes $\Kplus \leq 2|\V|$, and the algorithm terminates in less than $2|\V|$ iterations.

Likewise, for any $\barstate_0< \mathbb{E}_F[S_0]$, agents at node $0$ move to a subset of nodes $\Treeminus$, while the remaining agents do not move. In particular, as $\barstate_0$ decreases, nodes are included into $\Treeminus$ in decreasing order of $(1-\rate) (s_i - \beta_i m_i) -d_i$. 
In Appendix \ref{apx:characterization},  Algorithm \ref{alg:minus} computes the set $\Treeminus[k]$ for each regime $-k=-1, \dots, -\Kminus$, and the regime boundaries $\{\barstate_0[-k]\}_{k=1}^{\Kminus}$. Since each node is only added to $\Treeminus$ once, $\Kminus \leq |\V|$ and the algorithm terminates in less than $|\V|$ iterations. 



%


%

Based on the equilibrium characterization, we show that the platform's revenue function is a continuous piecewise linear function of $\barstate_0$ and the derivative of the linear piece $\frac{d \revenue(\barstate_0)}{d \barstate_0}[k]$ depends on the sets $\Treeplus[k]$ and $\Treedag[k]$ in regime $k$, and $\Treeminus[-k]$ in regime $-k$ (Proposition \ref{prop:eq_strategy} in Appendix \ref{apx:characterization}). 
We next generalize Theorem \ref{theorem:simple_partitional} to provide verifiable conditions that guarantee the existence of an optimal monotone partitional information mechanism under the following mild regularity condition of the revenue function $\revenue(\barstate_0)$.\footnote{
    It can be seen that this condition generically holds. More precisely,
    for any revenue function that violates Assumption \ref{as:regularity} with $k_a, k_b, k_c$, a slight perturbation of agent mass $m_i$ for $i \in \Treeplus[k_c]$ (or $\Treeminus[k_c]$ if $k_c<0$) will make $\{(\barstate_0[k_j], \revenue(\barstate_0[k_j]))\}_{j=a, b,c}$ not lie on the same line, and thus leads to satisfaction of this condition.
    }
    \begin{assumption}\label{as:regularity}
        There does not exist $k_a< k_b < k_c$ such that $\{(\barstate_0[k_j], \revenue(\barstate_0[k_j]))\}_{j=a, b,c}$ lie on the same line. 
    \end{assumption}

    We define \emph{concave intervals} of the revenue function $\revenue(\barstate_0)$ as $\{[\barstate_0[k_{\ell,a}], \barstate_0[k_{\ell, b}]]\}_{\ell=1}^{L}$, where $-\Kminus \leq k_{1,a} \leq  k_{1,b} < k_{2,a} \leq k_{2,b} < \cdots < k_{L,a} \leq k_{L,b} \leq \Kplus$. The platform's revenue function $\revenue(\barstate_0)$ is strictly concave (resp. strictly convex) at $\barstate_0[k]$ if $\frac{d \revenue(\barstate_0)}{d \barstate_0}[k-1] > (\text{resp.} <) \frac{d \revenue(\barstate_0)}{d \barstate_0}[k]$ and concave (resp. convex) at $\barstate_0[k]$ if $\frac{d \revenue(\barstate_0)}{d \barstate_0}[k-1] \geq (\text{resp.} \leq) \frac{d \revenue(\barstate_0)}{d \barstate_0}[k]$. For each interval $\ell$, the revenue function is strictly concave at $\barstate_0[k_{\ell,a}]$ and $ \barstate_0[k_{\ell,b}]$, and concave for all $\barstate_0[k]$ with $k_{\ell,a} < k< k_{\ell,b}$. Moreover, $\revenue(\barstate_0)$ is convex for all $k_{\ell,b} < k < k_{\ell+1,a}$ and strictly convex for at least one $k_{\ell,b} < \hat{k} < k_{\ell+1,a}$.\footnote{This condition is to ensure that  two cannot intervals cannot be combined into a single concave interval. }   

    Before presenting the result, we first introduce the four conditions (C1) -- (C4): \\
    \smallskip 
    \noindent \textbf{(C1)} $L \leq 1$, i.e. the revenue function has at most one concave interval.\\
    \smallskip 
    {\noindent \textbf{(C2)} There does not exist $\ell <  \ell'$ such that
\begin{subequations}\label{eq:tangent}
    \begin{align}
 &\frac{d \revenue(\barstate_0)}{d \barstate_0}[k_{\ell, a}-1]\left(\barstate_0- \barstate_0[k_{\ell, a}]\right)+ \revenue\left(\barstate_0[k_{\ell, a}]\right) - \revenue(\barstate_0) \geq 0, \quad \forall \barstate_0 \in \left[\barstate_0[k_{\ell', a}], \barstate_0[k_{\ell', b}]\right], \label{subeq:tangent_1}\\
 &\frac{d \revenue(\barstate_0)}{d \barstate_0}[k_{\ell', b}]\left(\barstate_0- \barstate_0[k_{\ell', b}]\right)+ \revenue\left(\barstate_0[k_{\ell', b}]\right) - \revenue(\barstate_0) \geq 0, \quad  \forall \barstate_0 \in \left[\barstate_0[k_{\ell, a}], \barstate_0[k_{\ell, b}]\right],\label{subeq:tangent_2}\\
        &\exists \barstate_0 \in \left[\barstate_0[k_{\ell', a}], \barstate_0[k_{\ell', b}]\right], \quad  s.t. 
 \quad \frac{d \revenue(\barstate_0)}{d \barstate_0}[k_{\ell, b}]\left(\barstate_0- \barstate_0[k_{\ell, b}]\right)+ \revenue\left(\barstate_0[k_{\ell, b}]\right) - \revenue(\barstate_0) < 0, \label{subeq:tangent_3}\\
        &\exists \barstate_0 \in \left[\barstate_0[k_{\ell, a}], \barstate_0[k_{\ell, b}]\right], \quad  s.t. 
 \quad  \frac{d \revenue(\barstate_0)}{d \barstate_0}[k_{\ell', a}-1]\left(\barstate_0- \barstate_0[k_{\ell', a}]\right)+ \revenue\left(\barstate_0[k_{\ell', a}]\right) - \revenue(\barstate_0) < 0.\label{subeq:tangent_4}
    \end{align}
\end{subequations}

    \normalfont

        The following lemma shows that when condition (C2) is not satisfied (i.e. there exists $\ell<\ell'$ that satisfy \eqref{eq:tangent}), there exists a unique linear function that is tangent to the concave intervals $\ell$ and $\ell'$ of $\revenue(\barstate_0)$. Otherwise, such linear tangent function does not exist. 
        \begin{lemma}\label{lemma:tangent}
        For any $\ell< \ell'$, there exists a unique linear function $g(z) = \gamma z+ \lambda$ such that 
        $g(\barstate_0[\kltangent]) = \revenue(\barstate_0[\kltangent]), \quad g(\barstate_0[\klptangent]) = \revenue(\barstate_0[\klptangent])$
        for some $k_{\ell, a} \leq \kltangent \leq k_{\ell,b}$ and $k_{\ell', a} \leq \klptangent \leq k_{\ell',b}$ and 
    $\gamma \in [\frac{d \revenue(\barstate_0)}{d \barstate_0}[\kltangent], \frac{d \revenue(\barstate_0)}{d \barstate_0}[\kltangent-1]] \cap [\frac{d \revenue(\barstate_0)}{d \barstate_0}[\klptangent], \frac{d \revenue(\barstate_0)}{d \barstate_0}[\klptangent-1]]$ 
        if and only if \eqref{eq:tangent} is satisfied. 
        \end{lemma}
        
\smallskip 
    \noindent \textbf{(C3)} For any pair of concave intervals $\ell< \ell'$ that satisfies \eqref{eq:tangent}, consider the associated affine function $g(z)$ and {$x= \tangentk$ (resp. $y= \tangentkp$)} 
    that satisfies $\revenue(x) = g(x)$, $\revenue(y)= g(y)$. Define $\zkl = \max\{z <x| \revenue(z) \geq g(z)\}$ and $       \zkph = \min\{z >y| \revenue(z) \leq g(z)\}$. Then, either $x, y < \mathbb{E}_F[S_0], \zkph = \sup \S_0$ or $x, y > \mathbb{E}_F[S_0], \zkl = \inf \S_0$.  

    \smallskip 
    \noindent \textbf{(C4)} For any pair of concave intervals $\ell< \ell'$ that satisfy \eqref{eq:tangent}, and $g(z)$, $x$, $y$, $\zkl$ and $\zkph$ defined as in (C3),  
    \begin{align} \label{subeq:C3}
        &\int_{[\zkl,y]} z dF(z) \geq x,
    \quad \int_{[x,\zkph]} z dF(z)\leq y, \quad 
\left(F^{-1}\left( \tilde{z}\right) - x\right) \tilde{z}
\geq  \int_{\zkl}^{F^{-1}\left( \tilde{z}\right)} (F(z) - F(\zkl)) dz,
    \end{align}
    where 

    \begin{align}\label{eq:zkl_zkph}
        \tilde{z}&= \frac{y F(\zkph) - xF(\zkl) - \zmean}{y-x}, \qquad \zmean = \int_{\zkl}^{\zkph} zdF(z). 
    \end{align}


\begin{proposition}\label{prop:three_conditions}
    There exists an optimal information mechanism that is monotone partitional if any of the four conditions (C1) -- (C4) is satisfied. 
\end{proposition}
Proposition \ref{prop:three_conditions} provides four verifiable conditions on the revenue function $R$ and the prior $F$ that guarantee the optimality of monotone partitional information mechanism. Among the four conditions, (C1) -- (C2) only concern the revenue function, while (C3) -- (C4) set conditions on $R$ and $F$ jointly. 

In particular, (C1) ensures that the revenue function has at most one concave interval so that the optimal information mechanism can have at most one pooling region -- recall that cases (i) -- (iii) in Theorem \ref{theorem:simple_partitional} satisfy this condition. (C2) ensures that we cannot find an affine function that is tangent with two concave intervals. The theory of optimal information design shows that the existence of such affine functions indicates that the optimal information mechanism may contain double-interval structures that are not monotone -- pooling high and low states in the interval to generate one posterior mean, and pooling the remaining states in the middle to generate another posterior mean. Specifically, the two posterior means take values of the two tangent points \cite{candogan2019persuasion, dworczak2019simple, arieli2023optimal}. Both (C1) and (C2) ensure the optimality of monotone partitional mechanism by ruling out the existence of such affine functions. {Thus, these two conditions are equivalent to the ``affine closure" condition of \cite{dworczak2019simple} for the optimality of monotone partitional mechanisms, and they are easy to verify given the piecewise linear payoff struture in our problem.}


Moreover, Proposition \ref{prop:three_conditions} extends the result in \cite{dworczak2019simple} by providing two other conditions (C3) -- (C4) that allow the revenue function not to be affine closed, but can still guarantee the optimality of monotone partitional mechanism. Here, (C3) ensures that any posterior mean distribution that generates the two posterior mean values associated with the tangent points cannot be mean preserving spread of the prior, and thus is infeasible. Additionally, (C4) characterizes conditions on the prior $F$ to ensure that no feasible double-interval structure can be constructed to generate posterior means associated with any pair of tangent points,
{and we show that in that case the optimal information mechanism is necessarily monotone partitional.}
 

\subsection{Computing optimal information mechanism}
In this section, we present the approach of computing an optimal information mechanism. We first show that any two posterior mean distributions lead to the same revenue if they induce the same expected posterior mean value and probability in each equilibrium regime, where the revenue function is linear. 
\begin{lemma}\label{lemma:equivalence}
   For any two information mechanisms with posterior mean distribution $G$ and $G'$ such that for all $k=-\Kminus, \dots, \Kplus$, 
   \begin{align*}
       \int_{\barstate_0[k-1]}^{\barstate_0[k]} z dG(z) = \int_{\barstate_0[k-1]}^{\barstate_0[k]} z dG'(z), \quad G(\barstate_0[k])- G(\barstate_0[k-1])= G'(\barstate_0[k])- G'(\barstate_0[k-1]),
   \end{align*}
   we have $R_G= R_{G'}$, where $R_G$ ($R_{G'}$) is the platform's revenue induced by the posterior mean distribution $G$ (resp. $G'$).  
\end{lemma} 
The proof of Lemma \ref{lemma:equivalence} leverages the fact that revenue function $\revenue(\barstate_0)$ is a continuous and piecewise linear function of $\barstate_0$. 
Thus, the revenue only depends on the   posterior mean of the state $y_k = \int_{\barstate_0[k-1]}^{\barstate_0[k]} z dG(z)/ (G(\barstate_0[k])- G(\barstate_0[k-1]))$ and 
the probability of inducing this posterior mean $p_k = G(\barstate_0[k])- G(\barstate_0[k-1])$ in each regime $k$, rather than the exact posterior mean distribution $G$. That is, any two posterior mean distributions that induce the same tuples $\{(p_k, y_k)\}_{k=-\Kminus}^{\Kplus}$ achieve the same revenue. The following proposition further shows that the optimal $\{(p_k, y_k)\}_{k=-\Kminus}^{\Kplus}$ can be computed by a convex optimization program. 
\begin{proposition}\label{prop:p_y}The tuple $\{(p_k, y_k)\}_{k=-\Kminus}^{\Kplus}$ associated with the optimal information mechanism can be computed as an optimal solution of the following convex program: 
    \begin{subequations}
\begin{align}
\max_{p, y} ~ &\sum_{k=-\Kminus}^{\Kplus} p_k \revenue(y_k/p_k), \notag\\
s.t. \quad & \sum_{j=-\Kminus}^{k} y_j \leq \int_{1-\sum_{j=-\Kminus}^k p_{j}}^{1} F^{-1}(x) dx, \quad \forall k=-\Kminus, \dots, \Kplus, \label{subeq:tight}\\
& \barstate_0[k-1] p_{k} \leq y_{k} \leq \barstate_0[k] p_{k}, \quad \forall k=-\Kminus, \dots, \Kplus, \\
& \sum_{k=-\Kminus}^{\Kplus} p_k =1, \quad p_k \geq 0, \quad \forall k=-\Kminus, \dots, \Kplus.
\end{align}
\end{subequations}
\end{proposition}
Proposition \ref{prop:p_y} builds on Lemma \ref{lemma:equivalence} and the convex optimization framework developed in \cite{candogan2019persuasion}. 
{In particular, \cite{candogan2019persuasion} demonstrated that the optimal information mechanism can be constructed from an optimal $\{(p_k, y_k)\}_{k=-\Kminus}^{\Kplus}$, where each $k$ with probability mass $p_k>0$ is associated with a posterior mean $y_k/p_k$ given the optimal information mechanism. Moreover, the tightness of constraint \eqref{subeq:tight} indicates whether or not each posterior mean is induced by a single pooling interval or by a double-interval structure, and the interval thresholds can be explicitly constructed from the optimal $\{(p_k, y_k)\}_{k=-\Kminus}^{\Kplus}$.}   
 The details closely follow \cite{candogan2019persuasion,candogan2021optimal}, and are omitted for brevity.





\bibliographystyle{plainnat}
\bibliography{library}

\newpage
\appendix
\section{Discussion: Computing best partitional mechanisms}\label{apx:discussion}
In earlier sections we identified conditions under which partitional mechanisms are optimal.
In this section we  focus on more general 
settings where they are not optimal. Due to their simple and intuitive features, 
from a practical point of view,
it may still be of interest 
to identify the best partitional mechanism in such settings.
We provide
an approach for obtaining (near-optimal) partitional mechanisms. 

To this end,
we first discretize the state space 
that induces  $\epsilon$ increments 
in the quantile space. 
We assume that we have access to an oracle that evaluates the inverse CDF function at the desired point, though the results go through  even if these can be only approximately computed.
We then argue that the dynamic programming (DP) approach
of \cite{candogan2020information}
applies in this setting to derive   the optimal partitional mechanism whose cutoffs are restricted to these points. It is easy to see that the payoffs are Lipshitz continuous in the cutoffs of the partitional mechanism. This, together with the fact that our DP runs in polynomial time in $1/\epsilon$ ensures that the resulting algorithm is a fully polynomial-time approximation scheme (FPTAS).
We first describe the approach 
for the single shock case, and then discuss its extensions to the multiple shock case.

Let $c_\ell=F^{-1}(\epsilon \ell)$ for $k\in {\cal I}:=\{0,\dots, \lfloor 1/\epsilon \rfloor \}$. 
If $c_\ell\neq 1$ for any $\ell$, then let $c_{\lfloor 1/\epsilon \rfloor+1}=1$.
Define the set of feasible cutoffs as $\{c_\ell\}, $ and the associated index set by ${\cal I}$.
Let $\max {\cal I}=L$
For any $\ell_1,\ell_2\in {\cal I}$ where $\ell_1\leq \ell_2$, let $p(\ell_1,\ell_2) = F(c_{\ell_2})-F(c_{\ell_1})$, and
 $w(\ell_1,\ell_2)=
R \left(\mathbb{E}[s_0|s_0\in [c_{\ell_1},c_{\ell_2}]\right)
\bigg/p(\ell_1,\ell_2)$.
Suppose that we choose indices
${\cal I}_c=\{\ell_0,\ell_1,\dots,\ell_K\}\subseteq {\cal I}$ with $\ell_0=0$, $\ell_K=L$. In this case, it can be seen that the expected payoff of the designer can be expressed as $\sum_{k=1}^K w(\ell_{k-1},\ell_k)$.
It can be readily seen that any feasible partitional mechanism (with cutoffs in $\{c_k\}_{k\in {\cal I}}$) can be associated with an index set of the type ${\cal I}_c$, the problem reduces to finding the best such set.

However, this problem can naturally be formulated as a DP. Specifically, suppose that we sequentially add indices to ${\cal I}_c$ starting with $\{ L\}$, at each step adding a new index that is smaller than the previously added ones, and stopping once index $0=\min {\cal I}$ is added to ${\cal I}_c$. 
Suppose we have a set ${\cal I}_c=\{\ell_{k+1},\dots, \ell_K\}$
with $\ell_{k+1}<\dots< \ell_K=L$,
of indices that are already chosen, and suppose that we next add index   $\ell_k<\ell_{k+1}$ to ${\cal I}_c$.
Denote the stage payoff for adding an index $\ell_k$ to the current set   by $w(\ell_k,\ell_{k+1})$, and denote the continuation payoff by $V(\ell_k)$. It can be readily seen that the objective value associated with the optimal mechanism 
(with cutoffs in $\{c_k\}_{k\in {\cal I}}$)
is $V(L)$, and we have the following Bellman equation:
\begin{equation} \label{eq:DP}
V(\ell_k)= \max_{\ell<\ell_k|\ell\in {\cal I}}V(\ell) + w(\ell, \ell_k).
\end{equation}
This equation can be solved via standard dynamic programming techniques, and yield the optimal partitional mechanism with cutoffs restricted to ${\cal I}$, which in turn yields an FPTAS as explained earlier.\footnote{See \cite{candogan2020information} for a detailed algorithm description. The aforementioned paper's algorithm also applies in richer settings where different groups of agents have access to different signals, and information can spill over from one agent to another according to an underlying communication network. }

In the discussion above, we did not specify which of the assumptions (Assumption \ref{as:complete_market_balance},\ref{as:no_depletion_max},
\ref{as:market_balance}) should be imposed. This is deliberate. If these assumptions hold, then the $R(\cdot)$ function can be obtained in closed form, and it can be used to construct $w(\cdot,\cdot)$ as described earlier. Otherwise, 
for any $\ell_1, \ell_2\in {\cal I}$ and the posterior mean level
$\mathbb{E}[s_0|s_0\in [c_{\ell_1},c_{\ell_2}] ]$
we can solve the optimization formulation in Proposition \ref{prop:potential} to obtain the distribution of agents for  different posterior mean levels
that are induced by partitional mechanisms with cutoffs in $\{c_\ell\}_{\ell\in {\cal I}}$.
We then evaluate the induced revenue. After solving these auxiliary optimization problems we evaluate the $R(\cdot)$  function at all relevant posterior mean levels and, in turn, construct the $w(\cdot,\cdot)$ function.

We next extend our approach to the multiple shock case. To start, we need to extend the definition of partitional information structures to the multiple shock case.
A natural idea is to choose partitions for each $\{S_i\}$ separately, and reveal the partition elements to which each $S_i$ belongs.
However, there is a fundamental difficulty: The number of possible signal realizations grows exponentially in the number of shock centers, thereby leading to computational difficulties when large number of nodes are impacted by shocks. This is to some extent expected, since in the aforementioned settings
the designer's payoff is piecewise linear with exponentially many pieces, and 
solving an optimization involving such a rich  class of functions is nontrivial.

That said, there is a practically relevant version of the problem that is  also tractable. Specifically,
suppose that we have finitely many scenarios $\{\sigma_k\}_{k\in \Sigma}$
and an associated probability distribution 
$\{\rho_k\}_{k\in \Sigma}$, where $\Sigma=\{1,\dots,D\}$.
Associated with  scenario $k$  is a fixed  shock vector $\nu^k$, and random variable $\theta^k$ with distribution $F^k$. When this scenario is realized the price curve in node $i$ takes the form \[p_i(q_i)= s_i + \theta^k \nu^k_i.\]
In other words, when this scenario is realized the intercept of the price curve in each node is perturbed according to the vector $\nu^k$. Note that we do not make any restrictions on the entries of $\nu^k$. For instance, multiple entries can be nonzero, e.g., capturing the fact 
that multiple nodes simultaneously experience varying degrees of shocks (which  in the ride-sharing application can capture 
some regions of the city jointly experiencing  demand shocks, e.g., due to weather events or special events leading to demand surge in a wide area). 
We also allow for negative entries/shocks (that in the ride-sharing application can capture reduced demand in some areas while others are experiencing increased demand, e.g., due to shocks in the public transit system).

As before, we discretize the support of each $\theta^k$ in a way that leads to $\epsilon$ increments in the quantile space.
Denote the corresponding cutoffs
for scenario $k$ by $\{c^k_{\ell}\}_\ell$, and the associated index set by ${\cal I}^k$.
For each scenario $k$, and  indices 
$\ell_1,\ell_2\in {\cal I}^k$, 
  we compute the induced payoff of the platform when the posterior mean vector is $\mathbb{E}_{F^k}[s + \theta^\k \nu^k | \theta^k\in [c_{\ell_1},c_{\ell_2}] ]$.
While, unlike the single shock case, this quantity is not possible to obtain in closed form,  Proposition \ref{prop:potential}  still applies to characterize equilibrium agent distribution for each scenario and pairs of discretization indices $(k,\ell_1, \ell_2 )$. 

We once again focus on partitional mechanisms.
However, note that without the knowledge of the underlying scenario, the partitional structure is not very intuitive or practical.\footnote{For instance, suppose that two scenarios represent demand surging in two different parts of the city. Without revealing which scenario is realized it is not meaningful to reveal that the demand is in top, say, 20\% of the possible levels.}
Motivated by this,
we restrict attention to mechanisms that (i) reveal the scenario that is realized, and (ii) use partitional information structures to map state realization to signals for each scenario. 

It can be readily seen that for such mechanisms, the design problem decouples over different scenarios. That is, the partition chosen for one scenario has no impact on the payoff obtained when another scenario is realized. As such, for each scenario we have a DP recursion of the type \eqref{eq:DP}, where $V(\cdot)$ and $w(\cdot,\cdot)$ (and more fundamentally the revenue function $R(\cdot)$) are now scenario-specific.

In short, the DP approached introduced earlier readily extends to multiple demand shock scenarios. This gives a simple and computationally efficient recipe for constructing practically relevant information disclosure mechanisms.

\section{Proofs of statements in Sections \ref{sec:model}}\label{apx:proof_two}
\noindent\emph{Proof of Proposition \ref{prop:potential}.} For any $t \in \T$, $\Phi(x|t)$ is a potential function since: 
\begin{align*}
\frac{d \Phi(x|t)}{d x_{ij(t)}} = (1-\rate) (\mathbb{E}[s_j|t]- \beta_j q_j(t)) - c_{ij} = u_{ij}(x|t), \quad \forall i, j \in V. 
\end{align*}
Therefore, following \cite{}, we know that $\xeq$ is a maximizer of the potential function. 

Moreover, given any $\qeq(t)$, since the associated equilibrium strategy $\xeq(t)$ maximizes the potential function, $\xeq(t)$ must be an optimal solution of the following problem: 
\begin{equation}
\begin{split}
    \min_x  \quad & \sum_{i, j \in V} c_{ij} x_{ij}(t), \\
    s.t. \quad &\sum_{i \in V} x_{ij}(t)= \qeq_j(t), \quad \forall j \in V, \\
    & \sum_{j \in V} x_{ij}(t) = m_i, \quad \forall i \in V, \\
    & x_{ij}(t) \geq 0, \quad \forall i, j \in V. 
 \end{split}
\end{equation}
From the duality theory, the optimal value $\psi(\qeq|t) = \sum_{i, j \in V} c_{ij} x^*_{ij}(t)$ is a convex function of $\qeq(t)$ (Theorem 5.1 on page 213 in \cite{bertsimas1997introduction}). Moreover, we denote $\phi(q|t)= (1-\rate)\sum_{i \in \V} \int_{0}^{q_i(t)}(\mathbb{E}[\state_i|\signal] - \beta_j z) dz$ note that 
\begin{align*}
\frac{\partial^2 \phi(q|t)}{\partial q_i \partial q_j} = \left\{
\begin{array}{ll}
- (1-\rate) \beta_i, & \quad i=j,\\
0, & \quad i \neq j.
\end{array}
\right.
\end{align*}
That is, $\phi(q|t)$ is strictly concave in $q$. We can re-write the potential function as a function of $q$, $\phi(q|t) - \psi(q|t)$, which is strictly concave in $q$. Thus, the equilibrium agents' distribution $\qeq(t)$ is unique for all $t \in \T$. \hfill $\square$

\medskip 
\noindent\emph{Proof of Proposition \ref{prop:ordering}.} We note that the feasible set of the optimization problem \eqref{eq:opt_single} associated with $F'$ is super set of that associated with $F$. Therefore, $\revenue_{G^{'*}} \geq \revenue_{G^{*}}$, where $G^{'*}$ (resp. $G^*$) is the optimal posterior mean distribution given prior $F'$ (resp. $F$). Moreover, since $\mathbb{E}_{F}[\barstate_0] = \mathbb{E}_{F'}[\barstate_0]$, we have $\revenue_F = \revenue(\mathbb{E}_F[\barstate_0]) = \revenue(\mathbb{E}_{F'}[\barstate_0]) = \revenue_{F'}$. Therefore, $V^*_F \leq V^{*}_{F'}$. \hfill $\square$

\section{Equilibrium characterization for the single shock case}\label{apx:characterization}
In this section, we provide a complete equilibrium characterization for the single shock case under Assumption \ref{as:market_balance}.

The following lemma demonstrates structural properties of equilibrium strategy profile and agents' distribution. In particular, for any $\barstate_0> \mathbb{E}_{F}[\state_0]$, that there exists an equilibrium strategy profile such that the nodes in the network are partitioned into three sets $\V= (V\setminus \Treeplus)\cup (\Treeplus \setminus \Treedag) \cup \Treedag$, where agents from $i \in (V\setminus \Treeplus)$ do not move, a fraction of agents from $i \in (\Treeplus \setminus \Treedag)$ move to node $0$ and the rest stay at node $i$, and all agents from $i \in \Treedag$ move to node $0$ in equilibrium. Additionally, for any $\barstate_0< \mathbb{E}_{F}[\state_0]$, nodes are partitioned into two sets $\V= (\V\setminus\Treeminus) \cup \Treeminus$ such that agents from node $i$ move to nodes in $\Treeminus$, and agents from the rest of the network do not move. 
\begin{lemma}\label{lemma:threshold_structure}
For any $\barstate_0> \mathbb{E}_{F}[\state_0]$, the unique agents' distribution in equilibrium $\qeq$ satisfies: 
\begin{subequations}\label{eq:qeq_positive}
\begin{align}
&\qeq_i = \frac{1}{\beta_i} \left( s_i - \barstate_0 + \frac{d_i}{1-\rate}\right)+ \frac{\beta_0}{\beta_i} \qeq_0 \in (0, m_i), \quad \forall i\in \Treeplus \setminus \Treedag,\label{eq:q_positive_relation}\\
&\qeq_i=0, \quad \forall i \in \Treedag, \\
&\qeq_i=m_i, \quad \forall i \in \V \setminus \{\Treeplus \cup \{0\}\}, 
\end{align}
\end{subequations}
where 
\begin{subequations}\label{eq:tree}
    \begin{align}
&\Treeplus = \{\V| (1-\rate) (s_i- \beta_i m_i) + \distance_i < (1-\rate) (\barstate_0 - \beta_0 \qeq_0)  \}, \label{subeq:Treeplus}\\
    &\Treedag = \{\V| (1-\rate) s_i + \distance_i \leq (1-\rate) (\barstate_0 - \beta_0 \qeq_0) \},
\end{align}
\end{subequations}and there exists an equilibrium strategy profile such that
\begin{align}\label{eq:xeq_positive}
\xeq_{00}=m_0, \quad \xeq_{i0}= m_i - \qeq_{i}, \quad \xeq_{ii}= \qeq_{i}, \quad \xeq_{ij}=0, \quad \forall i \in \V\setminus\{0\}, \quad \forall j \in \V\setminus \{0, i\}. 
\end{align}For any $\barstate_0< \mathbb{E}_{F}[\state_0]$, the unique agents' distribution in equilibrium $\qeq$ satisfies: 
\begin{subequations}\label{eq:qeq_negative}
\begin{align}
&\qeq_i = \frac{1}{\beta_i} \left( s_i - \barstate_0 - \frac{d_i}{1-\rate}\right)+ \frac{\beta_0}{\beta_i} \qeq_0 > m_i, \quad \forall i\in \Treeminus,\label{eq:q_negative_relation}\\
&\qeq_i=m_i, \quad \forall i \in \V \setminus\{ \Treeminus \cup \{0\}\}, 
\end{align}
\end{subequations}
where 
\begin{align}\label{subeq:Treeminus}
&\Treeminus = \{\V| (1-\rate) (s_i- \beta_i m_i) - \distance_i > (1-\rate) (\barstate_0 - \beta_0 \qeq_0)  \}, 
\end{align}
and there exists an equilibrium strategy profile such that
\begin{subequations}\label{eq:xeq_negative}
\begin{align}
\xeq_{0i}&= \qeq_{i} - m_i, \quad \forall i \in \Treeminus, \quad \xeq_{0i} =0, \quad \forall i \in \V \setminus \{0 \cup \Treeminus\}, \quad \xeq_{00} = \qeq_0, \\
\xeq_{ii}&=m_i, \quad \forall i \in \V\setminus \{0\}, \quad \xeq_{ij} = 0, \quad \forall j \in \V\setminus \{0, i\}. 
\end{align}
\end{subequations}
Furthermore, $\qeq_0$ increases in $\barstate_0$. 
\end{lemma}

\medskip 
\noindent\emph{Proof of Lemma \ref{lemma:threshold_structure}.} We first prove that given any $\barstate_0 > \mathbb{E}_F[\state_0]$, agents originating from all nodes do not have incentive to deviate from their equilibrium strategy $\xeq$ as in \eqref{eq:xeq_positive} given $\qeq$ in \eqref{eq:qeq_positive}. We first note that $\qeq$ as in \eqref{eq:qeq_positive} ensures that 
\begin{align}\label{eq:price_tree_positive}
(1-\rate) (s_i - \beta_i \qeq_i) = (1-\rate) (\barstate_0 - \beta_0 \qeq_0) - \distance_i, \quad \forall i \in \Treeplus \setminus \Treedag.
\end{align}

(1) For any $i \in V \setminus \{\Treeplus \cup \{0\}\}$, we note that the payoff of staying at node $i$ is $(1-\rate) (s_i -\beta_i \qeq_i) = (1-\rate) (s_i -\beta_i m_i)$, which is higher or equal to $(1-\rate)(\barstate_0- \beta_0 \qeq_0) - \distance_i$ -- the expected payoff of moving to node $0$. Thus, agents at node $i$ has no incentive to move to node $0$ in equilibrium. Additionally, following \eqref{eq:price_tree_positive}, we have
\begin{align*}
    (1-\rate) (s_j - \beta_j \qeq_j) - c_{ij} &=  (1-\rate)(\barstate_0- \beta_0 \qeq_0) - \distance_j - c_{ij} \stackrel{(a)}{\leq} (1-\rate)(\barstate_0- \beta_0 \qeq_0) - \distance_i\\
    &\leq (1-\rate) (s_i -\beta_i m_i) = (1-\rate) (s_i -\beta_i \qeq_i), \quad \forall j \in \Treeplus,
\end{align*}
where (a) is due to the triangular inequality of distances, i.e. $d_i \leq d_j + c_{ij}$ for any $i, j \in \V$. Therefore, agents at node $i \in \V \setminus \{\Treeplus\cup \{0\}\}$ have no incentive to move to any node in $\Treeplus \setminus \Treedag$. Furthermore, for any $j \in \Treedag$, 
\begin{align*}
    (1-\rate) (s_j - \beta_j \qeq_j) - c_{ij} &= (1-\rate) s_j - c_{ij} \leq  (1-\rate)(\barstate_0 - \beta_0 \qeq_0) - \distance_j - c_{ij} \leq (1-\rate)(\barstate_0 - \beta_0 \qeq_0) - \distance_j  \\
    & \leq (1-\rate) (s_i -\beta_i m_i) = (1-\rate) (s_i -\beta_i \qeq_i), \quad \forall j \in \Treedag. 
\end{align*}
Thus, agents originating from node $i$ have no incentive to move to any node $j \in \Treedag$. Additionally, we know from Assumption \ref{as:market_balance} that agents at $i$ have no incentive to move to any node $j \in \V \setminus \{\Treeplus \cup \{0\}\}$, where the equilibrium agent distribution $\qeq_j=m_j$. Thus, we conclude that agents originating from node $i \in \V \setminus \{\Treeplus \cup \{0\}\}$ have no incentive to move to any node in $\V$, i.e. have no incentive to deviate given the equilibrium strategy profile in \eqref{eq:qeq_positive} and \eqref{eq:xeq_positive}.

(2) For any $i \in \Treeplus \setminus \Treedag$, we know from \eqref{eq:price_tree_positive} and \eqref{subeq:Treeplus} that 
\[(1-\rate) (s_i - \beta_i \qeq_i) = (1-\rate) (\barstate_0 - \beta_0 \qeq_0) - \distance_i > (1-\rate) (s_i - \beta_i m_i),\]
and thus $\qeq_i < m_i$. Additionally, since $i \notin \Treedag$, 
\[(1-\rate) s_i  > (1-\rate) (\barstate_0 - \beta_0 \qeq_0) 
 - d_i = (1-\rate) (s_i - \beta_i \qeq_i),\]
 we have $\qeq_i > 0$. 
Thus, we have proved that $\xeq_i \in (0, m_i)$ indicating that agents originating from node $i \in \Treeplus \setminus \Treedag$ split between staying at $i$ and moving to node $0$. We know from \eqref{eq:price_tree_positive} that indeed agents are indifferent between staying at node $i$ and moving to $0$. It remains to show that agents at $i$ do not strictly prefer to move to any other node $j \in \V \setminus \{0, i\}$. We note that agents do not strictly prefer to move to any node $j \in \Treeplus \setminus\Treedag$ since 
\begin{align*}
    (1-\rate)(s_i - \beta_i \qeq_i) &= (1-\rate)(\barstate_0 - \beta_0 \qeq_0) - \distance_i = (1-\rate)(s_j - \beta_j \qeq_j) + \distance_j - \distance_i \\
    &\geq (1-\rate)(s_j - \beta_j \qeq_j) - |\distance_i - \distance_j| \geq (1-\rate)(s_j - \beta_j \qeq_j) - c_{ij},
\end{align*} 
where the last inequality arises from the triangular inequality. Similarly, agents do not strictly prefer to move to any $j \in \Treedag$ since 
\begin{align*}
    (1-\rate)(s_i - \beta_i \qeq_i) &= (1-\rate)(\barstate_0 - \beta_0 \qeq_0) - \distance_i \geq (1-\rate)s_j  + \distance_j - \distance_i= (1-\rate)(s_j - \beta_j \qeq_j) + \distance_j - \distance_i \\
    &\geq (1-\rate)(s_j - \beta_j \qeq_j) - |\distance_i - \distance_j| \geq (1-\rate)(s_j - \beta_j \qeq_j) - c_{ij}, \quad \forall j \in \Treedag.
\end{align*} 
Finally, since $\qeq_i\leq m_i$, we know that the payoff of staying at node $i$ with $\qeq_i$ is no less than the payoff with the original agent mass $m_i$. Since agents at node $i$ have no incentive to move to any $j \in \V \setminus \{\Treeplus \cup \{0\}\}$, and the price at node $j$ does not change after the repositioning, we know that agents at node $i$ have no incentive to move to node $j \in \V \setminus \{\Treeplus \cup \{0\}\}$. We can thus conclude that agents at node $i$ do not strictly prefer to move to any node $j \in \V \setminus \{0, i\}$, and hence $\xeq$ and $\qeq$ are equilibrium for agents at node $i$. 

(3) For any $i \in \Treedag$, \eqref{eq:qeq_positive} and \eqref{eq:xeq_positive} indicate that all agents move from $i$ to 0 in equilibrium. We note that agents do not strictly prefer to move to other nodes $j \in \Treeplus \setminus \Treedag$: 
\begin{align*}
    (1-\rate)(\barstate_0 - \beta_0 \qeq_0) - \distance_i =  (1-\rate)(s_j - \beta_j \qeq_j) + d_j - d_i \geq (1-\rate)(s_j - \beta_j \qeq_j) - c_{ij}, \quad \forall j \in \Treeplus \setminus \Treedag.
\end{align*} 
or $j \in \Treedag$: 
\begin{align*}
    (1-\rate)(\barstate_0 - \beta_0 \qeq_0) - \distance_i >  (1-\rate)(s_j - \beta_j \qeq_j) + d_j - d_i \geq (1-\rate)(s_j - \beta_j \qeq_j) - c_{ij}, \quad \forall j \in \Treedag.
\end{align*}
Moreover, agents also do not prefer to move to nodes $j \in \V \setminus \Treeplus$ since \begin{align*}
    (1-\rate)
(s_i - \beta_i \qeq_i) = (1-\rate)
s_i > (1-\rate)
(s_i - \beta_i m_i) \geq (1-\rate)
(s_j - \beta_j m_j) - c_{ij} \quad \forall j \in \V \setminus \Treeplus,
\end{align*}
where the last inequality follows from Assumption \ref{as:market_balance}. 

(4) For agents at node $0$, we know from \eqref{eq:price_tree_positive} that the payoff of staying at node $0$ is higher than moving to any node $i \in \Treeplus\setminus \Treedag$. Following analogous argument as in (1), we know that agents from node $i$ also do not have incentive to move to nodes in $\Treedag$ and nodes in $V\setminus \{\Treeplus\cup \{0\}\}$.

From (1) - (4), we conclude that no agents have strict incentive to deviate given $\qeq$ and $\xeq$ as in \eqref{eq:qeq_positive} and \eqref{eq:xeq_positive}. Thus, $\qeq$ and $\xeq$ are the agents' equilibrium distribution and equilibrium strategy profile.

Similarly, we prove that given any $\barstate_0 < \mathbb{E}_F[\state_0]$, $\qeq$ as in \eqref{eq:qeq_negative} is the agents' distribution in equilibrium and $\xeq$ in \eqref{eq:xeq_negative} is an associated equilibrium strategy profile by showing that agents originating from all nodes do not have incentive to deviate.

(i) For any $i \in V \setminus \Treeminus$, we note that the payoff of staying at node $0$, $(1-\rate) (\barstate_0 -\beta_0 \qeq_0)$, is higher or equal to $(1-\rate)(s_i - \beta_i \qeq_i) - \distance_i$, i.e. the expected payoff of moving to node $i$. Thus, agents from node $0$ have no incentive to move to node $i \in V\setminus \Treeminus$ in equilibrium. Additionally, we note that $\qeq$ as in \eqref{eq:qeq_negative} ensures that 
\begin{align}\label{eq:price_tree_negative}
(1-\rate) (s_i - \beta_i \qeq_i) - \distance_i = (1-\rate) (\barstate_0 - \beta_0 \qeq_0)  , \quad \forall i \in \Treeminus.
\end{align}
That is, agents from node $0$ are indifferent between staying at node $0$ or moving to a node $i \in \Treeminus$. Thus, $\xeq_0$ as in \eqref{eq:xeq_negative} is equilibrium strategy for agents from node $0$. 

(ii) For any $i \in \Treeminus$, agents from $i$ do not strictly prefer to move to any other nodes $j \in \Treeminus$ compared to staying at node $i$ since 
\begin{align*}
    (1-\rate) (s_j - \beta_j \qeq_j) - c_{ij} &=  (1-\rate)(\barstate_0- \beta_0 \qeq_0) + \distance_j - c_{ij} \leq (1-\rate)(\barstate_0- \beta_0 \qeq_0) + \distance_i\\
    &= (1-\rate) (s_i -\beta_i m_i) , \quad \forall j \in \Treeplus,
\end{align*}
Moreover, agents also do not strictly prefer to move to any $j \in \V \setminus \Treeminus$ since 
\begin{align*}
    (1-\rate) (s_j - \beta_j \qeq_j) - c_{ij} & \leq  (1-\rate)(\barstate_0- \beta_0 \qeq_0) + \distance_j - c_{ij} \leq (1-\rate)(\barstate_0- \beta_0 \qeq_0) + \distance_i\\
    &= (1-\rate) (s_i -\beta_i m_i) , \quad \forall j \in \V \setminus \Treeminus.
\end{align*}
Thus, agents originating from node $i$ have no incentive to move to any other node in $\V\setminus \{i\}$, i.e. $\xeq_{ii}=m_i$ is an equilibrium strategy. Additionally, since $(1-\rate)(\state_i - \beta_i m_i) > (1-\rate)(\barstate_0 - \beta_0 \qeq_0)+ d_i$ and $(1-\rate)(\state_i - \beta_i \qeq_i) = (1-\rate)(\barstate_0 - \beta_0 \qeq_0)+ d_i$, we know that $\qeq_i > m_i$ for all $i \in \Treeminus$.

(iii) For any $i \in \V\setminus \Treeminus$, we know from Assumption \ref{as:market_balance} that agents from $i$ do not strictly prefer to move to $j \in \V \setminus \Treeminus$. Additionally, since agents from $i$ do not strictly prefer to move to $j \in \Treeminus \cup \{0\}$ with the initial agent distribution $m_i$ and $\qeq_j> m_j$, agents also do not strictly prefer to move to $j \in \Treeminus \setminus \{0\}$ in equilibrium. Thus, $\xeq_{ii}=m_i$ is an equilibrium strategy for $i \in \Treeminus$. 

From (i) - (iii), we conclude that no agents have strict incentive to deviate given $\qeq$ and $\xeq$ as in \eqref{eq:qeq_negative} and \eqref{eq:xeq_negative}. Thus, $\qeq$ and $\xeq$ are indeed equilibrium agents' equilibrium distribution and equilibrium strategy profile. 

Finally, we prove that $\qeq_0$ is increasing in $\barstate_0$. Consider any $\barstate_0< \barstate'_0$. We denote the equilibrium agents' distribution at node $0$ associated with $\barstate_0$ and $\barstate'_0$ as $\qeq_0$ and $\qeqp_0$, respectively. For any $\barstate_0< \mathbb{E}_F[\state_0]< \barstate'_0$, we must have $\qeq_0< m_0 < \qeqp_0$. For any $\mathbb{E}_F[\state_0] < \barstate_0< \barstate'_0$, we assume for the sake of contradiction that $\qeqp_0 < \qeq_0$. From \eqref{eq:qeq_positive} and \eqref{subeq:Treeplus}, we know that $\Treeplus \subseteq \Treeplus'$, and $\qeq_i \geq \qeqp_i$ for all $i \in \Treeplus$. Therefore, 
\[\sum_{i \in \Treeplus} m_i = \qeq_0+ \sum_{i \in \Treeplus}\qeq_i > \qeqp_0+ \sum_{i \in \Treeplus} \qeqp_i \geq \sum_{i \in \Treeplus} m_i,\]
which is a contradiction. Thus, we must have $\qeqp_0 \geq \qeq_0$. 

Analogously, for any $\barstate_0< \barstate'_0< \mathbb{E}_F[\state_0] $, we assume for the sake of contradiction that $\qeqp_0 < \qeq_0$. From \eqref{eq:qeq_negative} and \eqref{subeq:Treeminus}, we know that $\Treeminus \supseteq \Treeminus'$, and $\qeqp_{i}< \qeq_i$ for all $i \in \Tree'_{-}$. Therefore, 
\[\sum_{i \in \Treeminus'} m_i \geq \qeq_0 + \sum_{i \in \Treeminus'}\qeq_i > \qeqp_0 + \sum_{i \in \Treeminus'}\qeqp_i = \sum_{i \in \Treeminus'} m_i,\]
which is a contradiction. Thus, we must have $\qeqp_0 \geq \qeq_0$. We conclude that $\qeq_0$ is increasing in $\barstate_0$. 
 \hfill $\square$

 \medskip

 From Lemma \ref{lemma:threshold_structure}, we know that as $\barstate_0$ increases above $\mathbb{E}_F[\state_0]$, $\qeq_0$ increases, and consequently the sets $\Treeplus$ and $\Treedag$ as in \eqref{eq:tree} are non-decreasing. On the other hand, as $\barstate_0$ decreases below $\mathbb{E}_F[\state_0]$, the set $\Treeminus$ is non-decreasing. Based on this lemma, we construct Algorithm \ref{alg:plus} to generate the sequence of $\Treeplus$ and $\Treedag$ associated with the increasing $\barstate_0$, and Algorithm \ref{alg:minus} to generate the sequence of $\Treeminus$ associated with the decreasing $\barstate_0$ below $\mathbb{E}_F[s_0]$. In each algorithm, we compute the interval of $\barstate_0$ such that $\Treeplus$, $\Treedag$, and $\Treeminus$ are associated with the support set of equilibrium strategies.

\begin{algorithm}[htp]             
    \DontPrintSemicolon
    $k \leftarrow 0$, $\Treeplus[0] \leftarrow \emptyset$, $\Treedag[0] \leftarrow \emptyset$, $\barstate_0[k]\leftarrow \mathbb{E}_F[s_0]$. \\
    \While{$\Treedag[k] \neq \V\setminus \{0\}$}{
    \begin{align}\label{eq:tau}\tau_1 \leftarrow \min_{i \in V\setminus \Treeplus[k]} \{(1-\rate)(s_i - \beta_i m_i)+ d_i\}, \quad \tau_2 \leftarrow \min_{i \in \Treeplus[k]\setminus \Treedag[k]} \{(1-\rate)s_i +d_i\}\end{align}
    \uIf{$\tau_1< \tau_2$}{
\footnotesize{\begin{align}\label{eq:barstate_threshold_1}\barstate_0[k+1] = \frac{\tau_1 \beta_0}{1-\rate}\left(\sum_{i \in \Treedag[k]}\frac{1}{\beta_i}+\frac{1}{\beta_0} \right) + \beta_0  \left(m_0 + \sum_{i \in \Treeplus[k]} m_i - \sum_{i \in \Treedag[k]} \frac{1}{\beta_i} \left(s_i + \frac{\distance_i}{1-\rate}\right)\right).
    \end{align}}
    $\Treeplus[k+1]\leftarrow \Treeplus[k] \cup \argmin_{i \in V\setminus \Treeplus[k]} \{(1-\rate)(s_i - \beta_i m_i)+ d_i\}$, $\Treedag[k+1] \leftarrow \Treedag[k] $
  }
  \uElseIf{$\tau_1> \tau_2$}{
\footnotesize{\begin{align}\label{eq:barstate_threshold_2}\barstate_0[k+1] = \frac{\tau_2 \beta_0}{1-\rate}\left(\sum_{i \in \Treedag[k]}\frac{1}{\beta_i}+\frac{1}{\beta_0} \right) + \beta_0  \left(m_0 + \sum_{i \in \Treeplus[k]} m_i - \sum_{i \in \Treedag[k]} \frac{1}{\beta_i} \left(s_i + \frac{\distance_i}{1-\rate}\right)\right).
  \end{align}}    $\Treeplus[k+1]\leftarrow \Treeplus[k]$, $\Treedag[k+1] \leftarrow \Treedag[k] \cup \argmin_{i \in \Treedag[k]} \{(1-\rate)s_i +d_i\}$
  }
  \Else{
\footnotesize{\begin{align}\label{eq:barstate_threshold_3}\barstate_0[k+1] = \frac{\tau_1 \beta_0}{1-\rate}\left(\sum_{i \in \Treedag[k]}\frac{1}{\beta_i}+\frac{1}{\beta_0} \right) + \beta_0  \left(m_0 + \sum_{i \in \Treeplus[k]} m_i - \sum_{i \in \Treedag[k]} \frac{1}{\beta_i} \left(s_i + \frac{\distance_i}{1-\rate}\right)\right)\end{align}}
    $\Treeplus[k+1]\leftarrow \Treeplus[k] \cup \argmin_{i \in V\setminus \Treeplus[k]} \{(1-\rate)(s_i - \beta_i m_i)+ d_i\}$\; $\Treedag[k+1] \leftarrow \Treedag[k] \cup \argmin_{i \in \Treedag[k]} \{(1-\rate)s_i +d_i\}$\;
  }
   $k \leftarrow k+1$\;
    }
    $\Kplus_{\mathrm{max}} \leftarrow k$\;
    \KwOut{$\Kplus_{\mathrm{max}}$, $\{\Treeplus[k], \Treedag[k], \barstate_0[k]\}_{k=1}^{ \Kplus_{\mathrm{max}}}$}
    \caption{Computation of $\{\Treeplus[k], \Treedag[k], \barstate_0[k]\}_{k=1}^{ \Kplus_{\mathrm{max}}}$.}\label{alg:plus}
\end{algorithm}

\begin{algorithm}[htp]             
    \DontPrintSemicolon
    $k \leftarrow 0$, $\Treeminus[0] \leftarrow \emptyset$, $\barstate_0[0] \leftarrow \mathbb{E}_F[s_0]$. \\
    \While{$\Treeminus[-k] \neq \V\setminus \{0\}$}{
   
    $\tau \leftarrow \max_{i \in V\setminus \Treeminus[-k]} \{(1-\rate)(s_i - \beta_i m_i)- d_i\}$\;
    \footnotesize{\begin{align}\label{eq:barstate_threshold_4}\barstate_0[-k-1] = \frac{\tau \beta_0}{1-\rate}\left(\sum_{i \in \Treeminus[-k]}\frac{1}{\beta_i}+\frac{1}{\beta_0} \right) + \beta_0  \left(m_0 + \sum_{i \in \Treeminus[-k]} m_i - \sum_{i \in \Treeminus[-k]} \frac{1}{\beta_i} \left(s_i - \frac{\distance_i}{1-\rate}\right)\right).\end{align}}\;
    \eIf{
    $\barstate_0[-k-1] \leq \frac{1}{\sum_{i \in \Treeminus[-k]}\frac{1}{\beta_i}} \left( -m_0 + \sum_{i \in \Treeminus[-k]} \frac{1}{\beta_i}\left(s_i - \frac{d_i}{1-\rate} - m_i \beta_i\right)\right)$
    }{$\barstate_0[-k-1] \leftarrow \frac{1}{\sum_{i \in \Treeminus[-k]}\frac{1}{\beta_i}} \left( -m_0 + \sum_{i \in \Treeminus[-k]} \frac{1}{\beta_i}\left(s_i - \frac{d_i}{1-\rate} - m_i \beta_i\right)\right)$\; 
    $k \leftarrow k+1$\;
    break}{$\Treeminus[-k-1]\leftarrow \Treeminus[-k] \cup \argmax_{i \in V\setminus \Treeminus[-k]} \{(1-\rate)(s_i - \beta_i m_i)- d_i\}$\;
   $k \leftarrow k+1$\;}
    }
    $\Kminus_{\mathrm{max}} \leftarrow k$\;
    \KwOut{$\Kminus_{\mathrm{max}}$, $\{\Treeminus[-k], \barstate_0[-k]\}_{k=1}^{\Kminus_{\mathrm{max}}}$}
    \caption{Computation of $\{\Treeminus[-k], \barstate_0[-k]\}_{k =1}^{\Kminus_{\mathrm{max}}}$.}\label{alg:minus}
\end{algorithm}

Based on the outputs of the two algorithms, we next characterize the equilibrium regimes in our problem.

\begin{proposition}\label{prop:eq_strategy}
Under Assumptions \ref{as:market_balance}, the equilibrium agent distribution $\qeq(\barstate_0)$ and platform's revenue $R(\barstate_0)$ are piecewise linear functions of the posterior state mean $\barstate_0$, and exhibit $\Kplus+\Kminus+1$ regimes, where 
\begin{align*}
\Kplus &= \max \left\{k=1, \dots, \Kplus_{\mathrm{max}} \left\vert ~ \barstate_0[k] \leq \sup S_0 \right.\right\}, \quad 
\Kminus  =  \max \left\{k=1, \dots, \Kminus_{\mathrm{max}} \left\vert ~ \barstate_0[-k] \geq \inf S_0 \right.\right\}.
\end{align*}

\medskip 
\noindent \underline{Regime 0}: $\barstate_0[-1]\leq \barstate_0 < \barstate_0[1]$. \begin{align}\label{eq:q_zero}
        \qwe_i(\barstate_0) &= m_i,  \quad \forall i \in \V, 
    \end{align}
    and in the interior of regime 0, 
    \[\frac{d \revenue(\barstate_0)}{d \barstate_0} = \rate m_0.\]

\medskip
\noindent \underline{Regime $\{k\}_{k=1}^{\Kplus}$}: $\barstate_0[k] \leq \barstate_0 < \barstate_0[k+1]$ with $\barstate_0[\Kplus+1] = \sup S_0$.    \begin{align}\label{eq:qzero_positive}
       \qeq_0(\barstate_0) &= \frac{1}{\beta_0 \left(\sum_{i \in \Treedag[k]}\frac{1}{\beta_i} + \frac{1}{\beta_0}\right)}\left( m_0+ \sum_{i \in \Treeplus[k]} m_i+ \sum_{i \in \Treedag[k]} \frac{1}{\beta_i} \left(\barstate_0 - s_i - \frac{\distance_i}{1-\rate}\right)\right),
    \end{align}
    and $\qeq_i(\barstate_0)$ is given by \eqref{eq:qeq_positive} for all $i \in V\setminus \{0\}$. Moreover, in the interior of regime $k$, 
    \begin{align}\label{eq:derivative_plus_k}
        \frac{d \revenue(\barstate_0)}{d \barstate_0} = \frac{m_0+ \sum_{i \in \Treeplus[k]}  \left(m_i + d_i/\beta_i (1-\rate) \right)}{\beta_0\left(\sum_{i \in \Treedag[k]}\frac{1}{\beta_i}+\frac{1}{\beta_0} \right)}.
    \end{align}

    \medskip 
    \noindent \medskip
\noindent \underline{Regime $\{-k\}_{k=1}^{\Kminus-1}$}: $\barstate_0[-k-1] \leq \barstate_0 < \barstate_0[-k]$.
  
    \begin{align}\label{eq:qzero_negative}
    \qeq_0(\barstate_0) &= \frac{1}{\beta_0 \left(\sum_{i \in \Treeminus[-k]}\frac{1}{\beta_i} + \frac{1}{\beta_0}\right)}\left( m_0+ \sum_{i \in \Treeminus[-k]} m_i+ \sum_{i \in \Treeminus[-k]} \frac{1}{\beta_i} \left(\barstate_0 - s_i + \frac{\distance_i}{1-\rate}\right)\right),
    \end{align}
    and $\qeq_i(\barstate_0)$ is given by \eqref{eq:qeq_negative} for all $i \in V\setminus \{0\}$. Moreover, in the interior of regime $-k$, 
    \begin{align}\label{eq:derivative_minus_k}
    \frac{d \revenue(\barstate_0)}{d \barstate_0} =  \rate \frac{m_0+ \sum_{i \in \Treeminus[k]}  \left(m_i - d_i/\beta_i (1-\rate) \right)}{\beta_0\left(\sum_{i \in \Treeminus[k]}\frac{1}{\beta_i}+\frac{1}{\beta_0} \right)}. \end{align}

 \noindent \underline{Regime $-\Kminus$}: $\barstate_0[-\Kminus-1] \leq \barstate_0 < \barstate_0[-\Kminus]$ with $\barstate_0[-\Kminus-1] = \inf S_0$. If \[\barstate_0[-\Kminus] > \frac{1}{\sum_{i \in \Treeminus[-k]}\frac{1}{\beta_i}} \left( -m_0 + \sum_{i \in \Treeminus[-k]} \frac{1}{\beta_i}\left(s_i - \frac{d_i}{1-\rate} - m_i \beta_i\right)\right),\] then $\qeq(\barstate_0)$ and $\frac{d \revenue(\barstate_0)}{d \barstate_0}$ are as in \eqref{eq:qzero_negative} and \eqref{eq:derivative_minus_k}. Otherwise,
 \begin{align}\label{eq:depleted}\qeq_0(\barstate_0)=0, \quad \qeq_i(\barstate_0) = \qeq_i(\barstate_0[-\Kminus]), \quad \forall i \in V\setminus \{0\}, \quad \frac{d \revenue(\barstate_0)}{d \barstate_0} = 0.\end{align}

\end{proposition}

\noindent\emph{Proof of Proposition \ref{prop:eq_strategy}.} In regime $k$, we know from Lemma \ref{lemma:threshold_structure} that a fraction of agents from nodes $\Treeplus[k] \setminus \Treedag[k]$ and all agents from nodes $\Treedag[k]$ move to node 0, and agents from the remaining nodes $\V \setminus \Treeplus[k]$ do not move. Therefore, the sum of $\qeq_i$ for $i \in \Treeplus[k] \cup \{0\}$ equals to the sum of the masses of agents initially locating at those nodes, i.e. 
\begin{align*}
&\qeq_0 + \sum_{i \in \Treeplus[k]} \qeq_i = \qeq_0 + \sum_{i \in \Treeplus[k] \setminus \Treedag[k]} \qeq_i = m_0 + \sum_{i \in \Treeplus [k]} m_i, \\
\stackrel{\eqref{eq:q_positive_relation}}\Rightarrow \quad & \qeq_0 = \frac{1}{\beta_0 \left(\sum_{i \in \Treedag[k]}\frac{1}{\beta_i} + \frac{1}{\beta_0}\right)}\left( m_0+ \sum_{i \in \Treeplus[k]} m_i+ \sum_{i \in \Treedag[k]} \frac{1}{\beta_i} \left(\barstate_0 - s_i - \frac{\distance_i}{1-\rate}\right)\right).
\end{align*}
Therefore, the equilibrium price of node 0 is given by: 
\begin{align*}
    \peq_0 &= \barstate_0 - \beta_0 \qeq_0 = \barstate_0 - \frac{1}{\left(\sum_{i \in \Treedag[k]}\frac{1}{\beta_i} + \frac{1}{\beta_0}\right)}\left( m_0+ \sum_{i \in \Treeplus[k]} m_i+ \sum_{i \in \Treedag[k]} \frac{1}{\beta_i} \left(\barstate_0 - s_i - \frac{\distance_i}{1-\rate}\right)\right)\\
    &= \frac{1}{\left(\sum_{i \in \Treedag[k]}\frac{1}{\beta_i}+\frac{1}{\beta_0} \right)}\left( \frac{1}{\beta_0} \barstate_0 + \sum_{i \in \Treedag[k]} \frac{1}{\beta_i} \left(s_i + \frac{\distance_i}{1-\rate}\right) -m_0 - \sum_{i \in \Treeplus[k]} m_i\right).
\end{align*}

We now prove that $[\barstate_0[k], \barstate_0[k+1]]$ is the range of $\barstate_0$ for regime $k$. Since $\qeq_0$ and the associated $\Treeplus$ and $\Treedag$ as in \eqref{subeq:Treeplus} are increasing in $\barstate_0$ (Lemma \ref{lemma:threshold_structure}), from regime $k$ to $k+1$, there are three possible cases: (a) agents from nodes in $\V \setminus \Treeplus[k]$ with the minimum $\{(1-\rate)(s_i - \beta_i m_i) + d_i\}$ start to move to node $0$, and thus these nodes are included in $\Treeplus[k+1]$; (b) agents from nodes in $\Treeplus[k] \setminus \Treedag[k]$ with the minimum $\{(1-\rate) s_i + d_i\}$ all move to node $0$, and thus these nodes are included in $\Treedag[k+1]$; (c) both (a) and (b) happen simultaneously. We denote $\tau_1$ and $\tau_2$ as in \eqref{eq:tau}. Then, case (a) corresponds to $\tau_1< \tau_2$. In this case, $(1-\rate)(\barstate_0[k+1] - \beta_0 \qeq_0) = \tau_1$, and $\barstate_0[k+1]$ is given by \eqref{eq:barstate_threshold_1}. Similarly, $\tau_2< \tau_1$ in case (b) so that $(1-\rate)(\barstate_0[k+1] - \beta_0 \qeq_0) = \tau_2$ and $\barstate_0[k+1]$ is given by \eqref{eq:barstate_threshold_2}. In case (c), $\tau_1=\tau_2$ and $\barstate_0[k+1]$ is given by \eqref{eq:barstate_threshold_3}. Furthermore, the lower bound threshold $\barstate_0[k]$ can be similarly computed given by $\Treeplus[k-1], \Treeplus[k], \Treedag[k-1], \Treedag[k]$. 

Additionally, in regime $k$, the experienced payoff of agents originating from $i \in \Treeplus[k]$ is $(1-\rate) \peq_0 - d_i$. In particular, for $i \in \Treeplus[k]\setminus \Treedag[k]$, the price of node $i$ is $\peq_i = \peq_0 - d_i/(1-\rate)$, and agents at node $i$ is indifferent between repositioning to node $0$ and staying at node $i$. For $i \in \Treedag[k]$, the price at node $i$ is $s_i < \peq_0 - \frac{d_i}{1-\rate}$, and agents from node $i$ move to node $0$, and the utility of these agents is $(1-\rate) \peq_0 - d_i$. We compute the total utility of all agents, denoted as $U$, as follows: 
\begin{align*}
    U^*= (1-\rate) \peq_0 m_0+ \sum_{i \in \Treeplus[k]} \left((1-\rate) \peq_0 - d_i\right) m_i + \sum_{i \in \V\setminus \{\Treeplus[k]\cup \{0\}\}} (1-\rate) (s_i - \beta_im_i) m_i. 
\end{align*}
Moreover, $U^*$ can be alternatively expressed as the total received service prices minus the cost of repositioning: 
\begin{align*}
U^*= (1-\rate) \sum_{i \in V} \peq_i \qeq_i - \sum_{i \in \Treeplus[k]} (m_i - \qeq_i) \distance_i. 
\end{align*}
Therefore, 
\begin{align*}
    \revenue(\barstate_0) = &\rate \sum_{i \in V} \peq_i \qeq_i = 
 \frac{\rate}{1-\rate} \left(U^*+ \sum_{i \in \Treeplus[k]} (m_i - \qeq_i) \distance_i\right) \\
 = &\rate\left(\peq_0 \left(m_0+ \sum_{i \in \Treeplus[k]} m_i \right)  + \sum_{i \in \V \setminus \{\Treeplus[k] \cup \{0\}\}} (s_i - \beta_i m_i)m_i- \sum_{i \in \Treeplus[k]} \frac{d_i \qeq_i}{1-\rate}\right).  
\end{align*}
Since $\peq_0$ and $\qeq_i$ are linear in $\barstate_0$, we know that $\revenue(\barstate_0)$ is also linear in $\barstate_0$. Particularly, 
\begin{align*}
    \frac{d \revenue(\barstate_0)}{d \barstate_0} &=\rate \frac{d \peq_0}{d \barstate_0} 
 \left(\sum_{i \in \Treeplus[k]}  m_i + m_0\right) - \frac{\rate}{1-\rate} \left(\sum_{i \in \Treeplus[k]} \distance_i \left(\frac{-\frac{1}{\beta_i}}{\beta_0\left(\sum_{i \in \Treedag[k]}\frac{1}{\beta_i}+\frac{1}{\beta_0} \right)}\right)\right)\\
 &= \rate \frac{\left(\sum_{i \in \Treeplus[k]}  m_i + m_0\right) }{\beta_0\left(\sum_{i \in \Treedag[k]}\frac{1}{\beta_i}+\frac{1}{\beta_0} \right)}  + \frac{\rate}{1-\rate}\left( \frac{\sum_{i \in \Treeplus[k]} d_i/\beta_i }{\beta_0\left(\sum_{i \in \Treedag[k]}\frac{1}{\beta_i}+\frac{1}{\beta_0} \right)}\right)\\
 &= \rate \frac{m_0+ \sum_{i \in \Treeplus[k]}  \left(m_i + d_i/\beta_i (1-\rate) \right)}{\beta_0\left(\sum_{i \in \Treedag[k]}\frac{1}{\beta_i}+\frac{1}{\beta_0} \right)}.  
\end{align*}

Analogously, in regime $-k$, we know from Lemma \ref{lemma:threshold_structure} that agents from node $0$ move to nodes in $\Treeminus[-k]$ and all agents from the remaining nodes do not move. Therefore, the sum of $\qeq_i$ for $i \in \Treeminus[-k]\cup \{0\}$ equals to the sum of the masses of agents initially locating at those nodes, i.e. 
\begin{align*}
&\qeq_0 + \sum_{i \in \Treeminus[-k]} \qeq_i = m_0 + \sum_{i \in \Treeminus[-k]} m_i, \\
\stackrel{\eqref{eq:qeq_negative}}{\Rightarrow} \quad & \qeq_0 = \frac{1}{\beta_0 \left(\sum_{i \in \Treeminus[-k]}\frac{1}{\beta_i} + \frac{1}{\beta_0}\right)}\left( m_0+ \sum_{i \in \Treeminus[-k]} m_i+ \sum_{i \in \Treeminus[-k]} \frac{1}{\beta_i} \left(\barstate_0 - s_i + \frac{\distance_i}{1-\rate}\right)\right).
\end{align*}
The equilibrium price of node 0 is given by: 
\begin{align*}
    \peq_0 &= \barstate_0 - \beta_0 \qeq_0 = \barstate_0 - \frac{1}{\left(\sum_{i \in \Treeminus[-k]}\frac{1}{\beta_i} + \frac{1}{\beta_0}\right)}\left( m_0+ \sum_{i \in \Treeminus[-k]} m_i+ \sum_{i \in \Treeminus[-k]} \frac{1}{\beta_i} \left(\barstate_0 - s_i + \frac{\distance_i}{1-\rate}\right)\right)\\
    &= \frac{1}{\left(\sum_{i \in \Treeminus[-k]}\frac{1}{\beta_i}+\frac{1}{\beta_0} \right)}\left( \frac{1}{\beta_0} \barstate_0 + \sum_{i \in \Treeminus[-k]} \frac{1}{\beta_i} \left(s_i - \frac{\distance_i}{1-\rate}\right) -m_0 - \sum_{i \in \Treeminus[-k]} m_i\right).
\end{align*}

We now prove that $[\barstate_0[-k-1], \barstate_0[-k]]$ is the range of $\barstate_0$ for regime $-k$. From Lemma \ref{lemma:threshold_structure}, we know that as $\barstate_0$ decreases from regime $-k$ to $-k-1$, agents from node $0$ start to move to nodes in $\V \setminus \Treeminus[-k]$ with the maximum $\{(1-\rate)(s_i - \beta_i m_i) - d_i\}$, and thus these nodes are included in $\Treeminus[-k-1]$. The threshold $\barstate_0[-k-1]$ corresponds to the case that $(1-\rate)(\barstate_0[-k-1] - \beta_i \qeq_0) \} = \tau =\max_{i \in V\setminus \Treeminus[-k]} \{(1-\rate)(s_i - \beta_i m_i)- d_i\}$, and $\barstate_0[-k-1]$ is given by \eqref{eq:barstate_threshold_4}. Similarly, we can compute the threshold $\barstate_0[-k]$ using $\Treeminus[-k]$ and $\Treeminus[-k-1]$. 

Moreover, we can check that if $\barstate_0= \hat{s}_0=\frac{1}{\sum_{i \in \Treeminus[-k]}\frac{1}{\beta_i}} \left( -m_0 + \sum_{i \in \Treeminus[-k]} \frac{1}{\beta_i}\left(s_i - \frac{d_i}{1-\rate} - m_i \beta_i\right)\right)$, then $\qeq_0$ as in \eqref{eq:qeq_negative} equals to 0 indicating that all agents leave node $0$. Then, for any $\barstate_0< \hat{s}_0$, $\qeq_0=0$ and $\qeq$ does not change with $\barstate_0$. Algorithm \ref{alg:minus} terminates when the computed threshold for the next regime $\barstate_0[-k-1] \leq \hat{s}_0$ (Line 5-8). If $\barstate_0[-\Kminus] \geq \hat{s}_0$, then the range of states $S_0$ does not include states that are smaller than $\hat{s}_0$, and $\qeq$ is the same as that in regime $-\Kminus+1$. Otherwise, $\barstate_0[-\Kminus]=\hat{s}_0$, and $\qeq$ is given by \eqref{eq:depleted}. 

In regime $-k$, 
For all $i \in \Treeminus[-k]$, the experienced payoff of agents originating from $i$ is $(1-\rate) \peq_0 + d_i$. The total welfare of all agents is given by:  
\begin{align*}
    U^*=(1-\rate) \peq_0 m_0+  \sum_{i \in \Treeminus[-k]} \left((1-\rate) \peq_0 + d_i\right) m_i + \sum_{i \in \V \setminus \Treeminus[-k]} (1-\rate)(s_i - \beta_i m_i)m_i. 
\end{align*}
Moreover, since $U^*$ can be alternatively expressed as: 
\begin{align*}
U^*= (1-\rate) \sum_{i \in \V} \peq_i \qeq_i - \sum_{i \in \Treeminus[-k]} (\qeq_i - m_i) \distance_i,
\end{align*}
we must have
\begin{align*}
    \revenue(\barstate_0) &= \frac{\rate}{1-\rate} \left(U^*+ \sum_{i \in \Treeminus[-k]} (\qeq_i - m_i) \distance_i\right)\\
    &= \rate\left(\peq_0 \left(m_0 + \sum_{i \in \Treeminus[-k]} m_i\right) + \sum_{i \in \V \setminus \Treeminus[-k]} (s_i - \beta_i m_i)m_i+ \sum_{i \in \Treeminus[-k]} \frac{d_i \qeq_i}{1-\rate}\right). 
\end{align*}
Since $\peq_0$ and $\qeq_i$ are linear in $\barstate_0$, we know that $\revenue(\barstate_0)$ is also linear in $\barstate_0$. Particularly, 
\begin{align*}
    \frac{d \revenue(\barstate_0)}{d \barstate_0} 
    &=\rate \frac{d \peq_0}{d \barstate_0} 
 \left(\sum_{i \in \Treeminus[-k]}  m_i + m_0\right) +\frac{\rate}{1-\rate} \left(\sum_{i \in \Treeminus[-k]} \distance_i \left(\frac{-\frac{1}{\beta_i}}{\beta_0\left(\sum_{i \in \Tree}\frac{1}{\beta_i}+\frac{1}{\beta_0} \right)}\right)\right)\\
 &= \rate \frac{\left(\sum_{i \in \Treeminus[-k]}  m_i + m_0\right) }{\beta_0\left(\sum_{i \in \Treeminus[-k]}\frac{1}{\beta_i}+\frac{1}{\beta_0} \right)}  - \frac{\rate}{1-\rate}\left( \frac{\sum_{i \in \Treeminus[-k]} d_i/\beta_i }{\beta_0\left(\sum_{i \in \Treeminus[-k]}\frac{1}{\beta_i}+\frac{1}{\beta_0} \right)}\right)\\
 &= \rate \frac{m_0+ \sum_{i \in \Treeminus[-k]}  \left(m_i - d_i/\beta_i (1-\rate) \right)}{\beta_0\left(\sum_{i \in \Treeminus[-k]}\frac{1}{\beta_i}+\frac{1}{\beta_0} \right)}. 
\end{align*}
In the last regime $-\Kminus$, if node $0$ is depleted, then $\frac{d \revenue(\barstate_0)}{d \barstate_0}=0$. \hfill $\square$

\section{Proofs of statements in Section \ref{sec:single_shock}}\label{apx:proof_three}

\noindent\emph{Proof of Proposition \ref{prop:eq_strategy_simple}.} In regime 0, since no agents move, we know that $\qeq_i(\barstate_0)=m_i$ for all $i \in \V$ and $d \revenue(\barstate_0)/ d \barstate_0=0$. 

Under Assumptions \ref{as:complete_market_balance}, we know from Proposition \ref{prop:eq_strategy} that in each regime $k=1, \dots, \Kplus$, the set of nodes where agents move to node $0$ is $\Treeplus[k]= \cup_{n=1}^{k} V_n$. Additionally, since no nodes are depleted given Assumption \ref{as:no_depletion_max}, $\Treedag[k]=\emptyset$ for all $k$. As a result, we know from \eqref{eq:qzero_positive} that 
\begin{align}
\qeq_0(\barstate_0) &= \frac{1}{\beta_0 \left(\sum_{i \in \Treedag[k]}\frac{1}{\beta_i} + \frac{1}{\beta_0}\right)}\left( m_0+ \sum_{i \in \Treeplus[k]} m_i+ \sum_{i \in \Treedag[k]} \frac{1}{\beta_i} \left(\barstate_0 - s_i - \frac{\distance_i}{1-\rate}\right)\right)\notag\\
&= \frac{1}{\beta_0 \left(\sum_{i \in \cup_{n=1}^{k}V_n}\frac{1}{\beta_i} + \frac{1}{\beta_0}\right)}\left( m_0+ \sum_{i \in \cup_{n=1}^{k}V_n}  \frac{1}{\beta_i} \left(\barstate_0+ m_i \beta_i - s_i - \frac{\distance_i}{1-\rate}\right)\right)\notag\\
&= \frac{1}{\beta_0 \left(\sum_{i \in \cup_{n=1}^{k}V_n}\frac{1}{\beta_i} + \frac{1}{\beta_0}\right)}\left( m_0+ \sum_{i \in \cup_{n=1}^{k}V_n}  \frac{1}{\beta_i} \left(\barstate_0 - \frac{\distance_i}{1-\rate}\right)\right).\label{eq:qzero_positive_calculation}
\end{align}
Additionally, $\qeq_i(\barstate_0)$ is given by \eqref{eq:q_positive_relation} following Lemma \ref{lemma:threshold_structure}. From \eqref{eq:derivative_plus_k}, we have: 
\begin{align*}
    \frac{d \revenue(\barstate_0)}{d \barstate_0} = \frac{m_0+ \sum_{i \in \Treeplus[k]}  \left(m_i + d_i/\beta_i (1-\rate) \right)}{\beta_0\left(\sum_{i \in \Treedag[k]}\frac{1}{\beta_i}+\frac{1}{\beta_0} \right)} =  \frac{m_0+ \sum_{i \in \cup_{n=1}^{k}V_n}  \left(m_i + d_i/\beta_i (1-\rate) \right)}{\beta_0\left(\sum_{i \in \cup_{n=1}^{k}V_n}\frac{1}{\beta_i}+\frac{1}{\beta_0} \right)}.
\end{align*}
In regime $-k$ with $k=1, \dots, \Kminus$, $\Treeminus[k] = \cup_{n=1}^{k}V_n$ under Assumption \ref{as:complete_market_balance}. Therefore, following \eqref{eq:qzero_negative}, 
\begin{align*}
     \qeq_0(\barstate_0) &= \frac{1}{\beta_0 \left(\sum_{i \in \cup_{n=1}^{k}V_n }\frac{1}{\beta_i} + \frac{1}{\beta_0}\right)}\left( m_0+ \sum_{i \in \cup_{n=1}^{k}V_n} m_i+ \sum_{i \in \cup_{n=1}^{k}V_n} \frac{1}{\beta_i} \left(\barstate_0 - s_i + \frac{\distance_i}{1-\rate}\right)\right)\\
     &= \frac{1}{\beta_0 \left(\sum_{i \in \cup_{n=1}^{k}V_n}\frac{1}{\beta_i} + \frac{1}{\beta_0}\right)}\left( m_0+ \sum_{i \in \cup_{n=1}^{k}V_n}  \frac{1}{\beta_i} \left(\barstate_0 + \frac{\distance_i}{1-\rate}\right)\right).
\end{align*}
and $\qeq_i(\barstate_0)$ is given by \eqref{eq:q_negative_relation}. From \eqref{eq:derivative_minus_k}, we have: 
\begin{align*}
    \frac{d \revenue(\barstate_0)}{d \barstate_0} =  \rate \frac{m_0+ \sum_{i \in \Treeminus[k]}  \left(m_i - d_i/\beta_i (1-\rate) \right)}{\beta_0\left(\sum_{i \in \Treeminus[k]}\frac{1}{\beta_i}+\frac{1}{\beta_0} \right)} =  \rate \frac{m_0+ \sum_{i \in \cup_{n=1}^{k}V_n}  \left(m_i - d_i/\beta_i (1-\rate) \right)}{\beta_0\left(\sum_{i \in \cup_{n=1}^{k}V_n}\frac{1}{\beta_i}+\frac{1}{\beta_0} \right)}.
\end{align*}
\hfill $\square$

\medskip

\noindent\emph{Proof of Lemma \ref{lemma:no_depletion}.} From \eqref{eq:qzero_positive_calculation}, we can check that under Assumption \ref{as:no_depletion_max}, $\qeq_0(\barstate_0) \geq 0$ for all $i \in \V$. \hfill $\square$

\medskip

Before proving Theorem \ref{theorem:simple_partitional}, we first present the following two lemmas:

\begin{lemma}\label{lemma:basic}
For any $a, b, c, d \in \mathbb{R}$ such that $c, d>0$. If $\frac{a}{c}< \frac{b}{d}$, then $\frac{a}{c}< \frac{a+b}{c+d}< \frac{b}{d}$. Moreover, if $\frac{a}{c}> \frac{b}{d}$, then $\frac{a}{c}> \frac{a+b}{c+d} > \frac{b}{d}$. 
\end{lemma}

\noindent\emph{Proof of Lemma \ref{lemma:basic}.} If $\frac{a}{c}< \frac{b}{d}$, then $ad< bc$ since $c, d >0$. By adding $ac$ and dividing $(c+d)c$, we have $\frac{a}{c}< \frac{a+b}{c+d}$. By adding $bd$ and dividing $(c+d)d$ on both sides, we have $\frac{a+b}{c+d}< \frac{b}{d}$. Similarly, if $\frac{a}{c}> \frac{b}{d}$, then $ad> bc$ since $c, d>0$. By adding $ac$ on both sides and divide by $(c+d)c$, we have $\frac{a}{c}> \frac{a+b}{c+d}$. By adding $bd$ and dividing $(c+d)d$ on both sides, we have $\frac{a+b}{c+d} > \frac{b}{d}$. \hfill $\square$

\begin{lemma}\label{lemma:convex_concave}
In case (i), the function $\revenue(\barstate_0)$ is convex in $\barstate_0$. In case (ii) (resp. case (iii)), there exists an interval $[\underline{s}_0, \bar{s}_0]$ with $\mathbb{E}_F[S_0] \leq \underline{s}_0 \leq \bar{s}_0 \leq \sup \S_0$ (resp. $\inf \S_0 \leq \underline{s}_0 \leq \bar{s}_0 \leq \mathbb{E}_F[S_0]$) such that $\revenue(\barstate_0)$ is concave in $[\underline{s}_0, \bar{s}_0]$ and convex in $\S_0 \setminus [\underline{s}_0, \bar{s}_0]$. 
\end{lemma}

\medskip 
\noindent\emph{Proof of Lemma \ref{lemma:convex_concave}.} In case (i), since all nodes have similar market sizes relative to their distances,
\begin{align}\label{eq:all_similar}
    \left\vert\frac{s_i- \barstate_0}{d_1}\right\vert \leq \frac{1}{1-\rate}, \quad  \forall i \in V_1, \quad \left\vert\frac{s_i - s_j}{d_{n}- d_{n-1}}\right\vert \leq \frac{1}{1-\rate}, \quad  \forall i \in V_n, j \in V_{n-1}, ~\forall n \leq \max\{\Kplus, \Kminus\}.
\end{align}
For $n=1$, we sort all nodes $i \in V_1$ in increasing order of $s_i$, and we denote the maximum index in $V_1$ as $\hat{i}$. Then, from \eqref{eq:all_similar}, we note that 
\[\barstate_0 \leq s_1+ d_1/(1-\rate) \leq  \cdots \leq  s_i+ d_1/(1-\rate) \leq  \cdots \leq s_{\hat{i}}+ d_1/(1-\rate).\]
From Assumption \ref{as:complete_market_balance}, we know that $s_i = \beta_i m_i$ for all $i \in \V$, and thus 
\[\frac{m_0}{\frac{1}{\beta_0}} \leq \frac{m_1 + d_1/\beta_1(1-\rate)}{\frac{1}{\beta_1}} \leq  \cdots \leq \frac{m_i + d_1/\beta_i(1-\rate)}{\frac{1}{\beta_i}} \leq \cdots \leq \frac{m_{\hat{i}} + d_1/\beta_{\hat{i}}(1-\rate)}{\frac{1}{\beta_{\hat{i}}}}.\]We denote the derivative of $\revenue(\barstate_0)$ in the interior of each regime $k$ as $\frac{d \revenue(\barstate_0)}{d \barstate_0}[k]$. From Lemma \ref{lemma:basic}, we have
\begin{align*}
    &\frac{d \revenue(\barstate_0)}{d \barstate_0}[0] = \rate m_0  = \frac{\rate}{\beta_0} \frac{m_0}{\frac{1}{\beta_0}} \\
    \leq&  \frac{\rate}{\beta_0} \frac{m_0 + (m_1 +d_1/\beta_1(1-\rate))}{\frac{1}{\beta_0} + \frac{1}{\beta_1}} \leq \cdots \leq \frac{\rate}{\beta_0} \frac{m_0 + \sum_{i \in V_1}(m_i + d_1/\beta_i(1-\rate))}{\frac{1}{\beta_0} + \sum_{i \in V_1}\frac{1}{\beta_i}}  = \frac{d \revenue(\barstate_0)}{d \barstate_0}[1].
\end{align*}
Following the same procedure, we can iteratively show that 
\[\frac{d \revenue(\barstate_0)}{d \barstate_0}[1] \leq \frac{d \revenue(\barstate_0)}{d \barstate_0}[2] \leq \cdots \leq \frac{d \revenue(\barstate_0)}{d \barstate_0}[\Kplus].\] 
On the other hand, 
\[\barstate_0 \geq s_1-  d_1/(1-\rate) \geq  \cdots \geq  s_i- d_1/(1-\rate) \leq  \cdots \geq s_{\hat{i}}- d_1/(1-\rate).\]
Thus, 
\[\frac{m_0}{\frac{1}{\beta_0}} \geq \frac{m_1 - d_1/\beta_1(1-\rate)}{\frac{1}{\beta_1}} \geq  \cdots \geq \frac{m_i - d_1/\beta_i(1-\rate)}{\frac{1}{\beta_i}} \geq \cdots \geq \frac{m_{\hat{i}} - d_1/\beta_{\hat{i}}(1-\rate)}{\frac{1}{\beta_{\hat{i}}}},\]
and 
\begin{align*}
    \frac{d \revenue(\barstate_0)}{d \barstate_0}[0] = \rate m_0 \geq   \cdots \geq \frac{\rate}{\beta_0} \frac{m_0 + \sum_{i \in V_1}(m_i - d_1/\beta_i(1-\rate))}{\frac{1}{\beta_0} + \sum_{i \in V_1}\frac{1}{\beta_i}}  = \frac{d \revenue(\barstate_0)}{d \barstate_0}[-1].
\end{align*}
Consequently, we have $\frac{d \revenue(\barstate_0)}{d \barstate_0}[-1] \geq \frac{d \revenue(\barstate_0)}{d \barstate_0}[-2] \geq \cdots \geq \frac{d \revenue(\barstate_0)}{d \barstate_0}[-\Kminus]$. Therefore, the function $\revenue(\barstate_0)$ is convex in $\barstate_0$ in case (i).

In case (ii), since nodes within distance $\dl$ have similar market sizes relative to distances, following the same argument as in case (i), we know that $\revenue(\barstate_0)$ is convex in $[\barstate_0[-\nl], \barstate_0[\nl]]$, where $\nl = \max\{n=1, \dots, N|d_n \leq \dl\}$, i.e. 
\begin{align*}
\frac{d \revenue(\barstate_0)}{d \barstate_0}[-\nl] \leq \frac{d \revenue(\barstate_0)}{d \barstate_0}[-\nl+1] \leq \cdots \leq \frac{d \revenue(\barstate_0)}{d \barstate_0}[0] \leq \cdots \leq \frac{d \revenue(\barstate_0)}{d \barstate_0}[\nl].
\end{align*}
Since nodes with distances between $\dl$ and $\dh$ have decreasing market sizes relative to distances as in \eqref{subeq:decrease_size}, we know that for nodes $i \in \cup_{i=\nl+1}^{\nh} V_n$, where $\nh = \max\{n=1, \dots, N|d_n \leq \dh\}$, satisfy: 
\begin{subequations}
\begin{align}
&s_i + d_i/(1-\rate) < s_j + d_j/(1-\rate), \quad i \in V_n, j \in V_{n-1}, \quad n=\nl+1, \dots, \nh,\label{subeq:decrease_interval}\\
\Rightarrow \quad &s_i - d_i/(1-\rate) < s_j - d_j/(1-\rate),  \quad i \in V_n, j \in V_{n-1}, \quad n=\nl+1, \dots, \nh.
\end{align}
\end{subequations}
Additionally, since nodes with distances higher than $\dh$ have similar market sizes relative to distances as in \eqref{eq:similar}, we have: 
\begin{align*}
&s_i - d_i/(1-\rate) \leq s_j - d_j/(1-\rate),  \quad i \in V_n, j \in V_{n-1}, \quad n> \nh.
\end{align*}
Following Lemma \ref{lemma:basic} and \eqref{eq:derivative_minus_k}, we have: 
\[\frac{d \revenue(\barstate_0)}{d \barstate_0}[-\Kminus] \leq \cdots \leq \frac{d \revenue(\barstate_0)}{d \barstate_0}[-\nl] \leq \frac{d \revenue(\barstate_0)}{d \barstate_0}[-\nl+1] \leq \cdots \leq \frac{d \revenue(\barstate_0)}{d \barstate_0}[0] \leq \cdots \leq \frac{d \revenue(\barstate_0)}{d \barstate_0}[\nl],\]
i.e. $\revenue(\barstate_0)$ is convex for $\barstate_0 \leq \barstate_0[\nl]$. 

Additionally, following \eqref{subeq:decrease_interval}, we know that 
\begin{align}
   \frac{\sum_{i \in V_n}(m_i + d_n/\beta_i(1-\rate))}{\sum_{i \in V_n}\frac{1}{\beta_i}} \leq \frac{\sum_{i \in V_{n-1}}(m_i + d_{n-1}/\beta_i(1-\rate))}{\sum_{i \in V_{n-1}}\frac{1}{\beta_i}}, \quad \forall n \in [\nl+1, \nh]
\end{align}
If 
\[\frac{d \revenue(\barstate_0)}{d \barstate_0}[\nl] > \frac{\sum_{i \in V_{\nl+1}}(m_i + d_{\nl+1}/\beta_i(1-\rate))}{\sum_{i \in V_{\nl+1}}\frac{1}{\beta_i}},\]
then we know from Lemma \ref{lemma:basic} that $\frac{d \revenue(\barstate_0)}{d \barstate_0}[\nl] > \frac{d \revenue(\barstate_0)}{d \barstate_0}[\nl+1] > \cdots > \frac{d \revenue(\barstate_0)}{d \barstate_0}[\nh]$. On the other hand, if 
\[\frac{d \revenue(\barstate_0)}{d \barstate_0}[\nl] \leq \frac{\sum_{i \in V_{\nl+1}}(m_i + d_{\nl+1}/\beta_i(1-\rate))}{\sum_{i \in V_{\nl+1}}\frac{1}{\beta_i}},\]
then $\frac{d \revenue(\barstate_0)}{d \barstate_0}[\nl] \leq \frac{d \revenue(\barstate_0)}{d \barstate_0}[\nl+1]$. We apply this step iteratively until either (1) we find $\hat{n} \leq \nh$ such that 
\[\frac{d \revenue(\barstate_0)}{d \barstate_0}[\hat{n}] > \frac{\sum_{i \in V_{\hat{n}+1}}(m_i + d_{\hat{n}+1}/\beta_i(1-\rate))}{\sum_{i \in V_{\hat{n}+1}}\frac{1}{\beta_i}},\]
or (2) such $\hat{n} \leq \nh$ does not exist. In scenario (1), we know that $\revenue(\barstate_0)$ is convex in $[\barstate_0[\nl], \barstate_0[\hat{n}]]$ and strictly concave in $[\barstate_0[\hat{n}], \barstate_0[\nh]]$. In scenario (2), the strictly concave interval is empty (i.e. $\underline{s}_0 = \bar{s}_0 = \sup \S_0$) 

Furthermore, since nodes with distances larger than $\dh$ have similar market sizes relative to distances, following the same argument as for nodes with distances less than $\dl$, we know that $\revenue(\barstate_0)$ is convex in $[\barstate_0[\nh+1], \sup \S_0]$. Therefore, we can conclude that $\revenue(\barstate_0)$ is strictly concave in an interval $[\barstate_0[\hat{n}], \barstate_0[\nh]]$ (which can be empty) and convex below $\barstate_0[\hat{n}]$ and above $\barstate_0[\nh]$. 

In case (iii), since nodes within distance $\dl$ have similar market sizes relative to distances, following the same argument as in case (i), we know that $\revenue(\barstate_0)$ is convex in $[\barstate_0[-\nl], \barstate_0[\nl]]$, where $\nl = \max\{n=1, \dots, N|d_n \leq \dl\}$. Since nodes with distances between $\dl$ and $\dh$ have increasing market sizes relative to distances as in \eqref{subeq:increase_size}, we know nodes $i \in \cup_{i=\nl+1}^{\nh} V_n$, where $\nh = \max\{n=1, \dots, N|d_n \leq \dh\}$, satisfy: 
\begin{subequations}
\begin{align}
&s_i - d_i/(1-\rate) > s_j - d_j/(1-\rate), \quad i \in V_n, j \in V_{n-1}, \quad n=\nl+1, \dots, \nh, \label{subeq:increase_interval_neg}\\
\Rightarrow \quad & s_i + d_i/(1-\rate) > s_j + d_j/(1-\rate), \quad i \in V_n, j \in V_{n-1}, \quad n=\nl+1, \dots, \nh, \label{subeq:increase_interval_pos}
\end{align}
\end{subequations}
and $\revenue(\barstate_0)$ is convex in $[\barstate_0[\nl], \barstate_0[\nh]]$. 
Additionally, since nodes with distances higher than $\dh$ have similar market sizes relative to distances as in \eqref{eq:similar}, $\revenue(\barstate_0)$ is convex in $[\barstate_0[\nh], \sup \S_0]$. Thus, $\revenue(\barstate_0)$ is convex in $[\barstate_0[-\nl], \sup \S_0]$.

Following Lemma \ref{lemma:basic} and \eqref{subeq:increase_interval_neg}, we know that 
\begin{align}
   \frac{\sum_{i \in V_n}(m_i - d_n/\beta_i(1-\rate))}{\sum_{i \in V_n}\frac{1}{\beta_i}} > \frac{\sum_{i \in V_{n-1}}(m_i - d_{n-1}/\beta_i(1-\rate))}{\sum_{i \in V_{n-1}}\frac{1}{\beta_i}}, \quad \forall n \in [\nl+1, \nh]
\end{align}
If 
\[\frac{d \revenue(\barstate_0)}{d \barstate_0}[-\nl] < \frac{\sum_{i \in V_{\nl+1}}(m_i - d_{\nl+1}/\beta_i(1-\rate))}{\sum_{i \in V_{\nl+1}}\frac{1}{\beta_i}},\]
then we know from Lemma \ref{lemma:basic} that $\frac{d \revenue(\barstate_0)}{d \barstate_0}[-\nl] <\frac{d \revenue(\barstate_0)}{d \barstate_0}[-\nl-1] <  \cdots < \frac{d \revenue(\barstate_0)}{d \barstate_0}[-\nh]$, i.e. $\revenue(\barstate_0)$ is strictly concave in $[\barstate_0[-\nh], \barstate_0[-\nl]]$. On the other hand, if 
\[\frac{d \revenue(\barstate_0)}{d \barstate_0}[-\nl] \geq \frac{\sum_{i \in V_{\nl+1}}(m_i - d_{\nl+1}/\beta_i(1-\rate))}{\sum_{i \in V_{\nl+1}}\frac{1}{\beta_i}},\]
then $\frac{d \revenue(\barstate_0)}{d \barstate_0}[-\nl] \geq \frac{d \revenue(\barstate_0)}{d \barstate_0}[-\nl-1]$. We apply this step iteratively until either (1) we find $\hat{n} \leq \nh$ such that 
\[\frac{d \revenue(\barstate_0)}{d \barstate_0}[-\hat{n}] < \frac{\sum_{i \in V_{-\hat{n}-1}}(m_i - d_{\hat{n}+1}/\beta_i(1-\rate))}{\sum_{i \in V_{-\hat{n}-1}}\frac{1}{\beta_i}},\]
or (2) such $\hat{n} \leq \nh$ does not exist. In scenario (1), we know that $\revenue(\barstate_0)$ is convex in $[\barstate_0[-\hat{n}], \barstate_0[-\nl]]$ and strictly concave in $[\barstate_0[-\nh], \barstate_0[-\hat{n}]]$. In scenario (2), the strictly concave interval is empty. 

Furthermore, since nodes with distances larger than $\dh$ have similar market sizes relative to distances, following the same argument as for nodes with distances less than $\dl$, we know that $\revenue(\barstate_0)$ is convex in $[\inf \S_0, ~ \barstate_0[-\nh-1]]$. Therefore, we can conclude that $\revenue(\barstate_0)$ is strictly concave in an interval $[\barstate_0[-\nh], \barstate_0[-\hat{n}]]$ (which can be empty) and convex below $\barstate_0[-\nh]$ and above $\barstate_0[-\hat{n}]$. 
\hfill $\square$

\begin{lemma}[\cite{dworczak2019simple}]\label{lemma:duality}
   If there exists a cumulative distribution function $G$ and a convex function $\nu: S_0 \to \mathbb{R}$, with $\nu(\barstate_0) \geq \revenue(\barstate_0)$ for all $\barstate_0 \in S_0$, that satisfy
   \begin{subequations}\label{eq:duality}
       \begin{align}
           &supp(G) \subseteq \{\S_0: \revenue(\barstate_0) = \nu(\barstate_0)\}, \label{subeq:support_set}\\
           &\int_{\S_0} \nu(z) dG(z) =  \int_{\S_0} \nu(z) dF(z), \label{subeq:integral}\\
         &\text{$F$ is a mean-preserving spread of $G$.} \label{subeq:mean_preserving} 
       \end{align}
   \end{subequations}
then $G$ is an optimal posterior distribution that maximizes the expected total revenue. 
\end{lemma}

\medskip 
\noindent\emph{Proof of Theorem \ref{theorem:simple_partitional}.}

In case (i), we set $\nu(\barstate_0) = \revenue(\barstate_0)$, which is convex, and $G= F$. We can check that $(\nu, G)$ satisfies the conditions in Proposition \ref{lemma:duality}, and thus $G=F$ is the optimal posterior distribution indicating that full information revelation is optimal. 

In case (ii), if the strictly concave interval is empty, then full information provision is optimal following case (i). On the other hand, when the strictly concave interval is nonempty, we construct a pooling region $\barstate_0 \in [\zl, \zh]$, where $\zl\leq \barstate[\hat{n}]$ and $\zh \geq \barstate[\nh]$ such that $\hat{n}$ (resp. $\tilde{n}$) is the regime where $\revenue(\barstate_0)$ changes from convex to strictly concave (resp. strictly concave to convex), and $z^* = \mathbb{E}_F[S_0|\zl \leq S_0 \leq  \zh] > \mathbb{E}_F[S_0]$. 

We show that such $\zl$ and $\zh$ exist. For every regime $\hat{n}-1\leq k \leq \nh$, we define the linear function 
 \begin{align}\label{eq:fk}
     g_{k}(z) = \frac{d \revenue(\barstate_0)}{d \barstate_0}[k] (z- \barstate_0[k]) + \revenue(\barstate_0[k]),
 \end{align}
 where $\frac{d \revenue(\barstate_0)}{d \barstate_0}[k]$ is the derivative of $\revenue(\barstate_0)$ in regime $k$. We define 
 \begin{subequations}\label{eq:zazb}
 \begin{align}
     z_a[k] &= \min\{s_0 \in \S_0| g_{k}(z) \geq \revenue(z) \quad s_0 \leq z \leq \barstate_0[k] \},\\
     z_b[k] &= \max\{s_0 \in \S_0| g_{k}(z) \geq \revenue(z), \quad \barstate_0[k] \leq z \leq s_0\}.
 \end{align}
 \end{subequations}
 Since $g_{k}(\barstate_0)$ is tangent to the strictly concave piece of $\revenue(\barstate_0)$, we know that $g_{k}(z) > \revenue(z)$ for all $z \in [z_a[k], z_b[k]] \setminus \barstate_0[k]$, and $g_{k}(\barstate_0[k]) = \revenue(\barstate_0[k])$. Moreover, $z_a[k] < \barstate_0[\hat{n}]$ and $z_b[k]> \barstate_0[\nh]$ for all $\hat{n}\leq k \leq \nh$. We also have \begin{align}\label{eq:twoend}
     z_a[\hat{n}-1] = \barstate_0[\hat{n}-1], \quad z_b[\nh] = \barstate_0[\nh+1].
 \end{align} 

We note that for any $z\leq \barstate_0[\hat{n}]$ and any $\hat{n}-1 \leq k \leq \nh$, 
\begin{equation}\label{eq:ordering}
\begin{split}
&g_k(z) - g_{k-1}(z)\\
=&\frac{d \revenue(\barstate_0)}{d \barstate_0}[k](z- \barstate_0[k])+ \revenue(\barstate_0[k]) - \left(\frac{d \revenue(\barstate_0)}{d \barstate_0}[k-1](z- \barstate_0[k-1])+ \revenue(\barstate_0[k-1])\right)\\
=& \left(\frac{d \revenue(\barstate_0)}{d \barstate_0}[k] - \frac{d \revenue(\barstate_0)}{d \barstate_0}[k-1]\right)z + \frac{d \revenue(\barstate_0)}{d \barstate_0}[k-1] \barstate_0[k-1] - \frac{d \revenue(\barstate_0)}{d \barstate_0}[k] \barstate_0[k] \\
&+  \revenue(\barstate_0[k])  -  \revenue(\barstate_0[k-1])\\
=& \left(\frac{d \revenue(\barstate_0)}{d \barstate_0}[k] - \frac{d \revenue(\barstate_0)}{d \barstate_0}[k-1]\right)z + \frac{d \revenue(\barstate_0)}{d \barstate_0}[k-1] \barstate_0[k-1] - \frac{d \revenue(\barstate_0)}{d \barstate_0}[k] \barstate_0[k] \\
&+   \revenue(\barstate_0[k-1])+ \frac{d \revenue(\barstate_0)}{d \barstate_0}[k-1] (\barstate_0[k]-  \barstate_0[k-1])  -  \revenue(\barstate_0[k-1]) \\
= & \left(\frac{d \revenue(\barstate_0)}{d \barstate_0}[k] - \frac{d \revenue(\barstate_0)}{d \barstate_0}[k-1]\right)z - \frac{d \revenue(\barstate_0)}{d \barstate_0}[k] \barstate_0[k] +  \frac{d \revenue(\barstate_0)}{d \barstate_0}[k-1]  \barstate_0[k]   \\
=& \left(\frac{d \revenue(\barstate_0)}{d \barstate_0}[k] - \frac{d \revenue(\barstate_0)}{d \barstate_0}[k-1]\right) (z - \barstate_0[k]) >0.
\end{split}
\end{equation}
where the last inequality is due to the fact that $\frac{d \revenue(\barstate_0)}{d \barstate_0}[k] - \frac{d \revenue(\barstate_0)}{d \barstate_0}[k-1]<0$ and $z - \barstate_0[k]<0$. Consequently, we know that $z_a[k] \leq z_a[k-1]$ for any $\hat{n}-1 \leq k \leq \nh$, and the inequality is strict if $z_a[k] > \inf \S_0$. 

Similarly, for any $z \geq \barstate_0[\nh]$ and any $\hat{n}-1 \leq k \leq \nh$, we have 
\[g_k(z) - g_{k-1}(z) = \left(\frac{d \revenue(\barstate_0)}{d \barstate_0}[k] - \frac{d \revenue(\barstate_0)}{d \barstate_0}[k-1]\right) (z - \barstate_0[k]) <0\]
since $\frac{d \revenue(\barstate_0)}{d \barstate_0}[k] - \frac{d \revenue(\barstate_0)}{d \barstate_0}[k-1]<0$ and $z - \barstate_0[k]>0$. Thus, $z_b[k] \leq z_b[k-1]$ for any $\hat{n}-1 \leq k \leq \nh$, and the inequality is strict if $z_b[k] < \sup \S_0$.

Therefore, 
\begin{align*}
& \mathbb{E}_F[S_0| z_a[\hat{n}-1] \leq S_0 \leq z_b[\hat{n}-1]] - \barstate_0[\hat{n}-1] > \mathbb{E}_F[S_0| z_a[\hat{n}] \leq S_0 \leq z_b[\hat{n}]] - \barstate_0[\hat{n}]\\
>& \cdots > \mathbb{E}_F[S_0|z_a[k] \leq S_0 \leq z_b[k]]- \barstate_0[k] > \mathbb{E}_F[S_0|z_a[k+1] \leq S_0 \leq z_b[k+1]]- \barstate_0[k+1] \\
>& \cdots > \mathbb{E}_F[S_0| z_a[\nh] \leq S_0 \leq  z_b[\nh]]- \barstate_0[\nh]
\end{align*}
Since $z_a[\hat{n}-1] = \barstate_0[\hat{n}-1]$, we know that $\mathbb{E}_F[S_0| z_a[\hat{n}-1] \leq S_0 \leq z_b[\hat{n}-1]] - \barstate_0[\hat{n}-1] >0$. 

Consider the case that
\[\mathbb{E}_F[S_0| z_a[\nh] \leq \barstate_0<  z_b[\nh]]- \barstate_0[\nh] \leq 0,\]
then there must exists $\hat{n} \leq k^* \leq \nh$ such that \[\mathbb{E}_F[S_0|z_a[k^*] \leq S_0 \leq z_b[k^*]]- \barstate_0[k^*]>0, \quad \mathbb{E}_F[S_0|z_a[k^*+1] \leq S_0 \leq z_b[k^*+1]]- \barstate_0[k^*+1]\leq 0.\]

There can be two cases: 
\begin{itemize}
\item[(i)] 
$\mathbb{E}_F[S_0|z_a[k^*] \leq S_0 \leq z_b[k^*]]- \barstate_0[k^*+1]\leq 0$. In this case, $\mathbb{E}_F[S_0|z_a[k^*] \leq S_0 \leq z_b[k^*]] \in [\barstate_0[k^*], \barstate_0[k^*+1]]$, and thus $\zl = z_a[k^*]$, $\zh = z_b[k^*]$, and $\zmean = \mathbb{E}_F[S_0|z_a[k^*] \leq S_0 \leq z_b[k^*]]$. 
\item[(ii)] $\mathbb{E}_F[S_0|z_a[k^*] \leq S_0 \leq z_b[k^*]]- \barstate_0[k^*+1]>0$. In this case, we consider a family of affine functions 
\[g_{\gamma}(z)= \gamma (z- \barstate_0[k^*+1]) + \revenue(\barstate_0[k^*+1]), \quad \gamma \in \partial \revenue(\barstate_0[k^*+1]) = \left[\frac{d \revenue(\barstate_0)}{d \barstate_0}[k^*], \frac{d \revenue(\barstate_0)}{d \barstate_0}[k^*+1]\right].\]
Similar to \eqref{eq:zazb}, we define $z_a[\gamma]=\min\{s_0 \in \S_0| g_{k}(z) \geq \revenue(z), s_0 \leq z \leq \barstate_0[k] \}$ and $z_b[\gamma] = \max\{s_0 \in \S_0| g_{k}(z) \geq \revenue(z), \barstate_0[k] \leq z \leq s_0\}$. Following similar procedure as in \eqref{eq:ordering}, we know that $z_a[\gamma]$ and $z_b[\gamma]$ increase in $\gamma$. Since
\[\mathbb{E}_F[S_0|z_a[k^*+1] \leq S_0 \leq z_b[k^*+1]]\leq  \barstate_0[k^*+1] \quad \mathbb{E}_F[S_0|z_a[k^*] \leq S_0 \leq z_b[k^*]]> \barstate_0[k^*+1] \]there must exists $\gamma^* \in \partial \revenue(\barstate_0[k^*+1])$ such that 
\[\mathbb{E}_F[S_0| z_a[\gamma^*] \leq S_0 \leq z_b[\gamma^*]] = \barstate_0[k^*+1].\]
Thus, we have also identified $\zl=z_a[\gamma^*]$, $\zh = z_a[\gamma^*]$, $\zmean = \barstate_0[k^*+1]$. 
\end{itemize}

On the other hand, consider the case that \[\mathbb{E}_F[S_0| z_a[\nh] \leq S_0 \leq  z_b[\nh]]- \barstate_0[\nh] > 0.\] 
As we know from \eqref{eq:twoend} that 
\[\mathbb{E}_F[S_0| z_a[\nh] \leq S_0 \leq  z_b[\nh]]- \barstate_0[\nh+1] \leq 0,\]
we must have $\zl = z_a[\nh]$,  $\zh = z_b[\nh]= \barstate_0[\nh+1]$, and $\zmean \in [\barstate_0[\nh], \barstate_0[\nh+1]]$. Therefore, we have found the pooling interval $[\zl, \zh]$ and $\zmean$ in all cases.

Finally, we construct the following $\nu$ function: 
\begin{align}\label{eq:nu}
    \nu(\barstate_0) = \left\{
    \begin{array}{ll}
     \revenue(\barstate_0), & \quad \barstate_0 \leq \zl,\\
     &\\
     \frac{\revenue(\zh) - \revenue(\zl)}{\zh -\zl} (\barstate_0- \zmean) + \revenue(\zmean), &\quad \barstate_0 \in [\zl, \zh], \\
     &\\
     \revenue(\barstate_0), & \quad \barstate_0 \geq \zh,
    \end{array}\right.
\end{align}
and $G$ as follows: 
\begin{align}\label{eq:G}
    G(\barstate_0) = \left\{
    \begin{array}{ll}
    F(\barstate_0), & \quad  \barstate_0 \leq \zl,\\
    F(\zl),  & \quad \zl< \barstate_0 < z^*, \\
    F(\zh), & \quad z^* \leq \barstate_0<  \zh, \\
    F(\barstate_0), &\quad \barstate_0 \geq \zh.
    \end{array}
    \right.
\end{align}

We first note that $\nu(\barstate_0) \geq \revenue(\barstate_0)$ since 
\[\frac{\revenue(\zh) - \revenue(\zl)}{\zh -\zl} (\barstate_0- \zmean) + \revenue(\zmean) \geq \revenue(\barstate_0), \quad \forall \barstate_0 \in [\zl, \zh].\]
Additionally, we note that for the intervals $[\inf \S_0, \barstate_0[\hat{n}]] \cup [\barstate_0[\nh], \sup \S_0]$, the function $\revenue(\barstate_0)$ is convex and $\nu(\barstate_0) = \max\{\revenue(\barstate_0), \frac{\revenue(\zh) - \revenue(\zl)}{\zh -\zl} (\barstate_0- \zmean) + \revenue(\zmean)\}$. Thus, $\nu(\barstate_0)$ is convex for $ \barstate_0 \in [\inf \S_0, \barstate_0[\hat{n}]] \cup [\barstate_0[\nh], \sup \S_0]$. Since $\nu(\barstate_0)$ is also linear for $\barstate \in [\zl, \zh] \supseteq [\barstate_0[\hat{n}], \barstate_0[\nh]]$. Thus, $\revenue(\barstate_0)$ is convex in $\barstate_0$. 

Additionally, $G$ satisfies \eqref{subeq:support_set} since $G$ pools states between $[\zl, \zh]$ to the mean $\zmean$, and $\nu(\zmean) = \revenue(\zmean)$. We next show that \eqref{subeq:integral} is satisfied: 
\begin{align*}
&\int_{S_0} \nu(z) dG(z) = \int_{\barstate_0 \leq \zl} \nu(z) dG(z) + \int_{\zl}^{\zh} \nu(z) dG(z) +  \int_{\barstate_0 \geq \zh} \nu(z) dG(z) \\
=&\int_{\barstate_0 \leq \zl} \revenue(z) dG(z) + \int_{\zl}^{\zh}  \left(\frac{\revenue(\zh) - \revenue(\zl)}{\zh -\zl} (\barstate_0- \zmean) + \revenue(\zmean)\right) dG(z) +  \int_{\barstate_0 \geq \zh} \revenue(z) dG(z) \\
=&\int_{\barstate_0 \leq \zl} \revenue(z) dG(z) +   \revenue(\zmean) 
(G(\zh) - G(\zl)) +  \int_{\barstate_0 \geq \zh} \revenue(z) dG(z) = \int_{S_0} \revenue(z) dG(z). 
\end{align*}
Finally, since $G$ is induced by a feasible information structure that pools stats between $\zl$ and $\zh$ and reveal the rest of states, we know that $G$ is a mean preserving spread of $F$. That is, $(\nu, G)$ satisfies all the conditions in Lemma \ref{lemma:duality}, and thus pooling $[\zl, \zh]$ and revealing the remaining states is an optimal information mechanism.  

Since $\zl \leq \barstate_0[\hat{n}]$ and $\zh \geq \barstate_0[\nh]$, we know that the posterior mean $\zmean > \mathbb{E}_F[S_0]$. Additionally, when $\dl=0$, we know from Lemma \ref{lemma:convex_concave} that $\hat{n}=1$. Thus, the strictly concave interval is non-empty, indicating that the pooling region is non-empty. Furthermore, when $\dh=D$, we have $\nh=\Kplus$. As a result, given any affine function $g_k(\barstate_0)$ that is tangent to $\revenue(\barstate_0)$ in the strictly concave interval, we must have $z_b[k]=\sup \S_0$. As a result, we know that $\zh= \sup \S_0$. 

In case (iii), analogous to case (ii), if the strictly concave interval is empty, then full information revelation is optimal. Otherwise, we construct a pooling region $[\zl, \zh]$ and the posterior mean $\zmean = \mathbb{E}_F[S_0|\zl \leq S_0 \leq \zh]$. In this case, $\zl \leq \barstate_0[-\nh]$ and $\zh \geq \barstate_0[-\hat{n}]$, where $-\nh$ (resp. $-\hat{n}$) is the regime boundary where $\revenue(\barstate_0)$ changes from convex to strictly concave (resp. strictly concave to convex) following Lemma \ref{lemma:convex_concave}. Following the same procedure as in case (ii), we can show that such $\zh, \zl, \zmean$ exist, and $(\nu(\barstate), G)$ as in \eqref{eq:nu} -- \eqref{eq:G} satisfy conditions in Lemma \ref{lemma:duality}. Thus, the optimal information mechanism is to pool states in $[\zl, \zh]$ and reveal the remaining states. Moreover, since $\zl \leq \barstate_0[-\nh]$ and $\zh \geq \barstate_0[-\hat{n}]$, $\zmean< \mathbb{E}_F[S_0]$. When $\dl=0$, we know from Lemma \ref{lemma:convex_concave} that $\hat{n}=-1$. Thus, the strictly concave interval is non-empty, indicating that the pooling region is non-empty. Furthermore, when $\dh=D$, we have $\nh=-\Kminus$, and $\zl= \inf \S_0$. \hfill $\square$

\medskip 
\noindent\emph{Proof of Corollary \ref{cor:two_regions}.} Analogous to the proof of Lemma \ref{lemma:convex_concave}, if all nodes have decreasing market sizes relative to distances, we can show that the function $\revenue(\barstate_0)$ is strictly concave for all $\state_0 = \barstate_0[k]$ with $k =1, \dots, \Kplus$, and strictly convex for all $\barstate_0= \state_0[-k]$ with $k=1, \dots, \Kminus$. In this case, we know from the proof of Theorem \ref{theorem:simple_partitional} that 
 the optimal information mechanism has a non-empty pooling interval. Moreover, following the construction of $g_k(z)$ as in \eqref{eq:fk} for each $k=1, \dots, \Kplus$, we can check that $z_b[k]=\sup \S_0$ as in \eqref{eq:zazb} since the derivative of $g_k(z)$ is larger than the derivative of all linear pieces associated with regimes $k'> k$ so that $g_k(z) > \revenue(z)$ for all $z> \barstate_0[k]$. Thus, we know from Theorem \ref{theorem:simple_partitional} that the optimal information mechanism fully reveal states below a threshold, and pool states above the threshold. 

 Similarly, if all nodes have increasing market sizes relative to distances, we can show that the function $\revenue(\barstate_0)$ is strictly convex for all $\state_0 = \barstate_0[k]$ with $k =1, \dots, \Kplus$, and strictly concave for all $\barstate_0= \state_0[-k]$ with $k=1, \dots, \Kminus$. The construction of $g_k(z)$ as in \eqref{eq:fk} for each $k=-1, \dots, -\Kminus$ satisfies that $z_a[k]=\inf \S_0$. The optimal information mechanism fully reveal states above a threshold, and pool states below the threshold. \hfill $\square$

\medskip 
\noindent\emph{Proof of Corollary \ref{cor:high_rate}.} We can check that when $\rate \geq \bar{\rate}$, for any $i, j$ with $d_i \neq d_j$, we have 
\[\left\vert \frac{s_{i}- s_j}{d_{i}- d_j} \right\vert \leq \frac{1}{1-\rate}.\]
Therefore, the function $\revenue(\barstate_0)$ is convex in $\barstate_0$, and full information revelation is optimal. \hfill $\square$

\medskip 
The proof of Proposition \ref{prop:computation} follows directly from the proof of Theorem \ref{theorem:simple_partitional}, and thus is omitted.

\section{Proof of statements in Section \ref{sec:single_general}}\label{apx:proof_4}

\noindent\emph{Proof of Lemma \ref{lemma:tangent}.} Following from \eqref{eq:ordering}, we know that any $k$ and function $g_k(z) = \frac{d \revenue(\barstate_0)}{d \barstate_0}[k](z- \barstate_0[k]) + \revenue(\barstate_0[k])$ satisfy 
\begin{align*}
    g_k(z) - g_{k-1}(z) \left\{\begin{array}{ll}
    <0, & \quad \forall z> \barstate_0[k]\\
    >0, & \quad \forall z< \barstate_0[k].
    \end{array}\right. 
\end{align*}
Therefore, we know from \eqref{subeq:tangent_1} and \eqref{subeq:tangent_3} that there must exists $k_{\ell, a} \leq \kltangent \leq k_{\ell, b}$ such that 
\begin{subequations}\label{eq:tangent_2}
    \begin{align}
 &g_{\kltangent-1} (\barstate_0) - \revenue(\barstate_0) > 0, \quad \forall \barstate_0 \in \left[\barstate_0[k_{\ell', a}], \barstate_0[k_{\ell', b}]\right],\label{subeq:tangent2_1}\\
        &\exists \barstate_0 \in \left[\barstate_0[k_{\ell', a}], \barstate_0[k_{\ell', b}]\right], \quad  s.t. 
 \quad g_{\kltangent} (\barstate_0)  - \revenue(\barstate_0) \leq 0.\label{subeq:tangent2_2}
    \end{align}
\end{subequations}
Similarly, following from \eqref{subeq:tangent_2} and \eqref{subeq:tangent_4}, there must exists $k_{\ell', a} \leq \klptangent \leq k_{\ell', b}$ such that 
\begin{subequations}
    \begin{align}
 &g_{\klptangent}(\barstate_0)- \revenue(\barstate_0) > 0, \quad  \forall \barstate_0 \in \left[\barstate_0[k_{\ell, a}], \barstate_0[k_{\ell, b}]\right],\\
        &\exists \barstate_0 \in \left[\barstate_0[k_{\ell, a}], \barstate_0[k_{\ell, b}]\right], \quad  s.t. 
 \quad  g_{\klptangent-1}(\barstate_0) - \revenue(\barstate_0) \leq 0. \label{subeq:tangent3_2}
    \end{align}
\end{subequations}
We construct the function $g(z)$ as follows 
\[g(z) = \frac{\revenue(\barstate_0[\klptangent])- \revenue(\barstate_0[\kltangent])}{\barstate_0[\klptangent]- \barstate_0[\kltangent]} (z- \barstate_0[\kltangent])+ \revenue(\barstate_0[\kltangent]).\]
Then, $g(\barstate_0[\kltangent]) = \revenue(\barstate_0[\kltangent])$ and $g(\barstate_0[\klptangent]) = \revenue(\barstate_0[\klptangent])$. It remains to prove that the following holds: 
        \[\gamma =\frac{\revenue(\barstate_0[\klptangent])- \revenue(\barstate_0[\kltangent])}{\barstate_0[\klptangent]- \barstate_0[\kltangent]}  \in \left[\frac{d \revenue(\barstate_0)}{d \barstate_0}[\kltangent], \frac{d \revenue(\barstate_0)}{d \barstate_0}[\kltangent-1]\right] \cap \left[\frac{d \revenue(\barstate_0)}{d \barstate_0}[\klptangent], \frac{d \revenue(\barstate_0)}{d \barstate_0}[\klptangent-1]\right]. \] 
        From \eqref{subeq:tangent2_1}, we know that $g_{\kltangent-1}(\barstate_0[\klptangent])> \revenue(\barstate_0[\klptangent])$, thus, $\gamma < \frac{d \revenue(\barstate_0)}{d \barstate_0}[\kltangent-1]$. Proving that $\gamma \geq \frac{d \revenue(\barstate_0)}{d \barstate_0}[\kltangent]$,  is equivalent to showing that $g_{\kltangent}(\barstate_0[\klptangent])\leq  \revenue(\barstate_0[\klptangent])$. For the sake of contradiction, we assume that $g_{\kltangent}(\barstate_0[\klptangent])>  \revenue(\barstate_0[\klptangent])$. Then, we must have: 
        \[\frac{d \revenue(\barstate_0)}{d \barstate_0}[\klptangent] \stackrel{(a)}{>} \gamma, \quad \frac{d \revenue(\barstate_0)}{d \barstate_0}[\klptangent-1] \stackrel{(b)}{<} \frac{d \revenue(\barstate_0)}{d \barstate_0}[\klptangent], \]
        where (a) is due to $g_{\klptangent}(\barstate_{0}[\kltangent])> \revenue(\barstate_{0}[\kltangent])$, and (b) is due to the fact that $\revenue(\barstate_0)$ is strictly concave at $\barstate_{0}[\klptangent]$. There are two cases (We note that $\frac{d \revenue(\barstate_0)}{d \barstate_0}[\klptangent-1]=\gamma$ violates Assumption \ref{as:regularity}): 

\medskip
   \noindent (Case 1): $\frac{d \revenue(\barstate_0)}{d \barstate_0}[\klptangent-1] \in \left(\gamma, \frac{d \revenue(\barstate_0)}{d \barstate_0}[\klptangent]\right)$. In this case, we argue that $\frac{d \revenue(\barstate_0)}{d \barstate_0}[\klptangent] < \frac{d \revenue(\barstate_0)}{d \barstate_0}[\kltangent]$. This is because
   \[\frac{d \revenue(\barstate_0)}{d \barstate_0}[\klptangent]= \frac{g_{\klptangent}(\barstate_0[\klptangent]) - g_{\klptangent}(\barstate_0[\kltangent])}{\barstate_0[\klptangent]- \barstate_0[\kltangent]}, \quad \frac{d \revenue(\barstate_0)}{d \barstate_0}[\kltangent] = \frac{g_{\kltangent}(\barstate_0[\klptangent]) - g_{\kltangent}(\barstate_0[\kltangent])}{\barstate_0[\klptangent]- \barstate_0[\kltangent]}, \]
   and $\revenue(\barstate_0[\klptangent]) = g_{\klptangent}(\barstate_0[\klptangent])< g_{\kltangent}(\barstate_0[\klptangent])$ and $g_{\klptangent}(\barstate_0[\kltangent]) > \revenue(\barstate_0[\kltangent])= g_{\kltangent}(\barstate_0[\kltangent])$. As a result, we know that 
    \[g_{\kltangent}(z) \stackrel{(a)}{>} g_{\klptangent}(z) \stackrel{(b)}{\geq}  \revenue(z), \quad \forall z \in [\barstate_0[\klptangent], \barstate_0[k_{\ell',b}]], \]
where (a) follows from $g_{\kltangent}(\barstate_0[\klptangent]) > \revenue(\barstate_0[\klptangent]) = g_{\klptangent}(\barstate_0[\klptangent])$ and $\frac{d \revenue(\barstate_0)}{d \barstate_0}[\klptangent] < \frac{d \revenue(\barstate_0)}{d \barstate_0}[\kltangent]$, and (b) is due to concavity of $\revenue(\barstate_0)$. 

Additionally, it follows from \eqref{subeq:tangent3_2} that there exists at least one $\hat{s}_0 \in [\barstate_0[k_{\ell, a}], \barstate_0[k_{\ell, b}]]$ such that $g_{\klptangent-1}(\hat{s}_0)\leq  \revenue(\hat{s}_0)$. Since $\revenue(\barstate_0)$ is concave in $[\barstate_0[k_{\ell, a}], \barstate_0[k_{\ell, b}]]$, we have $g_{\kltangent}(\hat{s}_0) \geq \revenue(\hat{s}_0)$. Therefore, $g_{\kltangent}(\hat{s}_0) \geq g_{\klptangent-1}(\hat{s}_0)$. Since we have assumed that $\revenue(\barstate_0[\klptangent]) = g_{\klptangent-1}(\barstate_0[\klptangent]) < g_{\kltangent}(\barstate_0[\klptangent])$, we must have $g_{\klptangent-1}(z) < g_{\kltangent}(z)$ for all $z \in [\barstate_0[k_{\ell', a}, \barstate_0[\klptangent]]$. Since $\revenue(\barstate_0)$ is concave in $[\barstate_0[k_{\ell', a}], \barstate_0[k_{\ell', b}]]$, we have $\revenue(z) \leq g_{\klptangent-1}(z)$ for all $z \in [\barstate_0[k_{\ell', a}, \barstate_0[\klptangent]]$. Therefore, $\revenue(z) < g_{\kltangent}(z)$ for all $z \in [\barstate_0[k_{\ell', a}, \barstate_0[\klptangent]]$, and hence  $g_{\kltangent}(z) > \revenue(z)$ for all $z \in [\barstate_0[k_{\ell', a}], \barstate_0[k_{\ell',b}]]$, which contradicts \eqref{subeq:tangent2_2}.

\medskip
    \noindent(Case 2): $\frac{d \revenue(\barstate_0)}{d \barstate_0}[\klptangent-1] < \gamma$. In this case, we note that $ \frac{d \revenue(\barstate_0)}{d \barstate_0}[\klptangent-1] <\gamma < \frac{d \revenue(\barstate_0)}{d \barstate_0}[\klptangent] $, and thus 
    \[\gamma (z- \barstate_0[\klptangent])+ \revenue(\barstate_0[\klptangent]) \geq \revenue(z), \quad \forall z \in [\barstate_0[k_{\ell',a}], \barstate_0[k_{\ell', b}]].\]
    Since we assumed that $\gamma < \frac{d \revenue(\barstate_0)}{d \barstate_0}[\kltangent]$, we know that 
    \[g_{\kltangent}(z) > \gamma (z- \barstate_0[\klptangent])+ \revenue(\barstate_0[\klptangent]) \geq \revenue(z), \quad \forall z \in [\barstate_0[k_{\ell',a}], \barstate_0[k_{\ell', b}]],\]
    which contradicts \eqref{subeq:tangent2_2}.

    We have derived contradiction in both cases. Therefore, we know that $\gamma \geq \frac{d \revenue(\barstate_0)}{d \barstate_0}[\kltangent]$. We have thus proved that $\gamma \in \left[\frac{d \revenue(\barstate_0)}{d \barstate_0}[\kltangent], \frac{d \revenue(\barstate_0)}{d \barstate_0}[\kltangent-1]\right]$. The proof of $\gamma \in \left[\frac{d \revenue(\barstate_0)}{d \barstate_0}[\klptangent], \frac{d \revenue(\barstate_0)}{d \barstate_0}[\klptangent-1]\right]$ is analogous, and thus is omitted.

    We next argue that such affine function $g(z)$ is unique. From the construction of $\kltangent$ and $\klptangent$, we know that $\kltangent$ and $\klptangent$ are unique. For any $\hat{s}_0 \in [\barstate_0[k_{\ell, a}], s_0[k_{\ell, b}]]$, we define affine function $\tilde{g}(z|\hat{s}_0)= \tilde{\gamma}(z- \hat{s}_0)  + \revenue(\hat{s}_0)$, where $\tilde{\gamma} = \frac{d \revenue(s_0)}{d s_0}[k]$ if $\hat{s}_0 \in (s_0[k], s_0[k+1])$ and $\tilde{\gamma} \in [\frac{d \revenue(s_0)}{d s_0}[k-1], \frac{d \revenue(s_0)}{d s_0}[k]]$ if $\hat{s}_0 = s_0[k]$. We can check that for any $s_0[k_{\ell, a}] \leq \hat{s}_0 < s_0[\kltangent]$, $\tilde{g}(z|\hat{s}_0) > \revenue (z)$ for all $z \in [s_0[k_{\ell', a}], s_0[k_{\ell', b}]]$. Additionally, for any $s_0[\kltangent]< \hat{s}_0 \leq s_0[k_{\ell, b}]$, $\tilde{g}(z|\hat{s}_0) < \revenue (z)$ for all at least one $z \in [s_0[k_{\ell', a}], s_0[k_{\ell', b}]]$. Thus, there does not exist another affine function that is tangent to $\revenue(s_0)$ in both $\ell$ and $\ell'$ concave intervals. 
    
    Finally, we prove that that if any one of the constraints \eqref{eq:tangent} is violated, then such affine function does not exist. Assume that constraint \eqref{subeq:tangent_1} is violated, i.e. there exists $s_0 \in [\barstate_0[k_{\ell', a}], s_0[k_{\ell', b}]]$ such that $g_{k_{\ell, a}}(s_0)<\revenue(s_0)$. Then, for any $\hat{s}_0 \in [s_0[k_{\ell, a}], s_0[k_{\ell, b}]]$, there must exist at least one $z \in [s_0[k_{\ell', a}], s_0[k_{\ell', b}]]$ such that $g(z| \hat{s}_0) < \revenue(z)$. Thus, the tangent affine function does not exist. We can analogously argue that the tangent affine function does not exist when any of the other three constraints is violated.  \hfill $\square$

\color{black}

    \medskip 

    \begin{definition}\label{def:mpc}
    $G$ is a mean-preserving contraction of $F$, i.e., $G\preceq F$, if \begin{subequations}
\begin{align}
    \int_{z \leq s_0} F(z) dz &\geq \int_{z\leq s_0} G(z) dz, \quad \forall \barstate_0 \in \S_0, \label{sub:mpc_interior}\\
    \int_{z \in \S_0} F(z) dz &= \int_{z \in \S_0} G(z) dz.\label{subeq:equal_boundary}
\end{align}
\end{subequations}
    \end{definition}
\begin{lemma}[\cite{blackwell1953equivalent, gentzkow2016rothschild}]\label{lemma:MPC}A posterior mean distribution $G$ is feasible given prior $F$ if and only if $G$ is a mean preserving contraction of $F$. 
\end{lemma}

\begin{lemma}[\cite{dworczak2019simple}]\label{lemma:existence_nu}
Suppose that $\revenue(\barstate_0)$ is Lipchitz continuous. Then, for every optimal posterior mean distribution $G^*$, there exists a convex and continuous function $\nu: \S_0 \to \mathbb{R}$ such that $\nu(\barstate_0) \geq \revenue(\barstate_0)$, and $(G^*, \nu)$ satisfies \eqref{subeq:support_set} -- \eqref{subeq:mean_preserving}. 
\end{lemma}

\begin{lemma}[\cite{candogan2021optimal}]\label{lemma:interval_decomposition}
For any optimal posterior mean distribution $G^{*}$, there exists a finite number of intervals $\{\intervalj\}_{j \in \J}$, where $\intervalj = [\zja, \zjb]$, such that 
\begin{itemize}
\item[-] States outside of $\cup_{j \in \J} \intervalj$ are fully revealed, i.e. $G^*(\barstate_0) = F(\barstate_0)$ for all $\barstate \in \S_0 \setminus \{\cup_{j \in \J} \intervalj\}$.
\item[-] Each interval is either pooled to a single atom $x= \mathbb{E}_F[S_0 | S_0 \in \intervalj]$ or pooled to two atoms $x, y$ according to a double-interval structure, where $x = \mathbb{E}_F[S_0 | S_0 \in [\zja, \zjp] \cup [\zjpp, \zjb]]$, $y = \mathbb{E}_F[S_0| S_0 \in [\zjp, \zjpp]]$ and $\zja < \zjp< \zjpp< \zjb$. 
\end{itemize}
\end{lemma}

\begin{lemma}\label{lemma:interval_boundary}
    Consider the sequence of intervals $\{\intervalj\}_{j \in \J}$, where $\intervalj = [\zja, \zjb]$, associated with the optimal posterior mean distribution $G^*$, we must have: 
    \begin{align}\label{eq:boundary}
        \int_{z \leq \zja} F(z) dz &= \int_{z\leq \zja} G^*(z) dz, \quad \int_{z \leq \zjb} F(z) dz = \int_{z\leq \zjb} G^*(z) dz \quad \forall j \in \J.
    \end{align}
\end{lemma}
\medskip 

\noindent\emph{Proof of Lemma \ref{lemma:interval_boundary}.} The proof builds on Lemmas \ref{lemma:MPC} and \ref{lemma:interval_decomposition}. We note that the posterior mean distribution $G^*$ is a mean-preserving contraction of $F$ in each interval $\intervalj$. Therefore, following from \eqref{subeq:equal_boundary}
\[\int_{\zja}^{\zjb} G^*(z) dz = \int_{\zja}^{\zjb} F(z) dz, \quad \forall j \in \J.\]
Since $F(\barstate_0)=G^*(\barstate_0)$ for all $\barstate_0 \in \S_0 \setminus \{\cup_{j \in J} \intervalj\}$, we can conclude that \eqref{eq:boundary} holds. \hfill $\square$

\begin{lemma}\label{lemma:double_interval}
Given the prior distribution $F$, if interval $[\zja, \zjb]$ with two atoms $\zja < x< y< \zjb$ is generated by a double interval structure given the optimal information mechanism, then there must exist $\px, \py$ that satisfies: 
\begin{subequations}\label{eq:px_py}
\begin{align}
&\px+\py = F(\zjb)- F(\zja), \label{subeq:sum_atoms_prob}\\
&x \px+ y \py = \int_{\zja}^{\zjb} z dF(z), \label{subeq:sum_atoms_expec}\\
& (F^{-1}(F(\zja) + \px) - x) \px  < \int_{\zja}^{F^{-1}(F(\zja) + \px)} (F(z) - F(\zja)) dz, \label{subeq:MPC}\\
&\px>0, \quad \py>0. \label{subeq:positive_p}
\end{align}
\end{subequations}
Moreover, $(\px, \py)$ that satisfies \eqref{eq:px_py} is unique. 
\end{lemma}
\medskip 
\noindent \emph{Proof of Lemma \ref{lemma:double_interval}.} If $[\zja, \zjb]$ has a double-interval information structure associated with atoms $x, y$, then $\px, \py$ must satisfy \eqref{subeq:sum_atoms_prob} -- \eqref{subeq:sum_atoms_expec}, and \eqref{subeq:positive_p}, {where \eqref{subeq:sum_atoms_prob} and \eqref{subeq:positive_p} ensure that $(\px, \py)$ is a valid probability vector given $F$ in the interval $[\zja, \zjb]$ and \eqref{subeq:sum_atoms_expec} ensures that $(\px, \py)$ is mean-preserving.} Additionally, we note that the optimal posterior mean distribution $G^*$ with $(\px, \py)$ is given by: 
\begin{align*}
G^*(\barstate)= \left\{\begin{array}{ll}
F(\zja), & \quad \forall \barstate_0 \in [\zja, x), \\
F(\zja)+ \px, & \quad \forall \barstate_0 \in [x, y), \\
F(\zjb) & \quad \forall \barstate_0 \in [y, \zjb].
\end{array}\right.
\end{align*}
We need to ensure that $\int_{z \leq s_0} F(z) dz - \int_{z \leq s_0} G^*(z) dz \geq 0$ for all $s_0 \in [\zja, \zjb]$. We define $\Delta(\barstate_0) = \int_{z \leq s_0} F(z) dz - \int_{z \leq s_0} G^*(z) dz$, and 
\begin{align}
    \frac{d \Delta(\barstate_0)}{d \barstate_0} = F(\barstate_0) - G^{*}(\barstate_0) \left\{
    \begin{array}{ll}
    =0, & \quad \barstate_0 =\zja,\\
    >0, & \quad \forall \barstate_0 \in (\zja, x)\\
    < 0, & \quad \forall \barstate_0 \in [x, F^{-1}(F(\zja)+\px)), \\
    =0, & \quad \barstate_0 = F^{-1}(F(\zja)+\px), \\
    >0, & \quad \forall \barstate_0 \in (F^{-1}(F(\zja)+\px), y), \\
<0, & \quad \forall \barstate_0 \in [y, \zjb)\\
=0, & \quad \barstate_0= \zjb. 
    \end{array}\right.
\end{align} 
Therefore, we only need to verify that $\Delta(\barstate_0) \geq 0$ for $\barstate \in \{\zja, F^{-1}(F(\zja)+\px), \zjb\}$. We know from Lemma \ref{lemma:interval_boundary} that $\Delta(\zja)= \Delta(\zjb)=0$. Thus, we need to satisfy: 
\begin{align*}
&\Delta(F^{-1}(F(\zja)+\px)) = \int_{\zja}^{F^{-1}(F(\zja)+\px)} F(z) dz-  \int_{\zja}^{F^{-1}(F(\zja)+\px)} G^*(z) dz \\
=& \int_{\zja}^{F^{-1}(F(\zja)+\px)} F(z) dz-\left( F(\zja) (x- \zja)+ (F(\zja)+ \px) (F^{-1}(F(\zja)+\px)- x)\right)\\
=& \int_{\zja}^{F^{-1}(F(\zja)+\px)} (F(z)- F(\zja)) dz-\px (F^{-1}(F(\zja)+\px)- x) \geq 0. 
\end{align*}
We note that when the above inequality is tight, the double-interval structure degenerates to two disjoint pooling intervals each with a single atom $x$ and $y$ separately. Therefore, $\px$ must satisfy \eqref{subeq:MPC}. 
\hfill $\square$ 

\begin{lemma}\label{lemma:extend_interval}
{Suppose $[\zjhata, \zjhatb]$ admits a double-interval partition with given atoms $x, y$.
Define \begin{subequations}
\begin{align}
   h(\epsilon) &= x (F(\zjhata) - F(\zjhata-\epsilon)) - \int_{\zjhata-\epsilon}^{\zjhata} zdF(z) - (y-x) \py, \label{subeq:epa}\\
   \tilde{h}(\epsilon) &=\int_{\zjhatb}^{\zjhatb+\epsilon} zdF(z) - y(F(\zjhatb+ \epsilon) - F(\zjhatb)) - (y-x)\px.\label{subeq:epb}
\end{align}
\end{subequations}
Let $\epsilon_a$ (resp. $\epsilon_b$) be defined as the smallest solution of $h(\epsilon)=0$ (resp. $\tilde{h}(\epsilon)=0$) when $h(\zja- \inf \S_0) \geq 0$ (resp. $\tilde{h}(\sup \S_0 - \zjb) \geq 0$), and $\zja - \inf \S_0$ (resp. $\sup \S_0 - \zjb$) otherwise. 
For any $\epsilon \in (0, \epa)$ (resp. $\epsilon \in (0, \epb)$), $[\zjhata- \epsilon, \zjhatb]$ (resp. $[\zjhata, \zjhatb+\epsilon]$) is also associated with a double-interval structure with atoms $x, y$.
Furthermore, {$\mathbb{E}_F[S_0 | \zjhata - \epa\leq S_0 \leq \zjhatb] = x$ when $\epa < \zja - \inf\S_0$, and $\mathbb{E}_F[S_0 | \zjhata \leq S_0 \leq \zjhatb+ \epb] = y$ when $\epb < \sup \S_0 - \zjb$.}}
\end{lemma}
\medskip 

\noindent\emph{Proof of Lemma \ref{lemma:extend_interval}.} 
We first prove for the interval $[\zjhata- \epsilon, \zjhatb]$. We denote the interval associated with $x$ (resp. $y$) as $[\zjhata, \zjhatp] \cup [\zjhatpp, \zjhatb]$ (resp. $[\zjhatp, \zjhatpp]$), and the probability as $\px$ (resp. $\py$). We define 
\[u(\epsilon) = F(\zjhata) - F(\zjhata-\epsilon), \quad v(\epsilon) = \int_{\zjhata-\epsilon}^{\zjhata} zdF(z).\] 
We construct 
$\px'(\epsilon) = \px+ u(\epsilon) + \delta(\epsilon)$ and $\py'(\epsilon)=\py(\epsilon)-\delta(\epsilon)$, where 
\[\delta(\epsilon) = \frac{xu(\epsilon) -v(\epsilon)}{y-x}.\]
We note that for any $\delta(\epsilon) \in (0, \py)$, there exists a sub-interval $[\tilde{z}^{'}_{\jhat}, \tilde{z}^{"}_{\jhat}] \subseteq [\zjhatp, \zjhatpp]$ such that $\mathbb{E}_F[S_0 | \tilde{z}^{'}_{\jhat} \leq S_0 \leq \tilde{z}^{"}_{\jhat}]=y$ and $F(\tilde{z}^{"}_{\jhat}) - F(\tilde{z}^{'}_{\jhat})= \py- \delta$. Thus, the double-interval structure $[\zjhata-\epsilon, \tilde{z}^{'}_{\jhat}] \cup [\tilde{z}^{"}_{\jhat}, \zjhatb]$ (resp. $[\tilde{z}^{'}_{\jhat}, \tilde{z}^{"}_{\jhat}]$) generates $x$ (resp. $y$) with probability $\px'(\epsilon)$ (resp. $\py'(\epsilon)$) for the extended interval $[\zjhata- \epsilon, \zjhatb]$. 

{For any $\epsilon < \inf \S_0 - \zja$, we can check that the value of $\delta(\epsilon)$ is non-decreasing in $\epsilon$, and strictly increasing if $f(\zja- \epsilon)>0$, where $f(\cdot)$ is the probability density function corresponding to $F(\cdot)$:  
\[\frac{d \delta(\epsilon)}{d \epsilon} = \frac{1}{y-x}\left(xf(\zja - \epsilon) - (\zja- \epsilon)f(\zja- \epsilon) \right) = \frac{f(\zja- \epsilon) (x- \zja + \epsilon)}{y-x} \geq 0. \]
We note that $\delta(0)=0$. Since $\delta(\epsilon)$ is non-decreasing in $\epsilon$, if $\delta(\zja - \inf \S_0) < \py$, then $\delta(\epsilon)< \py$ for any $\epsilon \in [0, \zja - \inf \S_0]$. This indicates that for any $\epsilon \in [0, \zja - \inf \S_0]$, the extended interval is also associated with a double-interval structure with atoms $x$ and $y$. 

On the other hand, if $\delta(\zja - \inf \S_0) \geq  \py$, then \eqref{subeq:epa}, which is equivalent to $\delta(\epsilon)=\py$, has a solution $\hat{\epsilon}$. This solution is unique when $f(\epsilon)>0$ in a local neighborhood of $\hat{\epsilon}$. In this case $\epa = \hat{\epsilon}$. Additionally, if there exists a local neighborhood of $\hat{\epsilon}$ such that $f(\epsilon)=0$, then the solution set of \eqref{subeq:epa} is a closed interval, and $\epa$ equals to the the smallest solution. In both cases, for any $\epsilon \in [0, \epa)$, the extended interval is also associated with a double-interval structure with atoms $x$ and $y$. Moreover, when  $\epsilon = \epa$, the double interval structure becomes degenerate and $\py'(\epsilon)=0$. Thus, $\mathbb{E}_F[S_0 | \zja- \epa, \zjb]= x$. }

Similarly, we can show that the interval $[\zjhata, \zjhatb+ \epsilon]$ is associated with $\px'(\epsilon) = \px- \tilde{\delta}(\epsilon)$ and $\py' = \py+ \tilde{u}(\epsilon)+ \tilde{\delta}(\epsilon)$, where 
\[\tilde{u}(\epsilon) = F(\zjhatb+ \epsilon) - F(\zjhatb), \quad \tilde{v}(\epsilon) = \int_{\zjhatb}^{\zjhatb+\epsilon} zdF(z), \quad \tilde{\delta}(\epsilon) = \frac{\tilde{v} - \tilde{u}y}{y-x}.\]
Additionally, there exists a double-interval structure $[\zjhata, \tilde{z}^{'}_{\jhat}] \cup [\tilde{z}^{"}_{\jhat}, \zjhatb]$ and $[\tilde{z}^{'}_{\jhat}, \tilde{z}^{"}_{\jhat}]$ associated atoms $x, y$, where $[\tilde{z}^{'}_{\jhat}, \tilde{z}^{"}_{\jhat}] \supseteq [\zjhatp, \zjhatpp]$. {Similarly, the value of $\tilde{\delta}(\epsilon)$ is non-decreasing in $\epsilon$. If $\tilde{\delta}(\sup \S_0 - \zjb) < \px$, then $\tilde{\delta}(\epsilon)< \px$ for any $\epsilon \in [0, \sup \S_0 - \zjb]$, indicating that for any $\epsilon \in [0, \sup\S_0 - \zjb]$, the extended interval is also associated with a double-interval structure with atoms $x$ and $y$. On the other hand, if $\tilde{\delta}(\sup \S_0 - \zjb) \geq \px$, then by taking $\epb$ equals to the smallest solution of \eqref{subeq:epb}, the extended interval is also associated with a double-interval structure with atoms $x$ and $y$ for any $\epsilon \in [0, \epb)$. Moreover, when  $\epsilon = \epb$, the double interval structure becomes degenerate and $\mathbb{E}_F[S_0 | \zja, \zjb+ \epb]= y$.}

\medskip 
\noindent\emph{Proof of Proposition \ref{prop:three_conditions}.} Assume for the sake of contradiction that there does not exist an optimal information mechanism that is monotone partitional. From Lemma \ref{lemma:interval_decomposition}, we know that the intervals $\{\intervalj\}_{\j \in \J}$ associated with the optimal posterior mean distribution $G^{*}$ must have at least one $\jhat$ such that $\intervaljhat= [\zjhata, \zjhatb]$ has a double-interval structure: pooling states in the intervals $[\zjhata, \zjhatp] \cup [\zjhatpp, \zjhatb]$ to generate atom $x$ and pooling states in the interval $[\zjhatp, \zjhatpp]$ to generate atom $y$, where $\zjhata < \zjhatp < \zjhatpp < \zjhatb$ and $\zjhata < x< y < \zjhatb$. Moreover, we denote the probability of atom $x$ as $\px$ and that of atom $y$ as $\py$. Then, we must have:
    \begin{subequations}
\begin{align}
\px&= \mathrm{Pr}(S_0 \in [\zjhata, \zjhatp] \cup [\zjhatpp, \zjhatb]) = \int_{\zjhata}^{\zjhatp} dF(z) + \int_{\zjhatpp}^{\zjhatb}dF(z), \\
\py&= \mathrm{Pr}(S_0 \in [\zjhatp,  \zjhatpp]) = \int_{\zjhatp}^{\zjhatpp} dF(z),\\
x &=\frac{1}{\px}\left(\int_{\zjhata}^{\zjhatp} z dF(z) + \int_{\zjhatpp}^{\zjhatb} z dF(z)\right), \label{subeq:x}\\
y&=\frac{1}{\py}\left(\int_{\zjhatp}^{\zjhatpp} z dF(z)\right). \label{subeq:y}
\end{align}
\end{subequations}
In our setting, $\revenue(\barstate_0)$ is a Lipchitz continuous function. Thus, we know from Lemma \ref{lemma:existence_nu} that for any optimal posterior mean distribution $G^{*}$, there must exist a convex upper closure function $\nu(\barstate_0)$ such that conditions $(G^*, \nu)$ satisfies \eqref{subeq:support_set} -- \eqref{subeq:mean_preserving}.  {We note that  
\begin{align}
&\int_{z \in [\zjhata, \zjhatb]} \nu(z) dF(z)= \int_{z \in [\zjhata, \zjhatp] \cup [\zjhatpp, \zjhatb]} \nu(z) dF(z) + \int_{z \in [\zjhatp, \zjhatpp]} \nu(z) dF(z) \notag \\
=& \px\int_{z \in [\zjhata, \zjhatp] \cup [\zjhatpp, \zjhatb]} \frac{\nu(z) dF(z)}{\px} +  \py  \int_{z \in [\zjhatp, \zjhatpp]} \frac{\nu(z) dF(z)}{\py}\notag\\
\stackrel{(a)}{\geq} & \px \nu\left(\frac{\int_{z \in [\zjhata, \zjhatp] \cup [\zjhatpp, \zjhatb]}  z dF(z)}{\px}\right)+ \py \nu\left( \frac{\int_{z \in [\zjhatp, \zjhatpp]} zdF(z)}{\py}\right)  \stackrel{(b)}{=} \px \nu\left(x\right)  + \py \nu(y) \notag \\
\stackrel{(c)} {=} &\int_{z \in [\zjhata,  \zjhatb]} \nu(z) dG(z),\label{eq:affine_side}
\end{align}where (a) is due to the fact that $\nu(\barstate_0)$ is convex, (b) follows from \eqref{subeq:x} -- \eqref{subeq:y} and (c) is due to the fact that $G^*$ assigns probability mass $\px$ to atom $x$ and $\py$ to $y$. 
From \eqref{subeq:integral}, we know that the inequality in \eqref{eq:affine_side} must be equality, indicating that the function $\nu(\barstate_0)$ is an affine function in $[\zjhata, \zjhatb]$. We denote this function as $\nu(\barstate_0) = \gamma \barstate_0 + \lambda$. }

Furthermore, we know from Lemma \ref{lemma:existence_nu} that $\nu(\barstate_0)\geq \revenue(\barstate_0)$ for $\barstate_0 \in [\zjhata, \zjhatb]$, and $\revenue(x)= \nu(x)$, $\revenue(y)=\nu(y)$. We argue that the following statements are true: 
\begin{itemize}
\item[(i)] $x, y$ are not in the same equilibrium regime, i.e. not belong to the same linear piece of $\revenue(\barstate)$, since otherwise the double interval can be replaced by either fully revealing the state in $[\zjhata, \zjhatb]$ or pooling all states in $[\zjhata, \zjhatb]$ without changing the expected revenue. 
\item[(ii)] $x, y$ are not interior points of two different regimes. This is because if $x$ (or $y$) is in the interior of a regime $k$ such that the derivative $\frac{d \revenue(\barstate_0)}{d \barstate_0}[k] \neq \gamma$, then there must exist $0< \epsilon< \min\{x- \zjhata, \zjhatb-y\}$ such that $\revenue(z) > \nu(z)$ for some $z \in (x-\epsilon, x+\epsilon)$ (or $z \in (y-\epsilon, y+\epsilon)$), contradicting to the fact that $\nu(\barstate_0) \geq \revenue(\barstate_0)$ for all $\barstate_0 \in [\zjhata, \zjhatb]$. On the other hand, if $\frac{d \revenue(\barstate_0)}{d \barstate_0}[k] = \gamma$, then Assumption \ref{as:regularity} is violated. 
\end{itemize}
Thus, $x, y$ must be two regime boundaries. We further show that $x, y$ are  the boundaries 
of $\barstate_0[\kltangent]$ and $\barstate_0[\klptangent]$ respectively, and they belong to two concave intervals $\ell< \ell'$. From Assumption \ref{as:regularity}, we know that $\revenue(\barstate_0)$ must be either strictly convex or strictly concave at each regime boundary. If $x$ (or $y$) is at a regime boundary that is strictly convex, then there must exist $\epsilon < \min\{x- \zjhata, \zjhatb - y\}$ such that $\revenue(z) > \nu(z)$ for some $z \in (x-\epsilon, x+\epsilon)$ (or $(y- \epsilon, y+\epsilon)$). Thus, $x$ and $y$ must be two regime boundaries such that $\revenue(\barstate_0)$ is strictly concave. Moreover, the two regime boundaries cannot belong to the same concave interval since otherwise $x, y$ must be associated with the same regime, which has been ruled out in (i). Therefore, we must have $x= \barstate_0[\kltangent]$, $y=\barstate_0[\klptangent]$, where $\ell< \ell'$ are two concave intervals. Furthermore, we must have: 
\begin{align*}
    \gamma \in \left[\frac{d \revenue(\barstate_0)}{d \barstate_0}[\kltangent], \frac{d \revenue(\barstate_0)}{d \barstate_0}[\kltangent-1]\right] \cap \left[\frac{d \revenue(\barstate_0)}{d \barstate_0}[\klptangent], \frac{d \revenue(\barstate_0)}{d \barstate_0}[\klptangent-1]\right].
\end{align*}

Under condition (C1), such two concave intervals $\ell<\ell'$ do not exist, and thus the optimal information mechanism cannot have a double-interval structure, indicating that the optimal information mechanism is monotone partitional. 

Under condition (C2), we know from Lemma \ref{lemma:tangent} that such affine function $\nu(\barstate_0)$ does not exist for any two concave intervals $\ell< \ell'$. Thus, an optimal information mechanism must be monotone partitional.

Under condition (C3), assume for the sake of contradiction that there exists a pair of concave intervals $\ell< \ell'$ such that the optimal information mechanism has a double-interval structure associated with the affine function $g(s_0)$ tangent to $\revenue(s_0)$ in the two concave intervals, and the two atoms $x= \barstate_0[\kltangent]$, $y=\barstate_0[\klptangent]$. Then, we know from Lemma \ref{lemma:existence_nu} that there must exist a convex function $\nu(s_0)$ such that $(G^*, \nu)$ satisfies \eqref{eq:duality}, where $G^*$ is the optimal posterior mean distribution. 

We know from the argument above that when the optimal information mechanism contains the double interval structure associated with $x, y$, there must exist an interval $[\underline{z}, \bar{z}] \subseteq [\zkl, \zkph]$ such that $\underline{z} < x< y< \bar{z}$ and $\nu(z) = g(z)$ for $z \in [\underline{z}, \bar{z}]$. Here, we recall that $\zkl = \max\{z <x| \revenue(z) \geq g(z)\}$ and $\zkph = \min\{z >y| \revenue(z) \leq g(z)\}$.

Consider the scenario where $x, y< \mathbb{E}_F[S_0]$ and $\zkph = \sup \S_0$. Since $\nu(s_0)$ is a convex function, we know that 
\begin{align}\label{eq:support_above}\nu(s_0) \stackrel{(a)}{\geq} g(s_0) \stackrel{(b)}{>} \revenue(s_0), \quad \forall s_0 > y,
\end{align}
where (a) is due to the convexity of $\nu(s_0)$ and (b) follows from the definition of $\zkph$ and the fact that $\zkph=\sup S_0$. {Consequently, we know that $\{\S_0 | \revenue(s_0) = \nu(s_0)\} \cap \{\S_0| s_0 > y\} = \emptyset$. Following from \eqref{subeq:support_set} $supp(G^*) \subseteq \{\S_0 | \revenue(s_0) = \nu(s_0)\}$ and the fact that $y< \mathbb{E}_F[S_0]$, we know that $supp(G^*) \cap \{\S_0| s_0 \geq \mathbb{E}_F[S_0]\} = \emptyset $, which contradicts  the fact that $G^*$ is a mean-preserving contraction of the prior $F$. }


Similarly, for the other scenario where $x, y> \mathbb{E}_F[S_0]$ and $\zkl = \inf \S_0$, we can show that $supp(G^*) \cap \{\S_0| s_0 \leq \mathbb{E}_F[S_0]\} = \emptyset$, which again contradicts   the fact that $G^*$ is a mean preserving contraction of $F$. 

Therefore, we can conclude that in both cases, the optimal information mechanism does not contain a double-interval structure associated with concave intervals $\ell< \ell'$. Since this argument holds for all such pairs, we know that the optimal information mechanism must be monotone partitional. 

\color{black}

We next prove that when the prior distribution $F$ satisfies condition (C4), then for any pair of $\ell<\ell'$ that violates (C2), the two atoms $x, y$ associated with the two concave intervals cannot be generated by a double interval structure. For the sake of contradiction, we assume that there exists a pair of $\ell < \ell'$ such that the associated two atoms $x= \barstate_0[\kltangent]$, $y=\barstate_0[\klptangent]$ are generated by a feasible double-interval $[\zjhata, \zjhatp] \cup [\zjhatpp, \zjhatb]$ and $[\zjhatp, \zjhatpp]$, respectively. Since $\nu(\barstate_0) \geq \revenue(\barstate_0)$, we know that the interval $[\zjhata, \zjhatb] \subseteq [\zkl, \zkph]$, where $\zkl$ and $\zkph$ are given by \eqref{eq:zkl_zkph}. Moreover,  from Lemma \ref{lemma:double_interval}, we know that there must exist $\px>0$ and $\py>0$ that satisfies \eqref{eq:px_py} given the interval $[\zjhata, \zjhatb] \subseteq [\zkl, \zkph]$.   

From Lemma \ref{lemma:extend_interval}, we know that there exists $\epsilon \in (0, \epa)$ (resp. $\epsilon \in (0, \epb)$) such that $[\zjhata- \epsilon, \zjhatb]$ (resp. $[\zjhata, \zjhatb+\epsilon]$) is also associated with a double-interval structure with atoms $x, y$. Consequently, the interval $[\zjhata, \zjhatb] \subseteq [\zkl, \zkph]$ being associated with a double-interval structure implies that one of the following three cases holds: 
\begin{enumerate}
\item[(a)] $\epa \leq \zjhata - \zkl$, i.e. there exists $\zdag \in (\zkl, \zjhata)$ such that 
\begin{align}\label{eq:case_a}
\mathbb{E}_F[S_0 | \zdag \leq S_0 \leq \zjhatb] = x. 
\end{align}
\item[(b)] $\epb \leq \zkph - \zjhatb$, i.e. there exists $\zddag \in (\zjhatb, \zkph)$ such that 
\[\mathbb{E}_F[S_0 | \zjhata \leq S_0 \leq \zddag] = y.\] 
\item[(c)] $\epa \geq \zjhata - \zkl$, $\epb \geq \zkph - \zjhatb$, and thus the interval $[\zkl, \zkph]$ can be generated by a double interval structure associated with atoms $x, y$. 
\end{enumerate} 

In case (a), \eqref{eq:case_a} implies that 
\[\mathbb{E}_F[S_0 | \zkl \leq S_0 \leq y] < \mathbb{E}_F[S_0 | \zdag \leq S_0 \leq \zjhatb] = x = \barstate_0[\kltangent],\]
which contradicts \eqref{subeq:C3} in condition (C4). Therefore, under condition (C4), case (a) does not hold. 

Similarly, in case (b), we have
\[\mathbb{E}_F[S_0 | x \leq S_0 \leq \zkph] > \mathbb{E}_F[S_0 | \zjhata \leq S_0 \leq \zddag] = y = \barstate_0[\klptangent],\]
which also contradicts \eqref{subeq:C3} in condition (C4), and thus case (b) does not hold. 

Moreover, case (c) implies that there exists $\px', \py'$ such that 
\begin{subequations}
\begin{align}
    &\px'+\py' = F(\zkph)- F(\zkl), \label{subeq:sum_atoms_prob_p}\\
&x \px'+ y \py' = \int_{\zkl}^{\zkph} z dF(z), \label{subeq:sum_atoms_expec_p}\\
& (F^{-1}(F(\zkl) + \px') - x) \px'  < \int_{\zkl}^{F^{-1}(F(\zkl) + \px')} (F(z) - F(\zkl))dz, \label{subeq:MPC_p}\\
&\px'>0, \quad \py'>0. 
\end{align}
\end{subequations}
We denote $\zmean = \int_{\zkl}^{\zkph} z dF(z)$. Then, from \eqref{subeq:sum_atoms_prob_p} and \eqref{subeq:sum_atoms_expec_p}, we have: 
\[\px'= \frac{y \left(F(\zkph) - F(\zkl) \right)-\zmean}{y-x}.\]
In case (c), we have $\px' \in (0, 1)$ since cases (a) and (b) do not happen. Thus, \eqref{subeq:MPC_p} indicates that 
\begin{align*}
&(F^{-1}(F(\zkl) + \px') - x) \px' = \left(F^{-1}\left( \frac{y F(\zkph) - xF(\zkl) - \zmean}{y-x}\right) - x\right) \frac{y \left(F(\zkph) - F(\zkl) \right)-\zmean}{y-x}\\
< & \int_{\zkl}^{F^{-1}\left( \frac{y F(\zkph) - xF(\zkl) - \zmean}{y-x}\right)} (F(z) - F(\zkl)) dz, 
\end{align*}
which contradicts \eqref{subeq:C3} of condition (C4). Thus, under (C4), all three cases cannot hold, which implies that there exists an optimal monotone partitional information mechanism.
\hfill $\square$

\medskip 
The proof of Proposition \ref{prop:p_y} directly builds on Lemma \ref{lemma:equivalence} and \cite{candogan2019persuasion}, and thus is omitted.

\end{document}